\RequirePackage{fix-cm}
\documentclass[smallcondensed]{svjour3}       
\smartqed  
\usepackage{graphicx}
\usepackage{natbib}
\usepackage{epsfig}
           \usepackage{color}

      \newcommand{\apj}{ApJ}                          
      \newcommand{\apjl}{ApJL}                        
      \newcommand{\aap}{A\&A}                   
      \newcommand{\araa}{ARA\&A}                      
      \newcommand{\aapr}{A\&A Rev.}             
      \newcommand{\apss}{Ap\&SS}                      
      \newcommand{\pasj}{PASJ}                        
      \newcommand{\nat}{Nature}                       
      \newcommand{\jaa}{JAA}                          
      \newcommand{\baas}{BAAS}                        
      \newcommand{\ssr}{Space Sci. Rev.}              
      \newcommand{\solphys}{Sol. Phys.}               
      \newcommand{\jgr}{JGR}
      \newcommand{\raa}{RAA}
      \newcommand{\memsai}{Mem. Soc. Astron. Ital.}
      \newcommand{\kms}{km\,s$^{-1}$}

      \newcommand{\as}{$^{\prime\prime}$}
      \newcommand{\lam}{$\lambda$}
      \def\ion#1#2{#1\,{\sc #2}}

\def\goes{{\sl GOES}}
\def\yohkoh{{\sl Yohkoh}}

\def\hinode{{\sl Hinode}}
\def\soho{{\sl SOHO}}
\def\trace{{\sl TRACE}}
\def\stereo{{\sl STEREO}}
\def\sdo{{\sl SDO}}
\def\rhessi{{\sl RHESSI}}
\def\iris{{\sl IRIS}}

\def\halpha{H$\alpha$}
\def\kms{km~s$^{-1}$}

\newcommand{\ltsim}{\raisebox{-1.0ex}{$\stackrel{\textstyle<}{\sim}$}}
\begin{document}

\title{Solar Coronal Jets: Observations, Theory, and Modeling}

\titlerunning{Solar Coronal Jets: Observations, Theory, \& Modeling}

\author{N. E. Raouafi$^1$ \and S. Patsourakos$^2$ \and E. Pariat$^3$ \and P. R. Young$^4$ \and A. C. Sterling$^5$ \and A. Savcheva$^6$ \and M. Shimojo$^7$ \and F. Moreno-Insertis$^8$ \and C. R. DeVore$^{9}$ \and V. Archontis$^{10}$ \and T. T\"or\"ok$^{11}$ \and H. Mason$^{12}$ \and W. Curdt$^{13}$ \and K. Meyer$^{14}$ \and K. Dalmasse$^{3,15}$ \and Y. Matsui$^{16}$ }

\authorrunning{Raouafi et al. 2016}

\institute{N. E. Raouafi \at
               \email{NourEddine.Raouafi@jhuapl.edu}        \\
           \and
             $^1$The Johns Hopkins University Applied Physics Laboratory, Laurel, MD 20723, USA \\
           $^2$Department of Physics, University of Ioannina, Ioannina, Greece\\
           $^3$LESIA, Observatoire de Paris, Meudon, France\\
           $^4$College of Science, George Mason University, Fairfax, VA, USA \\ 
           $\mbox{\ \ }$NASA/Goddard Space Flight Center, Code 671, Greenbelt, MD 20771, USA\\
           $^5$NASA/Marshall Space Flight Center, Huntsville, Alabama, USA\\
           $^6$Harvard-Smithsonian Center for Astrophysics, Cambridge, MA, USA\\
           $^7$National Astronomical Observatory of Japan, Mitaka, Tokyo, Japan\\
           $^8$Instituto de Astrofísica de Canarias, La Laguna, Tenerife, Spain \\
           $^9$Heliophysics Science Division, NASA Goddard Space Flight Center, Greenbelt, MD, USA\\
           $^{10}$School of Mathematics and Statistics, University of St. Andrews, St. Andrews, UK\\
           $^{11}$Predictive Science Inc., 9990 Mesa Rim Rd., Ste. 170, San Diego, CA 92121, USA\\
           $^{12}$DAMTP, Centre for Mathematical Sciences, University of Cambridge, Cambridge, UK\\
           $^{13}$Max-Planck-Institut f\"ur Sonnensystemforschung, G\"ottingen, Germany\\
           $^{14}$Division of Computing and Mathematics, Abertay University, Dundee, UK\\
           $^{15}$CISL/HAO, NCAR, P.O. Box 3000, Boulder, CO 80307-3000, USA\\
           $^{16}$Department of Earth and Planetary Science, University of Tokyo, Tokyo, Japan
           }

\date{Received: date / Accepted: date}

\maketitle

\begin{abstract}
Coronal jets represent important manifestations of ubiquitous solar transients, which may be the source of significant mass and energy input to the upper solar atmosphere and the solar wind. While the energy involved in a jet-like event is smaller than that of ``nominal" solar flares and coronal mass ejections (CMEs), jets share many common properties with these phenomena, in particular, the explosive magnetically driven dynamics. Studies of jets could, therefore, provide critical insight for understanding the larger, more complex drivers of the solar activity. On the other side of the size-spectrum, the study of jets could also supply important clues on the physics of transients close or at the limit of the current spatial resolution such as spicules. Furthermore, jet phenomena may hint to basic process for heating the corona and accelerating the solar wind; consequently their study gives us the opportunity to attack a broad range of solar-heliospheric problems.
\keywords{Plasmas \and Sun: activity \and Sun: corona \and Sun: magnetic fields \and Sun: UV radiation \and Sun: X-rays}
\end{abstract}


\section*{Abbreviations}
{\small
\begin{tabular}{ll || ll}
AIA		& Atmospheric Imaging				& PCH(s)		& Polar coronal hole(s) \\
		& Assembly \citep{2012SoPh..275...17L}	& QS		& Quiet Sun\\
AR(s)	& Active region(s)					& {\rhessi}		& Reuven Ramaty High \\
AU		& Astronomical unit					&			& Energy Solar Spectroscopic \\
BP(s)	& Bright point(s)					&			& Imager \citep{2002SoPh..210....3L} \\
CBP(s)	& Coronal bright point(s)				& {\sdo}		& Solar Dynamics Observatory \\
CDS		& Coronal Diagnostic				&			& \citep{2012SoPh..275....3P} \\
		& Spectrometer						& SECCHI	& Sun Earth Connection   \\
		& \citep{harrison95}					&			& Coronal and Heliospheric \\
CH(s)	& Coronal hole(s)					&			& Investigation \\
ECH(s)	&  Equatorial coronal hole(s)			&			& \citep{2008SSRv..136...67H} \\
EIS		&  EUV Imaging Spectrometer			& {\soho}		& Solar and Heliospheric  \\
		&  \citep{culhane07}					&			& Observatory \\
EIT		& EUV Imaging Telescope				&			& \citep{1995SoPh..162....1D} \\
		& \citep {1995SoPh..162..291D}		& {\stereo}		& Solar TErrestrial RElations \\
EUV		& Extreme ultraviolet					&			& Observatory \\
EUVI		& Extreme UV Imager				&			& \citep{2008SSRv..136....5K} \\
		& \citep{2004SPIE.5171..111W}		& SUMER		& Solar UV Measurements of \\
FOV		& Field of view						&			& Emitted Radiation \\
{\hinode}	& Solar-B pre-launch					&			& spectrometer \\
		& \citep[][]{2007SoPh..243....3K}		&			& \citep{1995SoPh..162..189W} \\
HMI		& Helioseismic and Magnetic			& SW		& Solar wind \\
		& Imager \citep{2012SoPh..275..207S}	& SXR(s)		& Soft X-ray(s) \\
HXR(s)	& Hard X-ray(s)						& SXT		& Soft X-ray Telescope \\
{\iris}		& Interface Region Imaging			&			& \citep{1991SoPh..136...37T} \\
		& Spectrometer						& {\trace}		& Transition Region And \\
		& \citep{2014SoPh..289.2733D}		&			& Coronal Explorer \\
ISSI		& International Space Science			&			& \citep{1999SoPh..187..229H} \\
		& Institute, Bern, Switzerland			& UV			& Ultraviolet \\
JBP(s)	& Jet-base bright point(s)				& UVCS  		& UV Coronagraph  \\		 
LASCO	& Large Angle and					&			& Spectrometer \\
		& Spectrometer COronagraph			&			& \citep {1995SoPh..162..313K} \\
		& \citep{1995SoPh..162..357B}			& WL		& White light \\
LOS		& Line of sight						& XRT		&  X-ray Telescope \\
MDI		& Michelson Doppler Imager			&			& \citep{2007SoPh..243...63G} \\
		& \citep{1995SoPh..162..129S}			& {\yohkoh}	& Solar-A pre-launch \\
MHD		& Magnetohydrodynamic				&			& \citep{1991SoPh..136....1O}
\end{tabular}
}

\section{Introduction: Brief Historical Aspect of Coronal Jets}

The wide variety of transient phenomena in the solar corona first became apparent in the 1970s with the discovery of coronal transients in ground-based, green-line observations \citep{1973SoPh...31..449D}; discovery of macro-spicules in Skylab EUV observations \citep{1975ApJ...197L.133B,1976ApJ...203..528W}; and the discovery of explosive events \citep{1980HiA.....5..557B}. These discoveries led to speculations on the role these transients, particularly coronal jets, play in the coronal heating and SW acceleration \citep{1978BAAS...10R.416B,1983ApJ...272..329B}. 

Coronal jets were seen by the U.S. Naval Research Laboratory (NRL)/UV telescope onboard the space shuttle in the 1980s and later by the Japanese spacecraft {\it{Yohkoh}} in the early 1990s. {\it{Yohkoh}}/SXT observations unveiled the largest, most energetic category of coronal jets \citep[e.g.,][]{1992PASJ...44L.173S,1992PASJ...44L.161S,1996PASJ...48..123S,1998SoPh..178..379S,2001ApJ...550.1051S}. Since then jet-like phenomena have occupied a center stage in coronal observational, theoretical, and state-of-the-art numerical analyses.

Coronal jets are a near-ubiquitous solar phenomenon regardless of the solar cycle phase. They are particularly prominent in CHs (e.g., open magnetic field regions) because of the darker background. X-ray and EUV observations reveal their collimated, beam-like structure, which are typically rooted in CBPs. Their signature can be traced out to several Mm in X-ray/EUV observations, up to several solar radii in WL images \citep[e.g.,][]{1998ApJ...508..899W}, and also at $>1$~AU in in-situ measurements \citep[e.g.,][]{2006ApJ...639..495W,2006ApJ...650..438N,2008ApJ...675L.125N,2012ApJ...750...50N}. The unceasing improvements in spatial and temporal resolution of data recorded over the last three decades by different space missions (e.g.,  {\it{Yohkoh}}, {\it{SOHO}}, {\it{STEREO}}, {\it{Hinode}}, {\it{SDO}}, {\it{IRIS}}) provide unprecedented details on the initiation and evolution of coronal jets. The recent imaging and spectroscopic observations unveiled jet characteristics that could not be observed with lower spatio-temporal resolution (e.g., morphology, dynamics, and their connection to other coronal structures).

Despite the major advances made on both observational and theoretical fronts, the underlying physical mechanisms, which trigger these events, drive them, and influence their evolution are not completely understood. Recent space missions (e.g., {\it{STEREO}}, {\it{Hinode}}, and {\it{SDO}}) represent important milestones in our understanding of the fine coronal structures, particularly coronal jets. The observations show that jets can be topologically complex and may contribute to the heating of the solar corona and the acceleration of the SW.

The present review is the result of work performed by the ISSI International Team on ``Solar Coronal Jets". We, the authors, met at ISSI twice (March 2013 and March 2014) and had intense discussions on the nature of coronal jets, their triggers, evolution, and contribution to the heating and acceleration of the coronal and SW plasma, from both observational and theoretical point of views. We do not claim that this review is in any way exhaustive but it presents a thorough overview on the wealth of observations available from different space missions as well as state-of-the-art models of these coronal structures. The work we accomplished addressed many questions regarding coronal jets, but also left many others unanswered and raised several other outstanding issues for these prominent structures. Future missions with better observational capabilities along with the maturing of existing numerical codes will help address these questions and may lead to a yet better understanding of coronal jets and their role as a component of the magnetic activity of the Sun.

In the present review, we mainly deal with observations from the {\it{SOHO}} era to the present.
{\it{Yohkoh}}/SXT observations led to important insights and laid the seeds of
major progress made during the later decades  \citep[see, e.g.,][]{1992PASJ...44L.173S}.
Chromospheric jets such as spicules may belong to the small-size end
of jet phenomena and may be related to our topic. We feel, however, that such studies
are beyond the objective of our review of coronal jets and should be excluded
here. The vast literature on spicules and other chromospheric jets includes
reviews of \citet{1968SoPh....3..367B,1972ARA&A..10...73B}, \citet{1974soch.book.....B}, \citet{1974IAUS...56....3M}, \citet{2000SoPh..196...79S}, and \citet{2012SSRv..169..181T}.

\section{Early Imaging and Coronagraphic Observations of Jets}

This section contains a description of jet observations carried out by EUV and SXR imagers and WL coronagraphs on-board various space-borne observatories since the early 90's. The improvements in terms of important instrumental parameters (e.g., spatial and temporal resolution, temperature coverage, etc.) have been and continue to be key factors in advancing our understanding of coronal jets. These instruments include {\it{Yohkoh}}/SXT; {\it{SOHO}}/EIT and LASCO; {\it{TRACE}}; {\it{RHESSI}}; {\it{Hinode}}/XRT; SECCHI/EUVI and COR1 and COR2 coronagraphs of  {\it{STEREO}}; {\it{SDO}}/AIA and HMI.
Key parameters of the imagers and coronagraphs are listed in Tables \ref{table:Instruments_image} and  \ref{tab:secchi}, respectively.

\begin{table}[!h]
\begin{center}
\caption{Summary of Imaging Instrument Capabilities for Jet Observations.}
\label{table:Instruments_image}
\begin{tabular}{lcccc}
\hline\noalign{\smallskip}
Instrument & Resolution & FOV & Cadence & Temperature \\
 & [$^{\prime\prime}$/pix] & [arcsec] & [s] & coverage [$\log{T/K}$] \\
\hline
{\it{Yohkoh}}/SXT &  2.5/5 & max full disk & min 20 & $6-7.5$ \\
{\it{SOHO}}/EIT & 2.5 & full disk & min 600 & $4.9-6.4$  \\
{\it{TRACE}} & 0.1 & $8.5\times8.5$ arcmin & $3 - 30$ & $3.60-7.41$  \\
{\it{RHESSI}} & $2-36$ & full disk & 2 &  $>7$ \\
{\it{Hinode}}/XRT &  1.028 & max full disk & min 10 & $6.1-7.3$  \\
{\it{Hinode}}/SOT/BFI & 0.0533 & max 218\as$\times$109\as & max 1.6 & $-$ \\
{\it{STEREO}}/EUVI & 1.6  &  full disk  & 150  & $4.9-6.4$   \\
{\it{SDO}}/AIA & 0.6 & full disk & 12 & $3.7-7.3$  \\
{\it{PROBA2}}/SWAP & 3.16 & full disk & $60$ & $\sim6$  \\
{\it{IRIS}} &  0.33-0.4 &  max 130\as$\times$175\as  &  2 &  $3.7-7.0$  \\
\noalign{\smallskip}\hline
\end{tabular}
\end{center}
\end{table}

\begin{table}[!h]
\begin{center}
\caption{Main Characteristics of  Coronagraphs on-board {\it{SOHO}} (C1, C2, \& C3) and {\it{STEREO}} (COR1 \& COR2).}
\label{tab:secchi}
\begin{tabular}{lllll}
\hline\noalign{\smallskip}
Instrument & Pixel size & FOV                & Bandpass & Cadence  \\
                  &  [arcsec]   & [R$_{\odot}$] &                 &  [min]  \\
\noalign{\smallskip}\hline\noalign{\smallskip}
C1   &  5.6 & 1.1-3  &  broad-band channel  & 10 \\
       &         &           &  and emission lines in visible &    \\
C2    &  11.4 & 1.5-6  &  broad-band channel in visible & 20\\
C3    &  56 & 3.7-30  &  broad-band channel in visible & 30 \\
COR1 &  3.75 & 1.4-4  &  broad-band channel in visible & 5\\
COR2 & 14.7 & 2.5-15 &  broad-band channel in visible  & 15\\
\noalign{\smallskip}\hline
\end{tabular}
\end{center}
\end{table}

\subsection{EIT and LASCO Observations}

\begin{figure*}
\centering
\parbox{0.45\textwidth}{  \includegraphics[width=0.4\textwidth]{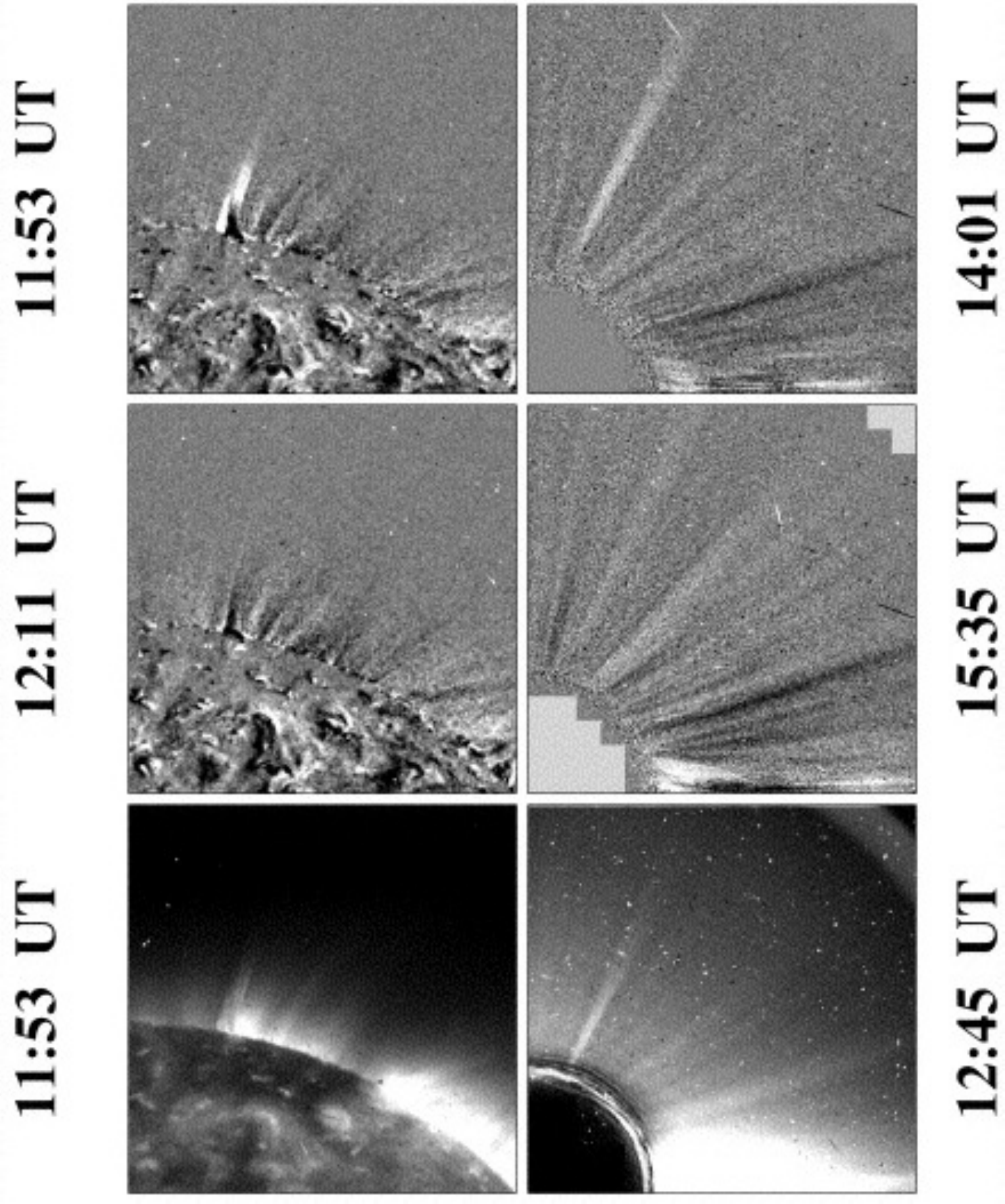}}
\parbox{0.45\textwidth}{\caption{EIT-LASCO observations of a PCH jet. Left (right) panels show EIT (LASCO/C2) images of the jet. The bottom row contains plain images whereas the remaining rows show difference images to enhance the jet visibility. Adapted from \citet{1998ApJ...508..899W}. \label{fig:1EITLASCO} }}
\end{figure*}

The first combined analysis of  EUV and WL coronal jets by instruments onboard {\it{SOHO}} was carried out by \citet{1998ApJ...508..899W}. A set of 27 PCH jets were analyzed using EIT and LASCO/C2 observations under solar minimum conditions (Fig.~\ref{fig:1EITLASCO}). The sources of these jets on the solar disk were near flaring BPs in the PCHs. On average there were $3-4$ such jets per day. The WL counterparts of these jets had angular extent in the range of $2^{\circ}-4^{\circ}$. These events were characterized by leading edge speeds in the range 400-1100~km~s$^{-1}$ and significantly lower centroid (i.e., bulk) speeds of $\approx\!250$~km~s$^{-1}$. The latter suggests jet deceleration to the ambient SW possibly due to the action of a drag-related force between $1-2~\mathrm{R_{\odot}}$. The SW drag hypothesis is also supported by kinematics fitting of five other coronal jets observed by EIT and LASCO \citep{1999SSRv...87..219K,1999ApJ...523..444W}.

\citet{2002ApJ...575..542W} analyzed  LASCO observations of WL jets during solar maximum conditions. Several important differences with respect to coronal jets observed during solar minimum conditions were found. Solar  maximum coronal jets originated from a wider range of latitudes compared to their solar minimum counterparts. The former did not only originate from polar regions but also from ARs and regions close to the boundaries of ECHs. In addition, during solar maximum coronal jets were wider ($3^{\circ}-7^{\circ}$) and brighter than solar minimum jets, which suggests that they could be more massive. Finally, the solar maximum jet average bulk speed in the LASCO/C2 FOV was  $\approx\!600$~km~s$^{-1}$.

\citet{2002ESASP.508..379B} determined the radial profiles of the electron density in four coronal jets observed during solar maximum conditions by LASCO/C2. The background-subtracted WL radiances of the observed jets were fitted with tube-like models of the jets' envelopes. The resulting density profiles of the observed jets in the range  $3-6~\mathrm{R_{\odot}}$ gave rise to densities of $\sim(2-10)\times10^5$ and $\sim(0.3-1.5)\times10^5$ $\mathrm{{cm}^{-3}}$ at 3 and 6 $\mathrm{R_{\odot}}$, respectively. These density values are significantly higher  (up to factor 50) than the densities of the ambient corona at the same heights.

\subsection{{\it{TRACE}} Observations}
\begin{figure*}
\begin{center}
\includegraphics[height=0.45\textwidth]{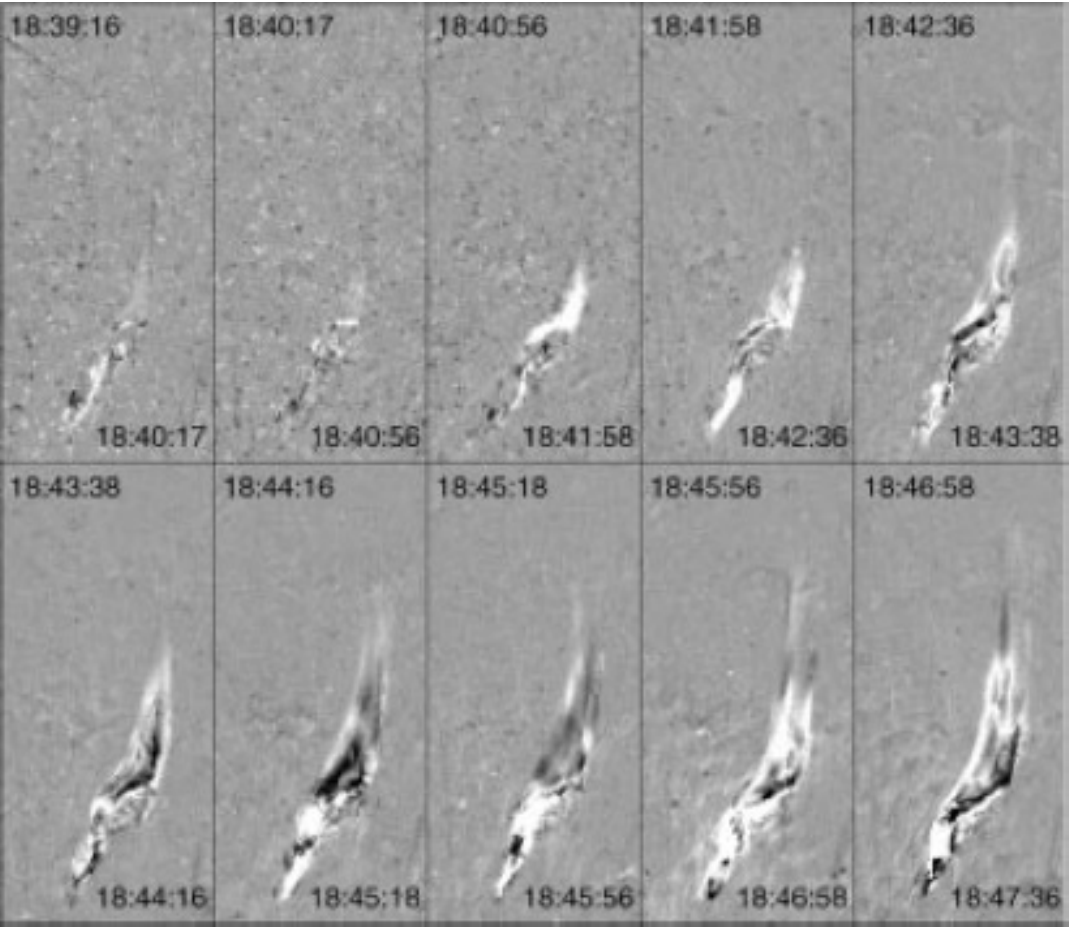}
  \includegraphics[height=0.45\textwidth]{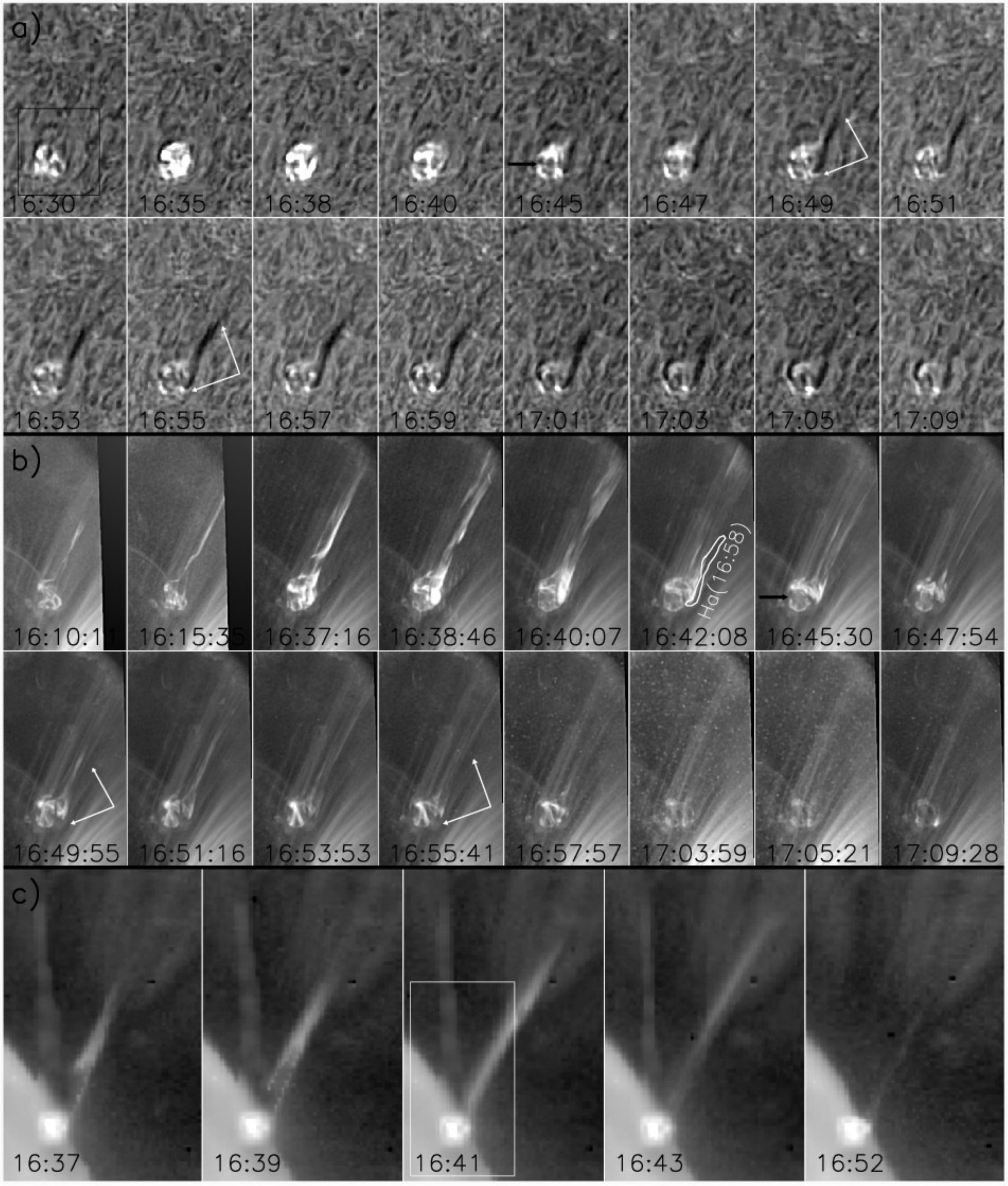}
 \caption{{\bf{(Left)}} running-difference snapshots during evolution of a two-sided coronal jet observed in the 171 \AA \, channel of {\it{TRACE}}. From \citet{1999SoPh..190..167A}. {\bf{(Right)}} evolution of coronal jet in H-$\alpha$ (two upper rows); in {\it{TRACE}} 171 \AA \, (two middle panels) and in SXT (lower panels; smaller FOV). From \citet{2007A&A...469..331J}.
\label{fig:1TRACE} }
\end{center}
\end{figure*}

The first detailed study of coronal jets observed by {\it{TRACE}} was reported by \citet{1999SoPh..190..167A}. The high temporal and spatial resolution as well as the multi-temperature coverage  of {\it{TRACE}} observations showed the co-existence of both cool and hot emitting plasmas in coronal jets. The cool material was  detected either in absorption in coronal channels (e.g., 171 \AA \,) or in emission in the Ly-$\alpha$ channel. The cool and hot emissions in the observed jets were not strictly co-spatial \citep[see also][]{2007A&A...469..331J}. Jets with both one-sided (single spire) anemone-type and two-sided (two spires) morphology (see for example Fig.~\ref{fig:1TRACE}) were observed. Finally, evidence of rotation and bifurcation was seen in one of the observed jets.

In a  series of studies, {\it{TRACE}} observations of chromospheric surge-like and coronal jets were combined with  co-temporal observations of the photospheric magnetic field \citep[e.g.,][]{2003ApJ...584.1084C,2004ApJ...610.1136L,2007A&A...469..331J, 2008A&A...478..907C,2009SoPh..255...79C}. Such studies supplied  important constraints on the magnetic environment and the formation mechanism(s) of the observed jets. EUV and UV jets were observed above sites of flux cancellation or emergence in the photosphere \citep[see, e.g.,][]{2008A&A...478..907C}. Both cool ($\approx\!{10}^{4}-{10}^{5}$ K) and hot ($\approx\!{10}^{6}$ K) plasma emissions were observed, which presumably resulted from the photospheric cancellation/flux-emergence episodes.  The photospheric cancellation events were associated with cool transition region jets carrying an estimated mass of $\approx1.7-4.6\times{10}^{13} $ g. Given a birth-rate of $\approx1$ jet per hour the mass of a typical prominence could have been accumulated in a matter of few days \citep{2003ApJ...584.1084C}.

\begin{figure*}
\begin{center}
\parbox{0.99\textwidth}{  \includegraphics[height=0.6\textwidth]{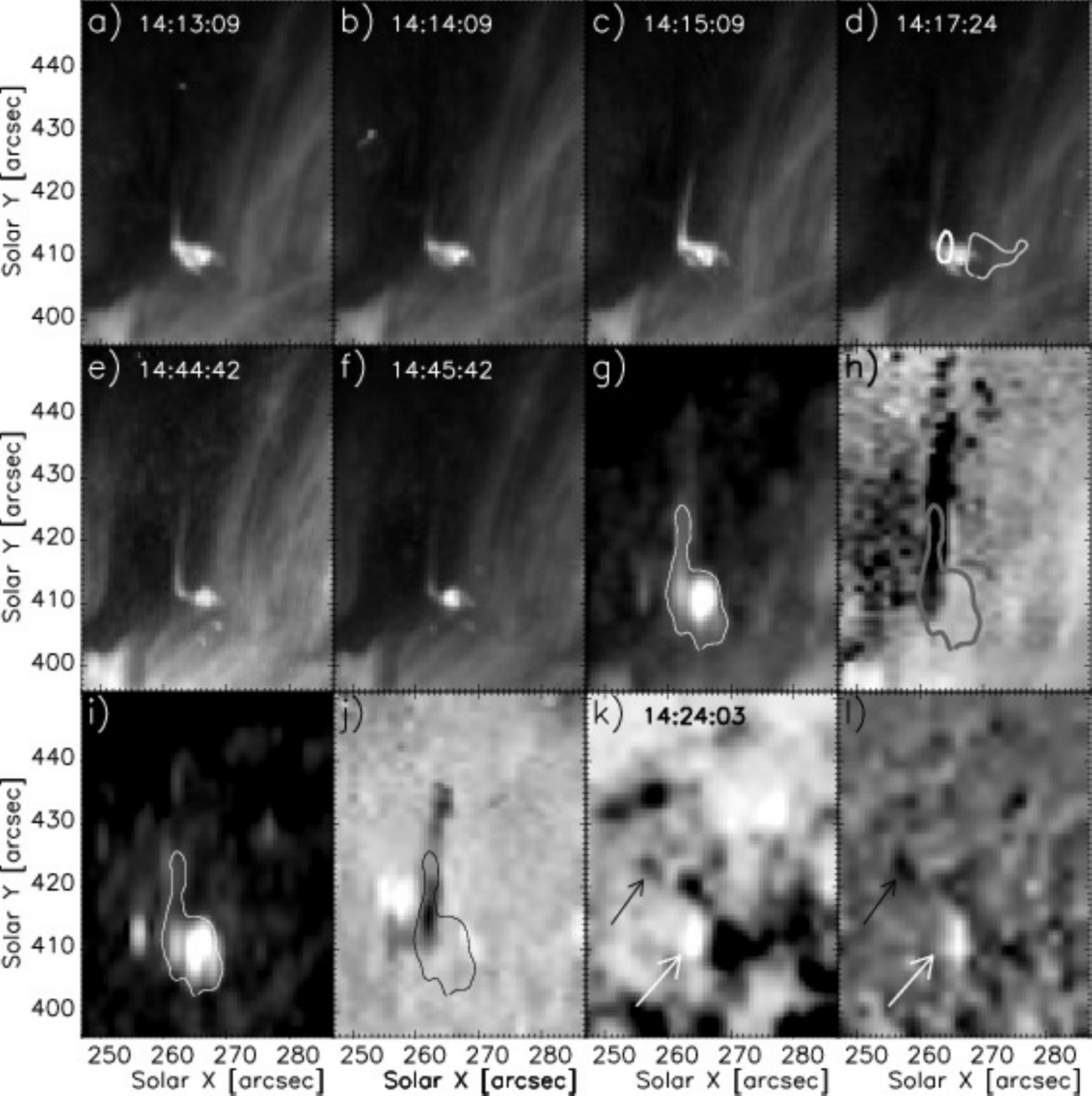}
  \includegraphics[height=0.6\textwidth]{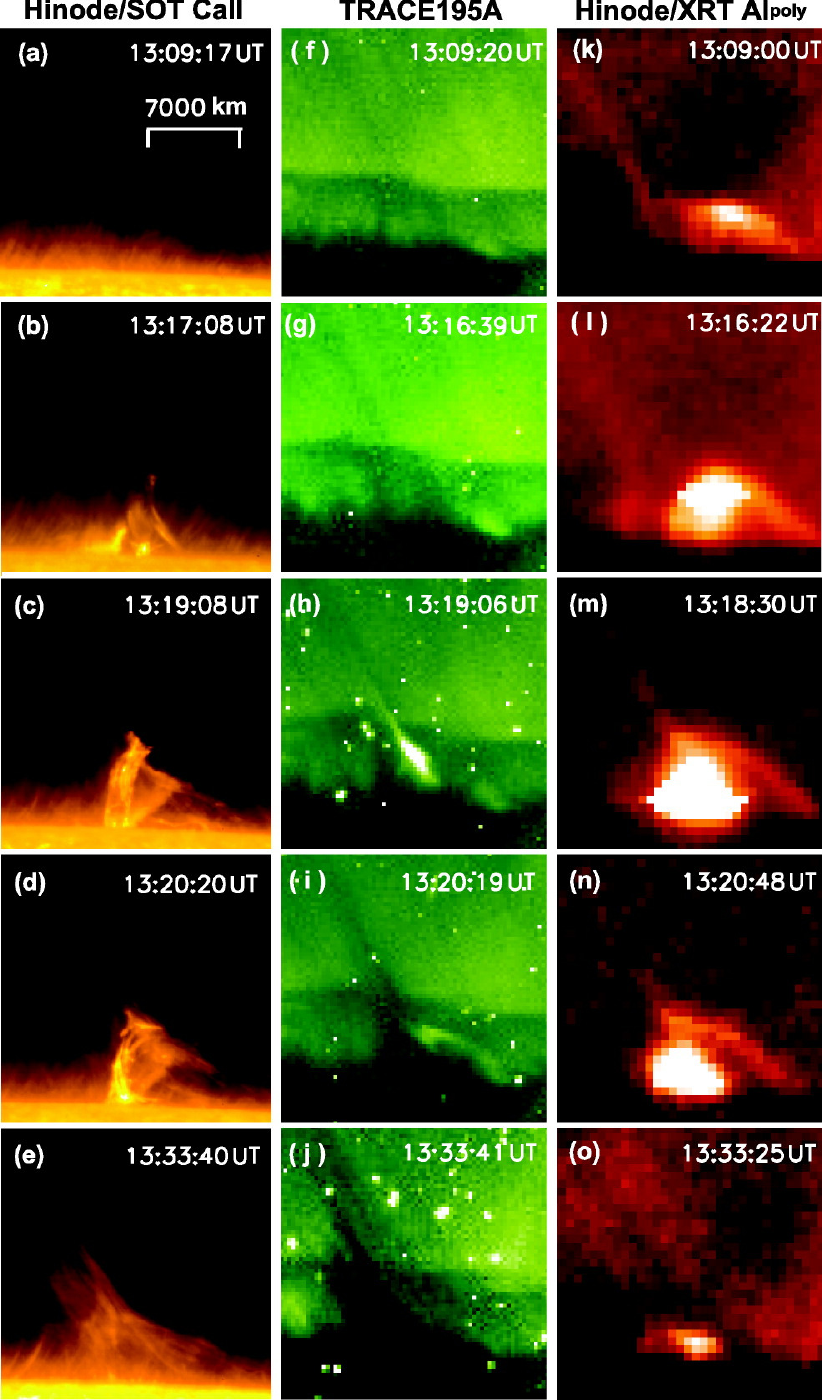}}
\caption{Left panel: evolution of a coronal jet observed in {\it{TRACE}} 171 \AA \, channel (a-f), SUMER (g-j) and MDI (k and l; the two arrows in this panel show the emerging magnetic flux). From \citet{2009A&A...506L..45G}. Right panel:evolution of a coronal jet observed by the {\it{Hinode}} Solar Optical Telescope (SOT) in Ca II (left column), {\it{TRACE}} 195 \AA \, (middle column) and {\it{Hinode}} XRT (right column). From \citet{2008ApJ...683L..83N}.
\label{fig:2TRACE}  }
\end{center}
\end{figure*}

EUV jets observed by {\it{TRACE}} in ARs often have SXR counterparts as observed by either SXT or XRT although there is not always a one-to-one correspondence \citep[e.g.,][] {1999SoPh..190..167A,2004ApJ...610.1136L,2007A&A...469..331J,2007PASJ...59S.763K,2008ApJ...683L..83N,2009A&A...506L..45G}. For the EUV events that do have SXR counterparts, a decent spatial correspondence is frequently observed (e.g., right panels of Figs.~\ref{fig:1TRACE}, \ref{fig:2TRACE}). As a matter of fact a joint {\it{TRACE}}-XRT study of coronal jets showed they have comparable speeds  of $ 90-310~\mathrm{{km~s}^{-1}}$, lifetimes of $100-2000~\mathrm{s}$ and sizes of $(1.1-5.0)\times{10}^{5}$~km \citep[][] {2007PASJ...59S.763K}.

{\it{TRACE}} jets were also observed over sites of microflares observed by {\it{RHESSI}} in ARs \citep[e.g.,][]{2004ApJ...604..442L,2008ApJ...680L.149C}. \cite{2004ApJ...604..442L} found that almost half of the studied {\it{RHESSI}} microflares were associated with a {\it{TRACE}} jet. These findings support the hypothesis of coronal jet formation by magnetic reconnection. Note that the HXR sources had a loop-like appearance and were observed at the feet of the EUV jets.

\subsection{{\it{STEREO}} Observations}

\begin{figure*}
\parbox{0.68\textwidth}{  \includegraphics[height=0.27\textwidth]{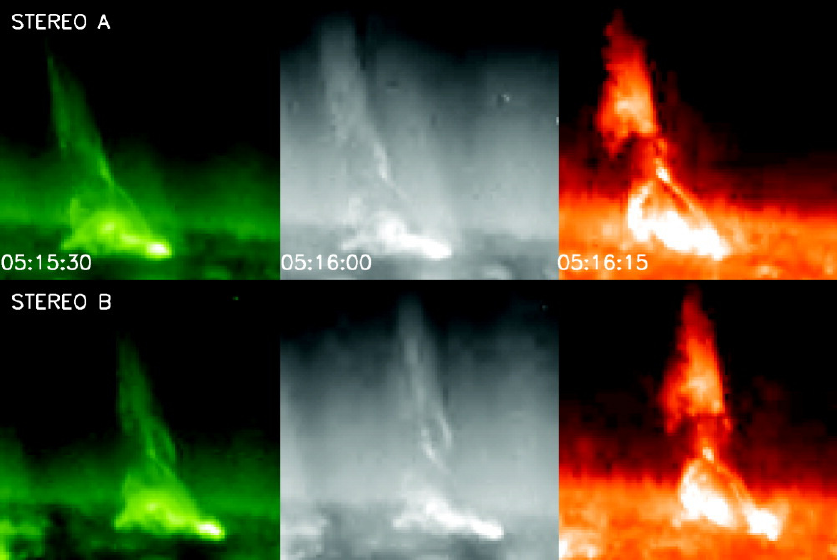}
\includegraphics[height=0.27\textwidth]{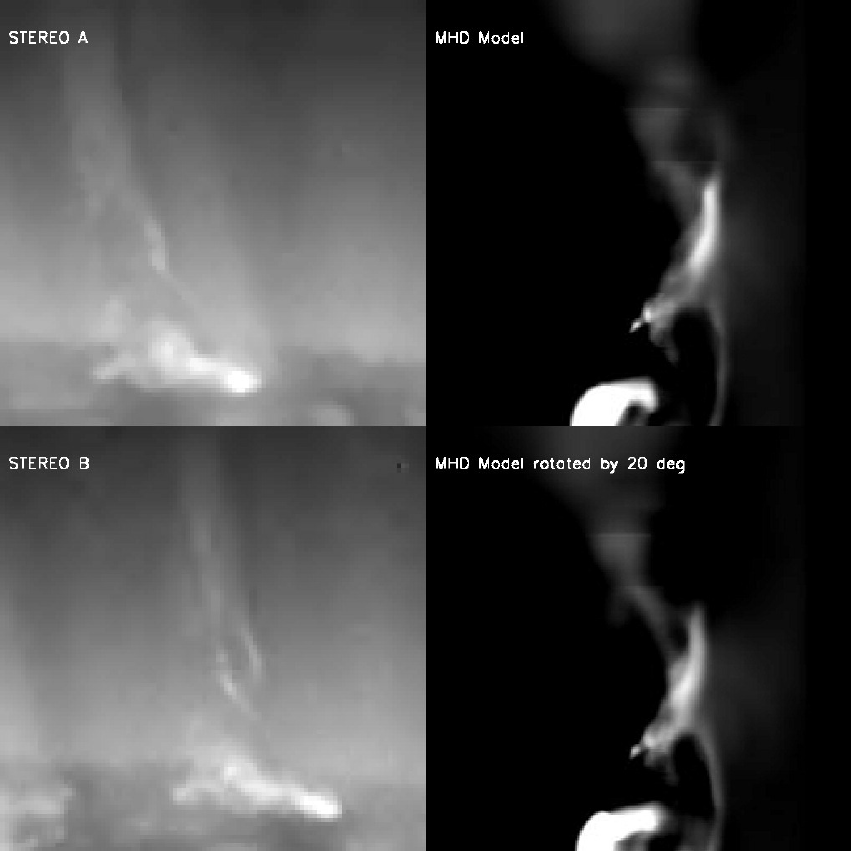}}
\parbox{0.3\textwidth}{ \caption{(Left) {\it{STEREO}} observations of a helical jet in a PCH (from left to right 195, 171 and 304~\AA \, EUVI images).  (Right) 171  \AA \, images of the jet compared with synthetic images from an MHD model. From \citet{2008ApJ...680L..73P}.
\label{fig:1STEREO}
}}
\end{figure*}

The first {\it{STEREO}} observations of jets were described by \citet{2008ApJ...680L..73P}. This study provided clear evidence of helical structure in a polar coronal jet observed by EUVI onboard STEREO-A and -B (STA and STB, respectively; see Fig.~\ref{fig:1STEREO}). The helical structure was observed edge-on and face-on from the two respective viewpoints during the untwisting of the rising jet structure. This supplied solid evidence for a ``true" helical structure something that was not possible to fully address with previous single-viewpoint observations. In addition, synthetic  images from a 3D MHD jet model \citep{2009ApJ...691...61P} based on magnetic twist were found in qualitative agreement with the reported {\it{STEREO}} observations (see right panels of Fig.~\ref{fig:1STEREO}).

\begin{figure*}
\begin{center}
\includegraphics[width=0.85\textwidth]{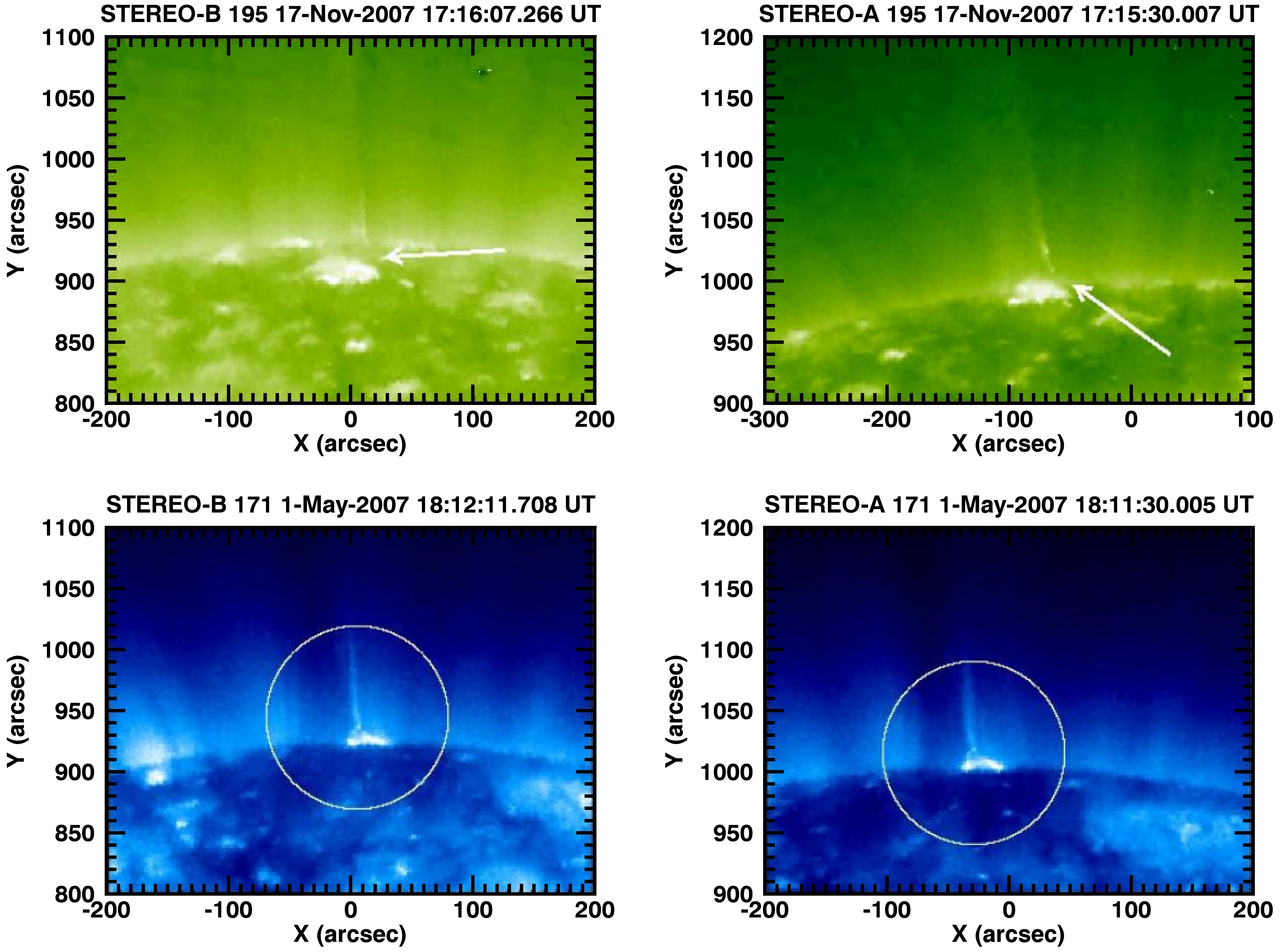}
\caption{EUVI {\it{STEREO}}-A and {\it{STEREO}}-B observations of an ``Eifel Tower"  (top) and of  ``$\lambda$" jet (bottom). Both jets occurred within a PCH. From \citet{2009SoPh..259...87N}.
\label{fig:2STEREO}
}
\end{center}
\end{figure*}

\citet{2009SoPh..259...87N} carried out a statistical survey of coronal jets (79 events) observed by SECCHI/EUVI and COR1 in both PCHs and ECHs. They found that about 40\% ($31/79$) of the observed jets by EUVI had a helical structure. Therefore, a helical structure can be considered a common element of coronal jets. Moreover, \textit{all} reported jets of this study were associated with a compact magnetic bipole with the resulting jets observed either on top or at the side of these bipoles (``Eiffel tower" and ``$\lambda$" jets, respectively; see Fig.~\ref{fig:2STEREO}). Note that SECCHI observations of jets suggest multipolar magnetic field settings \citep{2013SoPh..286..143F}.

A few (5/79) of the observed jets in the \citet{2009SoPh..259...87N} sample  had the appearance of a ``micro-CME", i.e., contained a small loop that eventually erupted while straightening  giving rise to a jet-like appearance \citep[see also][who termed them blow-out jets]{2010ApJ...720..757M}. Such blow-out jets often contain filament-like material in the erupting jet core as observed in the 304~\AA \, channel.  The association between blow-out jets and small CME-like eruptions was extended in a series of studies which combined {\it{STEREO}} with {\it{SDO}}, {\it{Hinode}} or PROBA2 observations. Typically one instrument provides a disk-view and the other a limb-view of the same event. \citet{2012ApJ...745..164S} found for a blow-out jet observed on disk by AIA which exhibited a bubble-like morphology when viewed off-limb by  {\it{STEREO}}. The bubble morphology is frequently observed in CMEs. \citet{2013ApJ...766....1L}  analyzed EUVI observations showing an EUV dimming  left behind by a jet observed off-disk with {\it{Hinode}}. The association between twisted mini-filament eruptions and blow-out jets was  also shown in \citet{2011ApJ...738L..20H,2013RAA....13..253H}. All these findings suggest that the blow-out jets have significant similarities with the larger-scale CMEs and hint at a scale-invariant eruptive solar phenomenon.

The width of EUV jets observed by EUVI ranges from down the instrument's spatial resolution (i.e., 1.6\as\ $\approx\!1150$~km) to few times ${10}^{3}-{10}^{4}$ km. By jet width we mean here the transverse spatial scale of the analyzed jet's envelope and this does not incorporate any of the  omni-present fine  structure seen in jet observations.

{\it{STEREO}} observations not only allowed to establish high correlations between EUV and WL jets \citep[$73-78\%$ of 10,912 jets observed by COR1,][]{2010SoPh..264..365P} but also to provide insight into the kinematics and speeds of  these events. Various methods were used  for this task: triangulation \citep[e.g.,][]{2008ApJ...680L..73P}, image stack-plots  \citep[e.g.,][]{2013ApJ...776...16P} and jet transit-times through the FOV of a given instrument \citep[e.g.,][]{2009SoPh..259...87N}. The resulting jet speeds are in the range of $\approx\!250-400$~km~s$^{-1}$ and of $\approx\!100-400$~km~s$^{-1}$ for EUV and WL jets, respectively.  From the statistical studies of \citet{2009SoPh..259...87N} and \citet{2010SoPh..264..365P} the average speeds of EUVI and COR1 jets are both around $\approx\!300-400$~km~s$^{-1}$. Note that most  of the speeds quoted above correspond to the propagation phase of jets (i.e., after their initiation). Before reaching the typical cruising speeds of few hundred km~s$^{-1}$, the magnetic structure that eventually gives rise to a jet ascending at a much smaller speed of typically few 10~km~s$^{-1}$  \citep[e.g.,][]{2008ApJ...680L..73P}. This kinematic behavior (slow rise followed by impulsive acceleration) is a characteristic of an instability taking place in a quasi-statically driven MHD system \citep{2009ApJ...691...61P}. PCH and ECH EUV jets have similar speeds as shown in \citet{2010AnGeo..28..687N}.

Ratios of EUVI channel intensities have been used to estimate jet temperatures. Temperatures of $0.8-1.3$~MK were found from 171/195 and 195/284 ratios \citep{2011AdSpR..48.1490N}, while 284/195 and  SXR ratios provided relatively higher temperatures of $1.6-2.0$~MK \citep{2013ApJ...776...16P}. This may in fact show that jets are not monolithic structures, but rather consist of  different plasma components at different temperatures.

Brightness evolution has been utilized along with a kinematic particle model based on  the ballistic assumption to infer the energetics of a large PCH jet observed by EUVI, COR1 and COR2. The jet kinetic energy and mass are found in the range $(0.21-2.4)\times{10}^{29}$~erg (i.e., microflare range) and $(0.32 - 1.8)\times{10}^{15}$~g, respectively \citep{2012A&A...538A..34F}. The initial jet density was estimated in the range $(0.8 - 5)\times{10}^{10}$~cm$^{-3}$. Another analysis using SECCHI and XRT observations of two jets, one blow-out and one standard, provided an energy budget (mechanical+radiative+enthalpy) of  $\approx\!2.0\times{10}^{27}$~erg for the blow-out jet, which is about an order of magnitude larger than that of the standard jet \citep{2013ApJ...776...16P}.

In a recent article, \citet{2015arXiv150801072N} studied the deflections of 79 PCH jets, at $\approx$1 and 2  $\mathrm{R_{\odot}}$ as observed by EUVI and COR1 respectively, and found that their propagation was not radial and larger in the north than in the south. These properties were used to constrain models of the large-scale configuration of the coronal magnetic field.

\subsection{{\it{RHESSI}} Observations}

\begin{figure*}
\begin{center}
\parbox{.68\textwidth}{\includegraphics[width=0.33\textwidth]{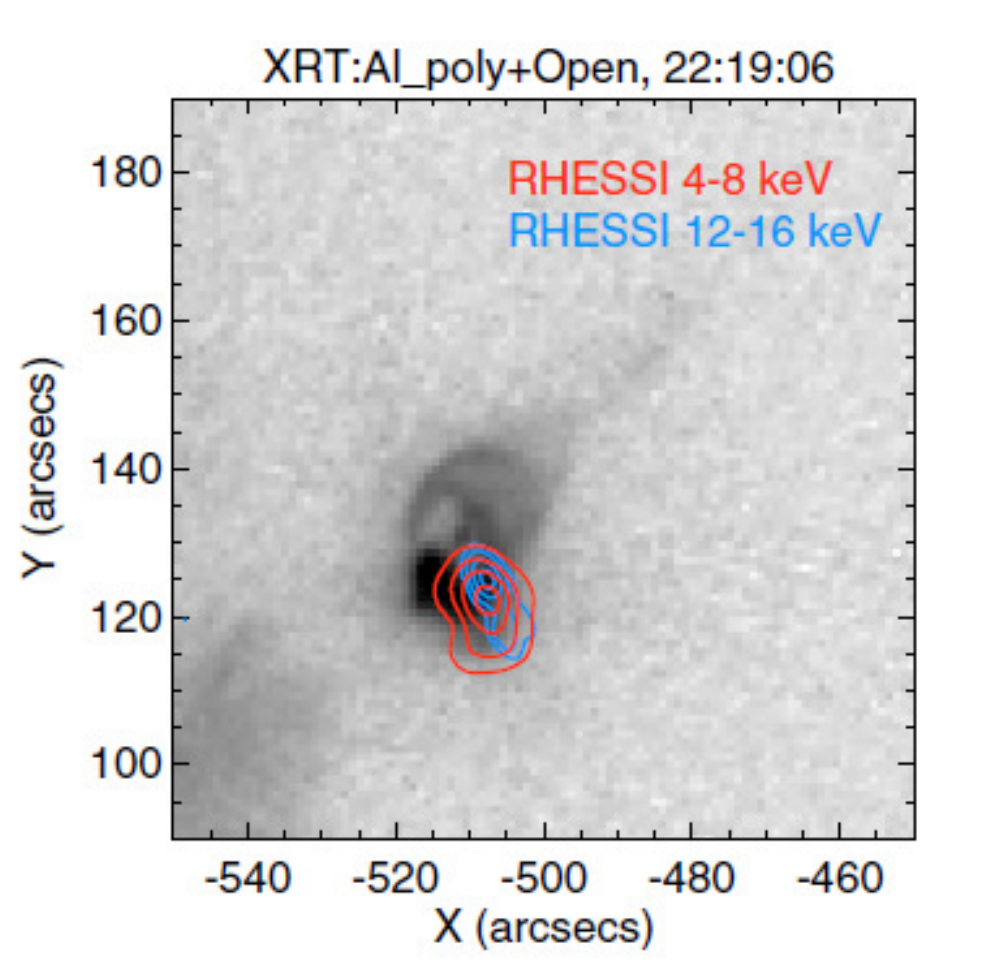}
                                  \includegraphics[width=0.33\textwidth]{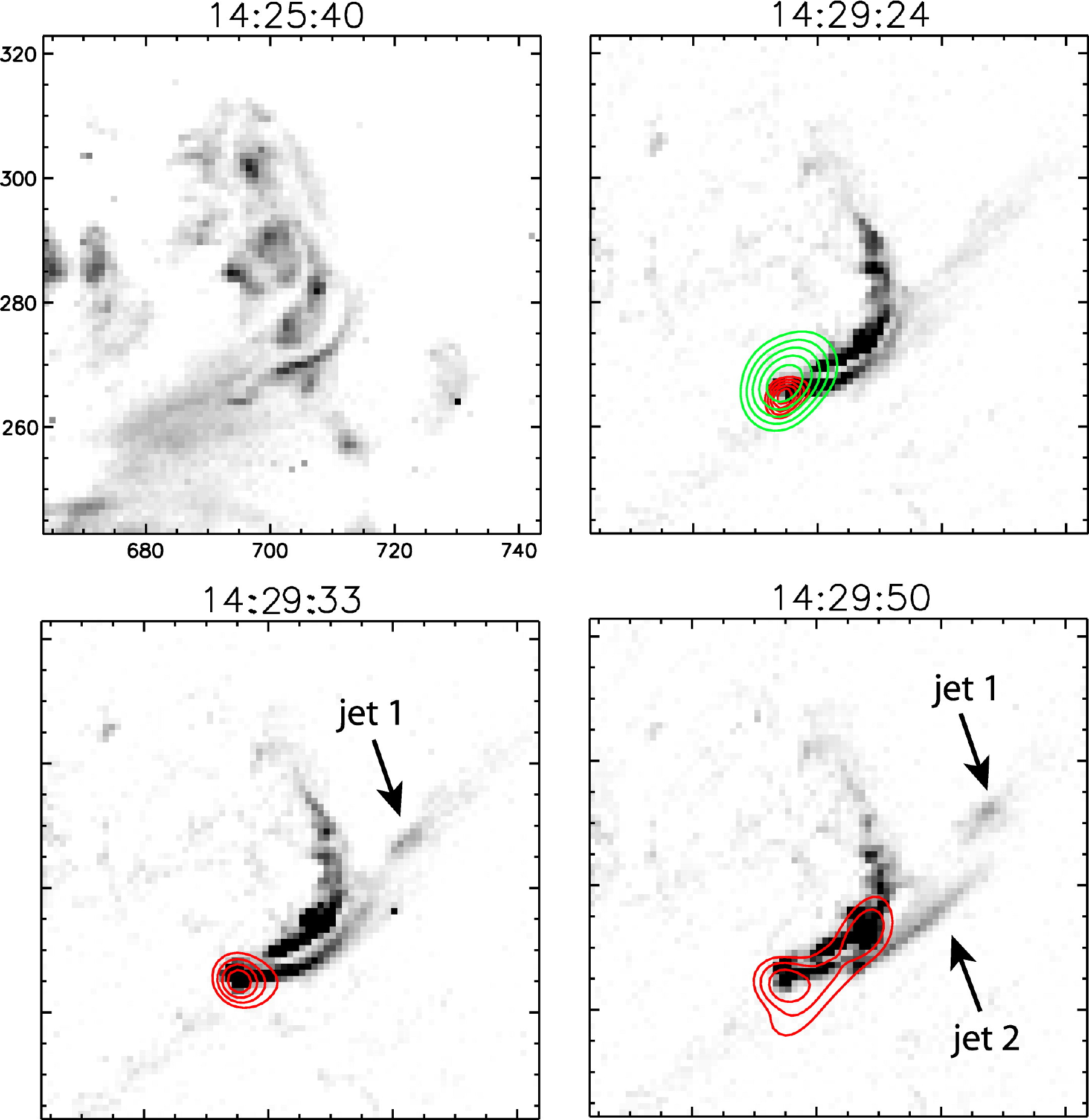}  }
\parbox{.3\textwidth}{\caption{{\bf{(Left)}} {\it{RHESSI}} observations (colored contours) overlayed on co-temporal XRT observations (reverse color-table) of a coronal jet. From \citet{2008A&A...491..279C}.  {\bf{(Right)}} {\it{RHESSI}} observations (colored contours) overlayed on co-temporal {\it{TRACE}} observations (reverse color-table) of a coronal jet.  From \citet{2008ApJ...680L.149C}.
\label{fig:1_RHESSI}  }}
\end{center}
\end{figure*}

There exist several {\it{RHESSI}} observations of coronal jet counterparts taking place during microflares or even standard flares \citep[e.g.,][]{2004ApJ...604..442L, 2008ApJ...680L.149C,2008A&A...491..279C, 2009SoPh..255...79C, 2013ApJ...769...96C}. The HXR emissions are typically limited to the base of the jets and presumably correspond to small loops which were energized by the microflares (e.g., Fig.~\ref{fig:1_RHESSI}), which is suggestive of the important role of magnetic reconnection in generating both phenomena.

\begin{figure*}
\begin{center}
\parbox{.6\textwidth}{ \includegraphics[width=0.6\textwidth]{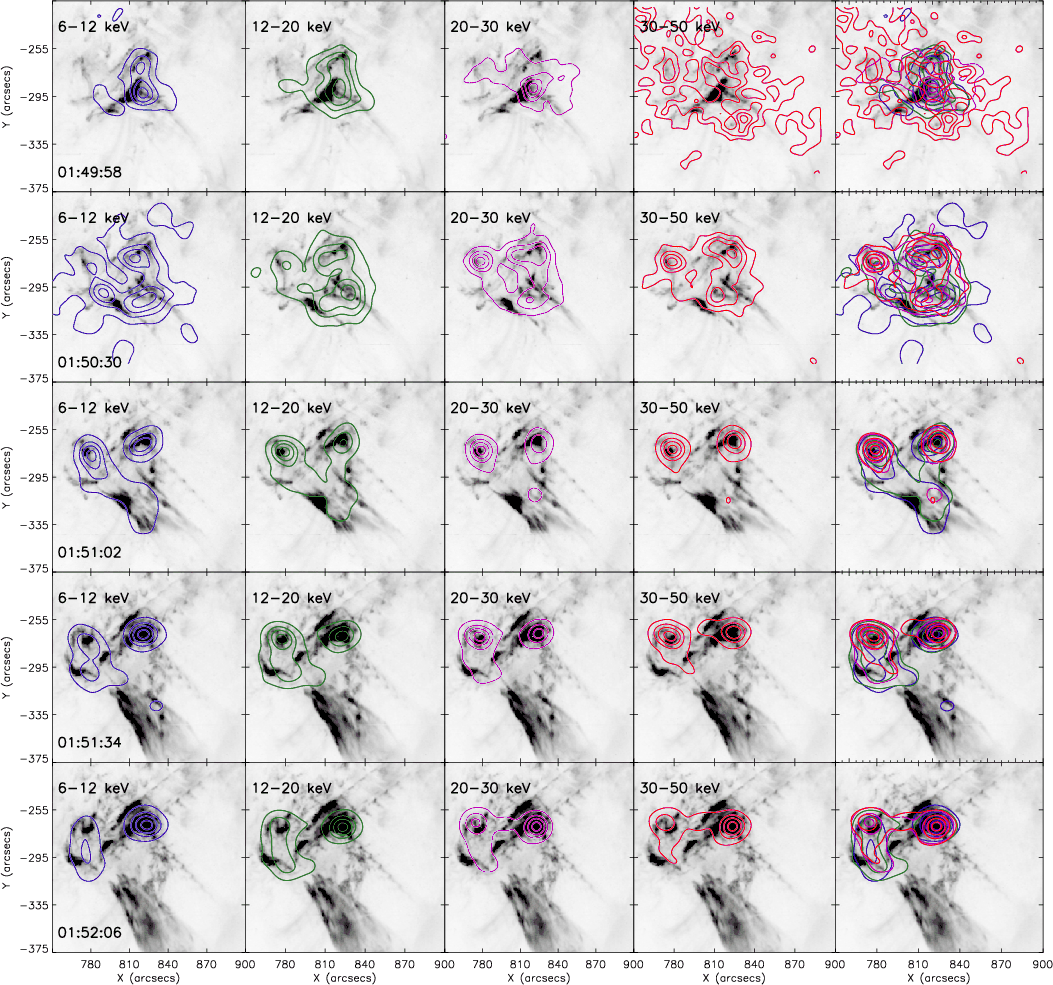}
                                   \includegraphics[width=0.6\textwidth]{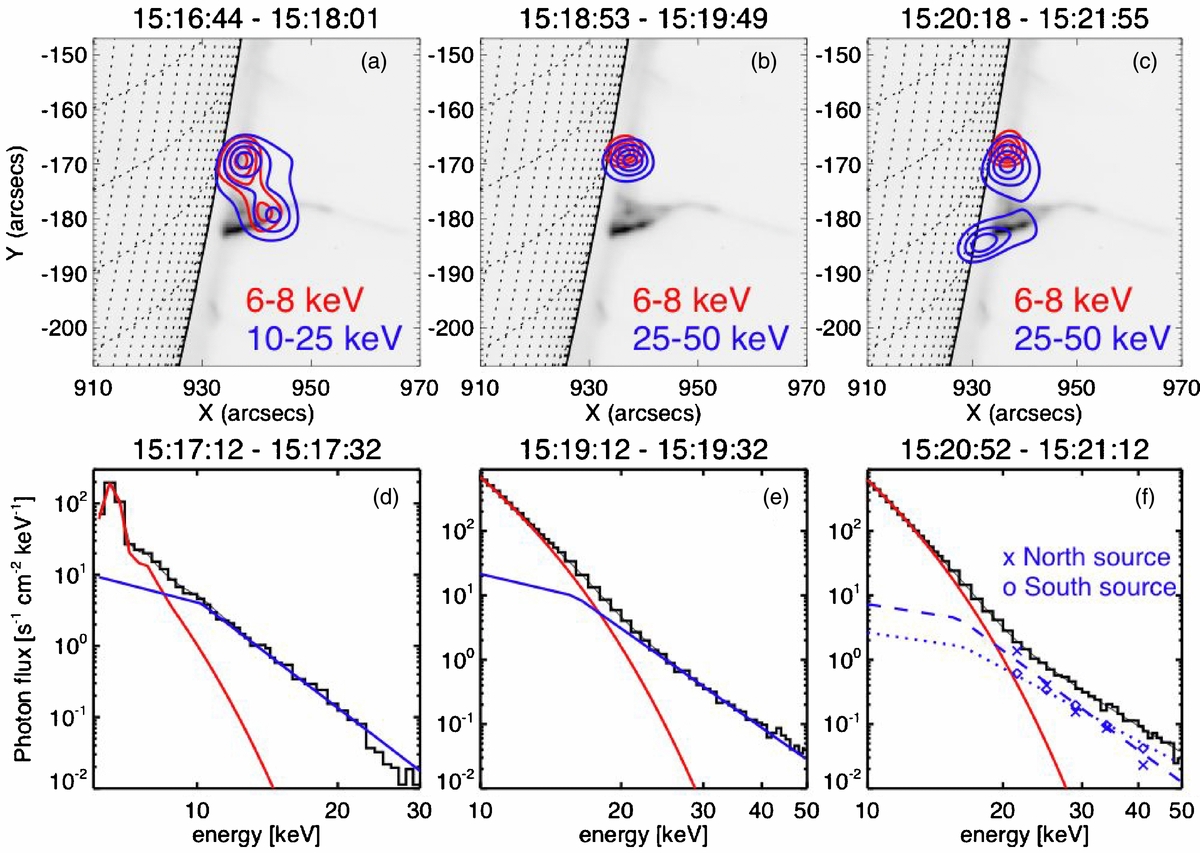}  }
\parbox{.38\textwidth}{\caption{{\bf{(Top)}} {\it{RHESSI}} observations (colored contours) overlayed on co-temporal {\it{TRACE}} observations (reverse color-table) of a coronal jet. From \citet{2009A&A...508.1443B}. {\bf{(Bottom)}} {\it{RHESSI}} observations (colored contours) overlayed on co-temporal {\it{TRACE}} observations (reverse color-table) of a coronal jet. The red (blue) contours correspond to thermal (non-thermal) sources as determined from the corresponding spectral fittings displayed at the bottom of this panel. From \citet{2012ApJ...754....9G}.}
\label{fig:2_RHESSI} }
\end{center}
\end{figure*}

Two studies during coronal jets supplied evidence of HXR emissions not only from the bases of the observed jets but also from their spires \citep[][see Fig.~\ref{fig:2_RHESSI}]{2009A&A...508.1443B,2012ApJ...754....9G}. \cite{2009A&A...508.1443B} showed that the HXR emission corresponds to energies of 20-30 keV and the fitting of the {\it{RHESSI}} spectrum provides evidence for a jet temperature of $\approx28$~MK and the non-thermal nature of the emission, which was also corroborated by multi-frequency imaging observations in the microwaves by the Nobeyama radioheliograph. Off-limb observations by \citet{2012ApJ...754....9G} of a footpoint-occulted coronal jet showed faint HXR coronal sources along the jet spires reaching heights of $\approx50$~Mm above the limb. The  spectral analysis of the jet HXR source showed that collisional losses  either in the corona  or at the occulted chromospheric footpoints by accelerated electrons  can supply the thermal and mechanical energy of the jet. Note that theoretical calculations by \cite{2009ApJ...696..941S} placed limits on the number of non-thermal electrons accelerated along open magnetic field  (e.g., $\approx 3\times {10}^{36}$ for {\it{RHESSI}}) to allow for their detection in HXRs or SXRs.

Frequently during coronal jets  the temporal profile of the associated HXRs matches the associated type III radio burst. This suggests that the magnetic configuration associated with jets (i.e., transient magnetic field opening) released and accelerated electrons which escaped into the interplanetary space. The close temporal associations between coronal jets, HXRs, and type III radio bursts has been reported in a number of  studies  \citep[e.g.,][]{2008A&A...491..279C, 2008ApJ...681..644K,2009A&A...505..811B,2009A&A...508.1443B, 2012ApJ...754....9G,2013ApJ...769...96C}.

\section{\textbf{\textit{Hinode}}/XRT and \textbf{\textit{SDO}}/AIA Imaging: Morphology of Coronal Jets}

Most of the works of this Section include analyses of {\it{SDO}}/AIA data. We separate the discussion into subsections covering general morphological observations of jets. These divisions however are, in many cases, largely artificial, and there can be substantial overlap in the categories into which a particular study should fall. For example, some of the jets studied outside of the subsection on twisting jets also displayed helical motions.  Therefore the divisions are best considered as a method to give a rough order to the substantial body of literature on coronal jets.

\subsection{Standard and blow-out Jets} \label{subsubsec-standard_blow-out}

\cite{2010ApJ...720..757M} introduced the concept of ``blow-out jets,'' along with the terminology ``blow-out'' and ``standard'' jets. These terms were originally based on morphological descriptions of  coronal jets when viewed in {\it{Hinode}}/XRT movies.  They observed that the spires of some X-ray jets remained  thin and narrow during their entire lifetime, and their bases remained relatively dim, except for the commonly observed compact JBP on one side of the jet's base. They also observed that other X-ray jets evolve such that the spire begins narrow, as in the narrow-jet case, but then broadens out with time until it is of size comparable to the width of the jet's base.  For this second class of jets, the JBP again starts as a compact feature off to one side of the jet's spire, but eventually the entire base brightens to become about as bright as the compact JBP. Analysis by the same authors of {\it{STEREO}} observations also suggested that most of the narrow-spire jets had no counterpart in the 304~{\AA} images, while most of the broad-spire events had accompanying 304~{\AA} jets.

\begin{figure}
\begin{center}
\parbox{.48\textwidth}{\includegraphics[width=0.45\textwidth]{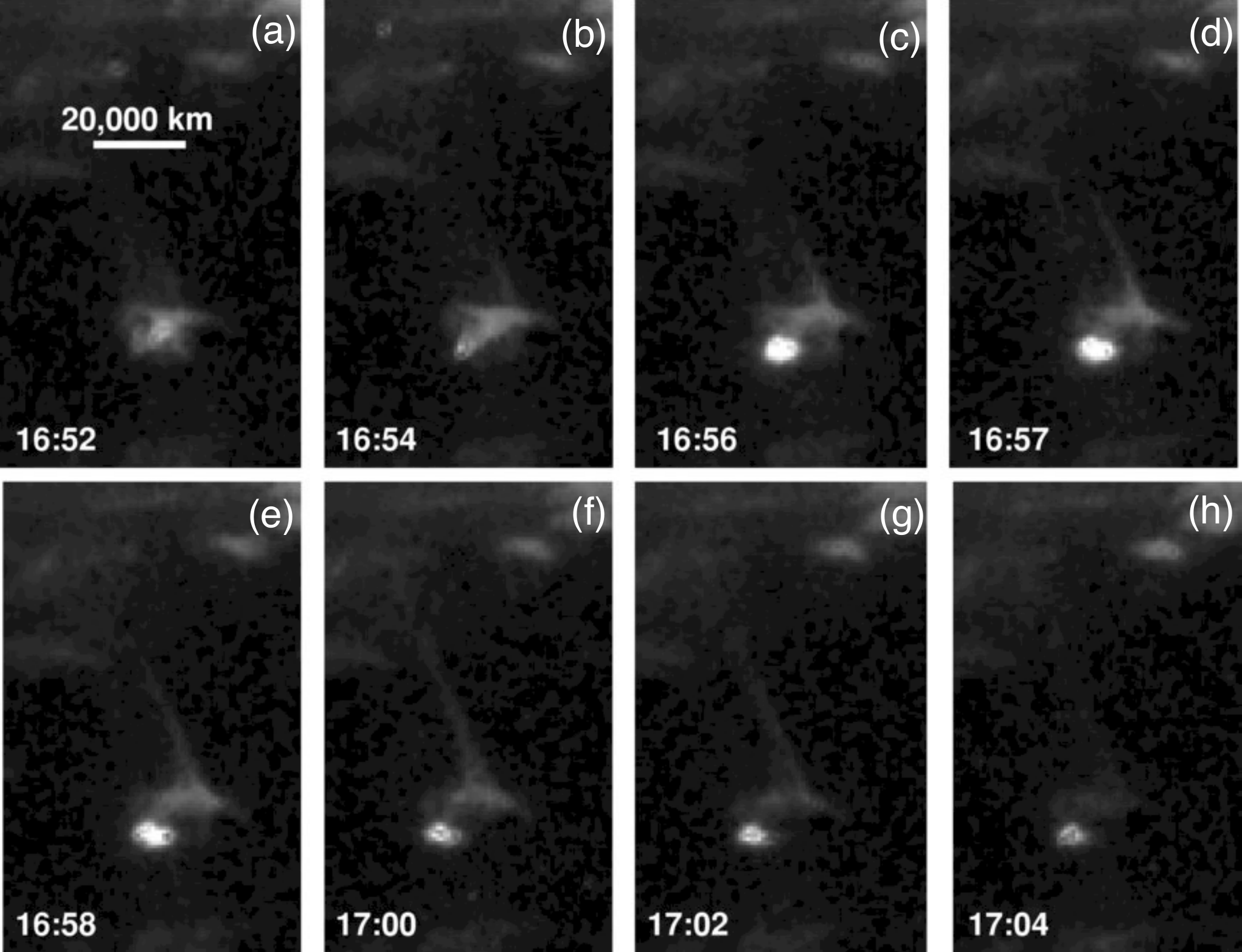} }
\parbox{.5\textwidth}{\includegraphics[width=0.5\textwidth]{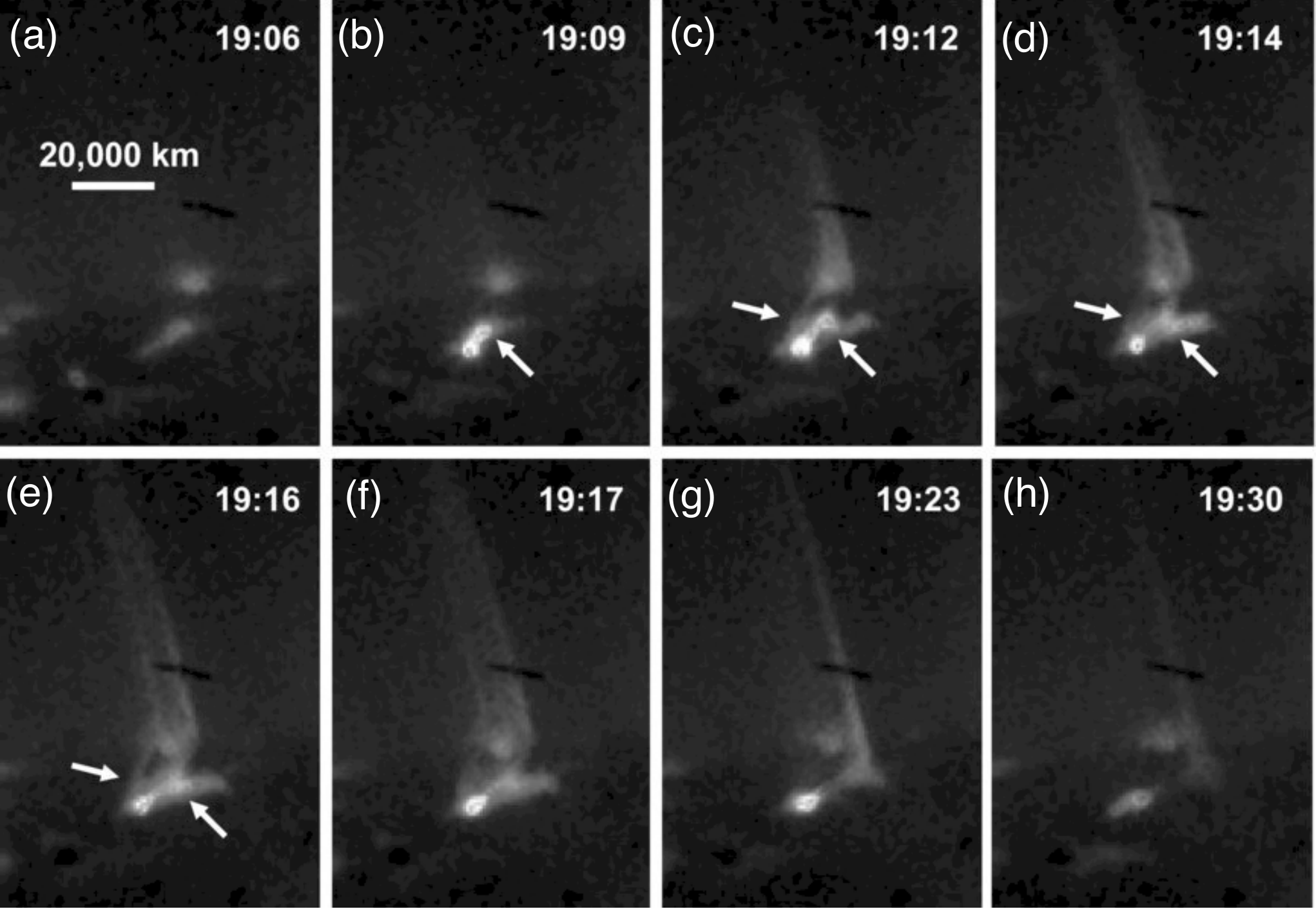} }
\caption{Examples of {\it standard} {\bf{(Left)}} and {\it blow-out} {\bf{(Right)}} jets observed by {\it{Hinode}}/XRT\@. Times are UT times on Sep. 22, 2008, and Sep. 20, 2008, respectively. The defining characteristics of standard jets are: narrow spire, compact JBP (c), and the absence of cool (chromospheric-temperature) emission in {\it{STEREO}}/EUVI~304~{\AA} images. Blow-out jets are, on the other hand, characterized by initially narrow spire (c) that later broadens to span nearly the width of the base region (e,f); initial compact brightening (b) that spread to the whole jet-base (c--e); and a strong cool (chromospheric-temperature) component visible in EUV~304~{\AA} images. Adapted from \citet[][see also \citeauthor{2013ApJ...769..134M}~\citeyear{2013ApJ...769..134M}]{2010ApJ...720..757M}.}
\label{fig:moore.et10_fig2_standard_jet}
\end{center}
\end{figure}

\begin{figure}[!ht]
\begin{center}
\parbox{.45\textwidth}{\includegraphics[width=.45\textwidth]{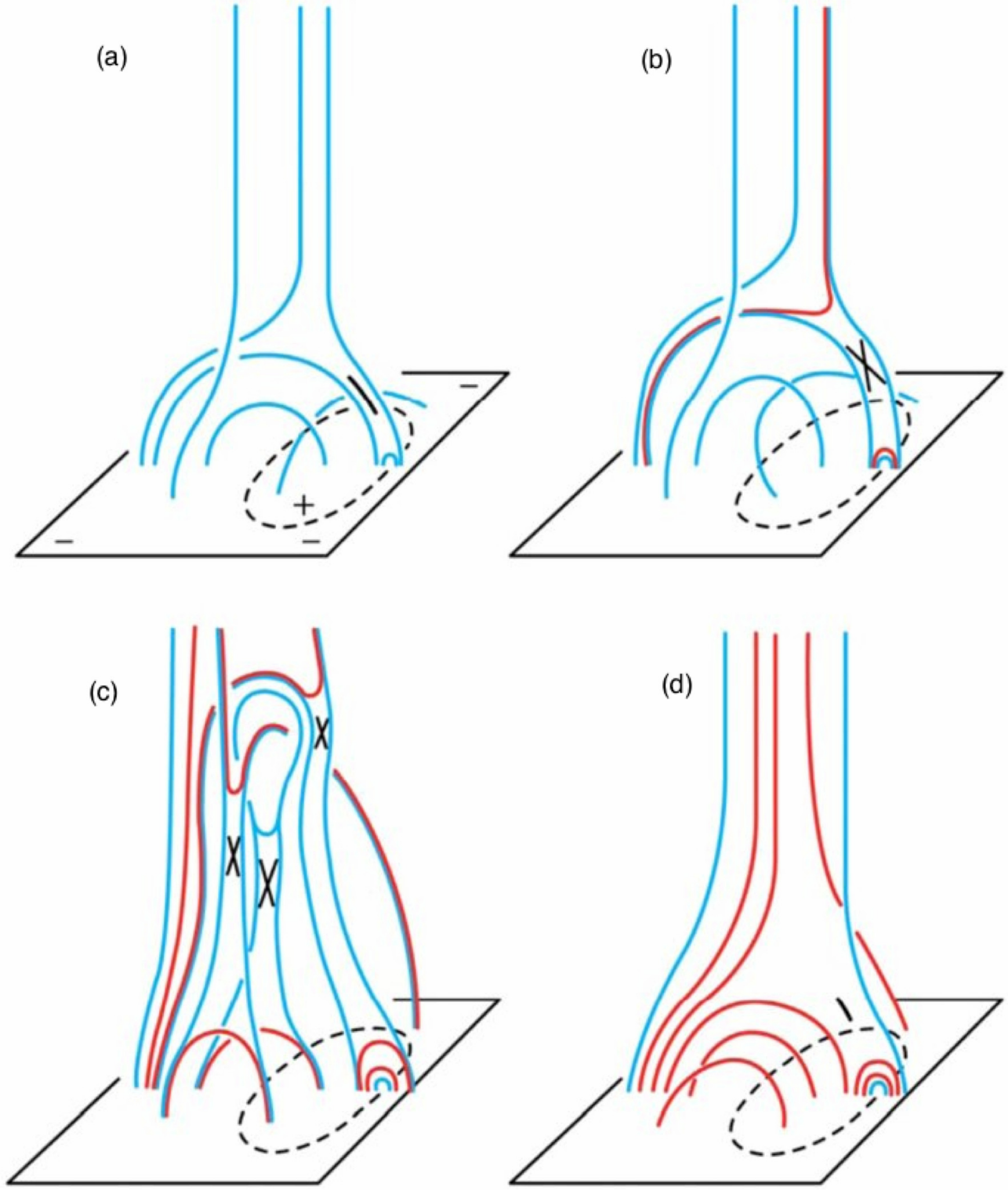}}
\parbox{.45\textwidth}{
\caption{Proposed process for blow-out jets, according to \citet{2010ApJ...720..757M,2013ApJ...769..134M}.  (a) Initial set up is as in the case of standard jets: ambient coronal magnetic field (open) and emerging or emerged bipolar field (closed).  (b) Magnetic reconnection (X) occurs at the location of the current sheet (short black-line arch) shown in (a).  A narrow jet spire resembling standard jets forms along the new open field lines.  (c) Destabilization of the bipolar field leading to full eruption and various reconnections (crosses). Cool material originally entrained inside of the bipole is carried outward with the eruption. (d) Late stages of the bipole's eruption: new reconnection-produced loops form and brightening the base region of the jet. Adapted from \citet{2010ApJ...720..757M}.\label{fig:moore.et10_jets}  }   }
\end{center}
\end{figure}

Based on these observations, they suggested that the narrow-spire jets were produced as in the original jet-production model due to Shibata \citep[][see \S\ref{ModelSection9}]{1992PASJ...44L.173S,shibata01}; thus they dubbed these types of jets ``{\it standard jets,}" since the jets seemed to obey that original ``standard'' picture (Fig.~\ref{fig:moore.et10_fig2_standard_jet}).   In contrast, they suggested that the broad-spire jets were generated by a variation of the standard picture.  In this case, the emerging (or emerged) flux would have much more free magnetic energy than in the case where a standard jet was formed. They suggested that these jets started out the same as standard jets, with an emerging bipole reconnecting with ambient open field.  That reconnection resulted in a narrow spire, as in the standard-jet case.  But during this reconnection process, the emerging bipole is triggered unstable and erupts outward. This eruption blows out the bipole field and the surrounding field, carrying outward the cool (chromospheric-temperature) material entrained in those fields.  Thus they named these types of jets ``{\it blow-out jets.}"  This eruption of the bipole results in much-more widespread reconnections than in the standard-jet case, where the emerging bipole remains inert throughout the jetting process.  This scenario for blow-out jets could explain why the jet spire can grow from narrow to broad, and also why cool material, visible in 304~{\AA} EUV images, would often accompany the blow-out jets (Fig.~\ref{fig:moore.et10_fig2_standard_jet}). An alternative possibility allowed for by \cite{2010ApJ...720..757M} is that the bipole starts erupting before the reconnection with the ambient field begins; in this view, the eruption of the bipole would drive the reconnection with the ambient field. In both the standard and the blow-out cases, the suggestion was that the reconnection between the emerging or emerged bipole field and the ambient coronal field created the compact JBP. Fig.~\ref{fig:moore.et10_jets} shows the basic picture for standard and blow-out jets.

\cite{2013ApJ...769..134M} expanded upon the earlier work on standard and blow-out jets \citep{2010ApJ...720..757M} by examining 54 X-ray jets found in {\it{Hinode}}/XRT data, and they also observed them in AIA 304~{\AA} images.  They identified 32 of the jets as blow-out, 19 as standard, and 3 were ambiguous.  When these newer results are combined with the previous work \citep{2010ApJ...720..757M}, the total number of X-ray jets examined is 109, and among these, 53 are standard, 50 are blow-out, and 6 are ambiguous. This new work \citep{2013ApJ...769..134M} found that almost all blow-out jets (29 out of 32)\footnote{A typo in paragraph~2 of \S5 of \cite{2013ApJ...769..134M} says ``all 29 blow-out X-ray jets displayed a cool component.'' This instead should read:  ``29 out of 32 blow-out X-ray jets displayed a cool component.''  This typo does not affect the general discussion and conclusions of that paper.} had corresponding jets observable in 304~{\AA} images. They also found that almost none of the standard jets had such a cool jet visible in 304~\AA; only 3 of the 19 standard jets had such a corresponding cool-component jet.

\subsection{General Morphological Observations of Jets}  \label{subsubsec-general_observations}

In the last few years, analyses of coronal jets within CHs, QS, and in the vicinity of ARs based on multi-instrument observations ({\it{SDO}}/AIA and HMI, {\it{Hinode}}/XRT and EIS, and {\it{STEREO}}/EUVI) provided valuable insights into the morphology and causes of coronal jets.  It has frequently been assumed or inferred that magnetic reconnection is the cause of impulsive eruptions of jets following either flux emergence and/or cancellation. Different events showed different behavior and dynamics.

\begin{figure} [!ht]
\begin{center}
\parbox{.62\textwidth}{\includegraphics[width=0.6\textwidth]{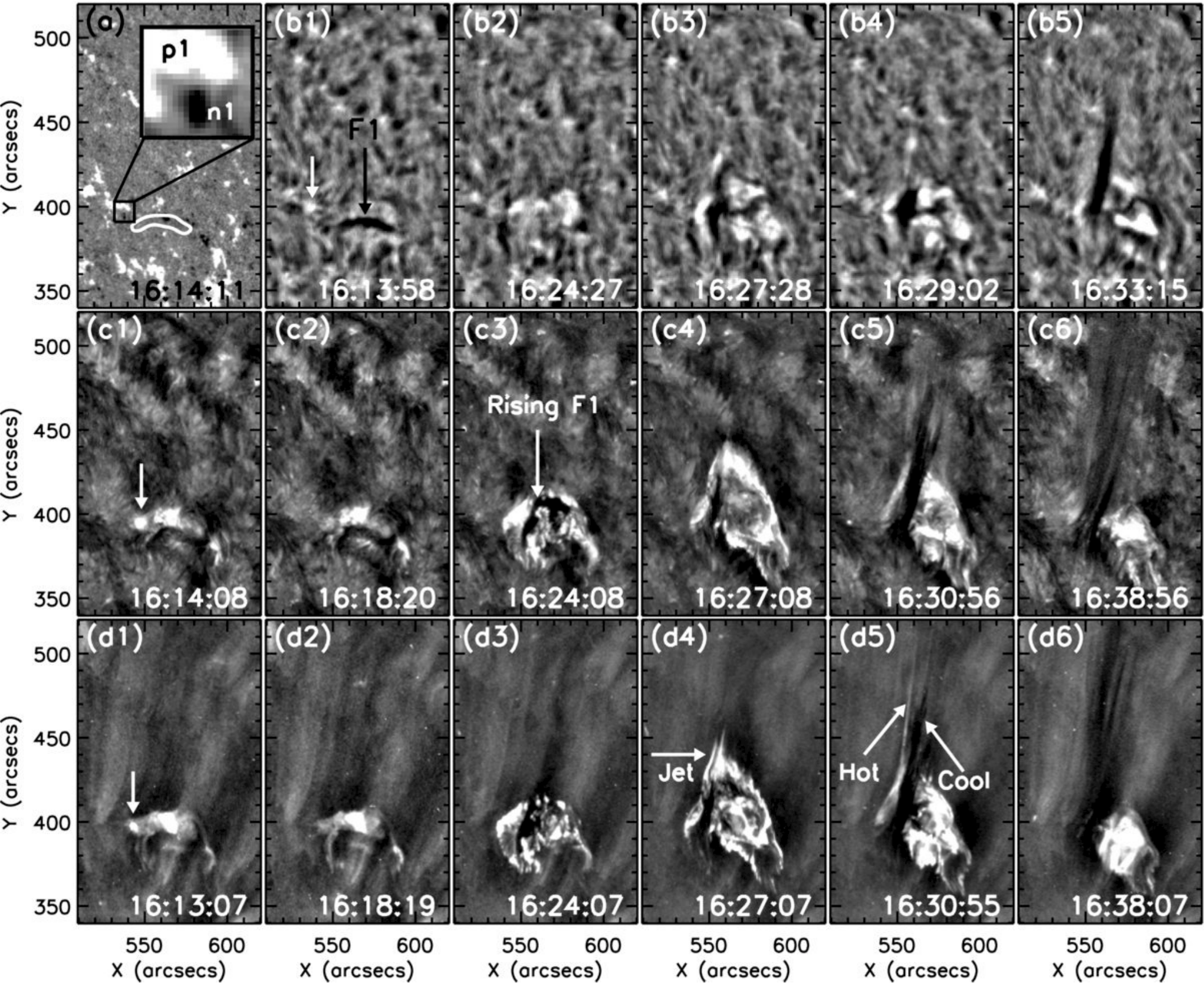} }
\parbox{.35\textwidth}{
\caption{On-disc observations of a blow-out jet, in (b1--b5) \halpha\ from  Big Bear Solar Observatory, (c1--c6) AIA 304~\AA, and (d1--d6) AIA 193~{\AA}.  Panel (a) shows an HMI magnetogram, with a close up showing positive (p1) and negative polarities, and the contour of the of  profile of the filament labeled F1 in panel~(b1).  Vertical arrows show a bright patch prior to ejection of the filament, and the two arrows in (d5) show the hot and cool components of the jet. From \citet{2012ApJ...745..164S}. }
\label{fig:shen.et12_fig_2_blow-out_jet}}
\end{center}
\end{figure}

Regarding the morphological aspect of jets, studies of different events reported cases of fan-spine magnetic field topology following flux emergence \citep[e.g.,][]{2011ApJ...728..103L}, evolution from standard to blowout type jet \citep{2011ApJ...735L..18L}, blowout resulting from mini-filament eruption \citep[][see Fig.~\ref{fig:shen.et12_fig_2_blow-out_jet}]{2011ApJ...738L..20H,2012ApJ...745..164S,2014ApJ...783...11A}, blobs and quasi-periodic small-scale plasma-ejection events along the jet spire \citep{2014A&A...567A..11Z,2014A&A...561A.104C,2014A&A...562A..98C}, and hot and cold loops expanding out from a bright base forming a blowout jet \citep{2014SoPh..289.3313Y}.

\citet{2014ApJ...783...11A} showed that the mini-filament they observed initially moved slowly ($\sim$15~km~s$^{-1}$) prior to jet formation, and then more rapidly ($\sim$80~km~s$^{-1}$) as the mini-filament material ejected along open field lines to form the jet; they also identified a faint, faster component ($\sim$200~km~s$^{-1}$) in 193~{\AA} images. The slow rise followed by a fast rise pattern is similar to that frequently observed in larger-scale filament eruptions \citep{1975SoPh...42..425R, 2005ApJ...630.1148S}. It is also noteworthy that \citet{2011ApJ...738L..20H} show that their jet had characteristics of large-scale CME-producing eruptions, including a small flare-like brightening, a small coronal dimming region, and a micro-CME. One is left to wonder whether solar activity is scale-invariant as noted by \citet{2010ApJ...718..981R}.

Very recently \cite{sterling.et15} observed 20 near-limb PCH X-ray jets, using both {\it{Hinode}}/XRT and {\it{SDO}}/AIA images. They reported that all 20 jets originated from mini-filament eruptions, and with the JBP being flaring loops occurring in the wake of the eruption.  Based on this, they suggested that the variety of coronal jet observed by \cite{2014ApJ...783...11A}, rather than being exceptional, is in fact the {\it{predominant}} variety of coronal jet (at least in PCHs). They further suggested that standard jets and blow-out jets are fundamentally the same phenomenon, with either ``standard'' or ``blow-out'' morphology ensuing depending upon particulars of the mini-filament eruption.  As of the time of this writing, it is too early to tell how well the \cite{sterling.et15} observations and inferences describe coronal jets in general.


The connection between CBPs and jets was discussed in studies by \cite{2014ApJ...796...73H} based on {\it{SDO}}/AIA and HMI. These CBPs are long-lasting features that are different from the transient jet base brightenings (i.e., the JBPs) that occur in conjunction with the jets themselves. From a study of 30 CBPs, they find that $\sim$25--33\% of them experience one or more mini-filament eruptions, consistent with the blow-out-jet concept. They report that the mini-filament eruptions possibly result from flux convergence and cancellation. 

Most of the above studies also reported on what magnetic field behavior caused the different events, often based on {\it{SDO}}/HMI data. Although the event observed by \citet{2011ApJ...728..103L} was at the limb so that direct magnetic information was not available, the authors argue that observations of a growing set of loops near the start of the event is consistent with emerging flux in an open-field region leading to magnetic reconnection and jet formation. \citet{2014A&A...562A..98C}, using HMI magentograms, did not find flux emergence in the jetting region, but suggested that emerging flux might be present but relatively weak. \citet{2012ApJ...745..164S} identified flux changes they interpreted as a series of flux emergences and cancellations that resulted in jet onset, and found indications of ``impulsive cancellation between the opposite polarities during the ejection of the blow-out jet."  As far as we know, all other studies of the magnetic configuration (using HMI data) at the bases of observed jets reported flux cancellation as the cause of the jets \citep[e.g.,][] {2011ApJ...738L..20H,2014ApJ...783...11A,2014SoPh..289.3313Y,2014PASJ...66S..12Y}. So overall, we found that several observational studies provide evidence that flux cancellation leads to jets, while relatively few observational studies provide evidence that emerging flux leads to jets.


\citet{2014A&A...567A..11Z} suggest that the observed blobs are plasmoids ejected during reconnection resulting from tearing-mode instability in current sheets occurring with the jets.

We do not yet know with certainty the interrelationship between jets seen at different wavelengths. From \cite{2010ApJ...720..757M,2013ApJ...769..134M}, we know that some X-ray jets have corresponding cooler-counterparts visible in AIA 304~\AA, and other X-ray jets do not have such a cool counterpart. Apparently many if not all X-ray jets have EUV counterparts \citep[e.g.,][]{2008ApJ...682L.137R}, but again a full study of the correspondence has not yet been undertaken. Therefore caution should be exercised to not generalize results from studies of jets seen at one wavelength to jets seen at substantially-different wavelengths. Thus for example, jets seen at, e.g., 171~{\AA} and 211~{\AA} should not be assumed to have counterparts at 304~{\AA} or in X-rays; rather, data in those wavelengths should be checked before drawing conclusions.

\section{Spectroscopic Observations of Jets}\label{SpectroscopySect}

Spectrometers can observe the LOS bulk flow of plasma as if they were in-situ instruments
and they can do this -- in analogy to spectroscopic binaries -- even for unresolved features.
As such they ideally complement imaging instruments.
In this Section we describe jet observations carried out by several spectrometers starting from the mid-90's.
These include observations by {\it{SOHO}}/CDS, SUMER, and UVCS, {\it{Hinode}}/EIS, and {\it{IRIS}} (see Table~\ref{tbl.uv}).

\begin{table}[!h]
\caption{Solar ultraviolet spectrometers.}
\begin{tabular}{llllll}
\hline
Name     & Duration         & Wavelength                    & Spatial    & Spectral & Slits \\
     &          &  [\AA]                   & Resolution    & Resolution &  \\
\hline
CDS       & 1996--2014    & 308--381,          & 6--10\as\ &   0.3-0.5~\AA & 2\as, 4\as\ \\
       &    & 515--632$^a$          &  &    &  \\
SUMER & 1996--2014    &  660-1610$^a$      & 1.5\as\     &  0.1~\AA    &   0.3\as, 1\as, 4\as           \\
UVCS    & 1996--2012    & 984--1080$^a$,      &   20\as   &   0.15-0.23~\AA            &    3-100\as           \\
    &     & 1100--1361     &      &         &              \\
EIS        & 2006--present & 170--212,   & 3--4\as\   &   60~m\AA            & 1\as, 2\as\   \\
        &  & 246--292  &   &              &  \\
{\it{IRIS}}      & 2013--present  & 1332-1358,   & 0.33-0.4\as\      & 26-53~m\AA    &  0.33\as              \\
      &   & 1389-1407,   &       &     &                \\
      &   & 2783-2835   &     &     &                \\
\hline
\end{tabular}\\
{$^a$ 2$^{nd}$ order lines are also seen.}
\label{tbl.uv}
\end{table}

CDS consisted of two spectrometers (the normal incidence, NIS,  and grazing incidence, GIS)
fed by the same telescope that observed the 150--800~{\AA} wavelength range. The
NIS was far more widely used than the GIS, and so we focus only on results from the NIS.
Two wavelength bands 308-381~{\AA} and 515--632~{\AA} were observed with a spatial resolution of 6--8\as,
although following the temporary loss of {\it{SOHO}} in 1998 the spatial resolution worsened to around 10\as\
and line profiles developed extended wings\footnote{see http:/http://solar.bnsc.rl.ac.uk/software/uguide/NIS\_PSF/}.
The wavebands consist mostly of emission lines from the upper transition region and
corona (temperatures $\ge 10^5$~K), with the important exception of the strong \ion{He}{i} \lam584.3 and \ion{He}{ii} \lam303.8
emision lines (the latter observed in the  second spectral order).

In the SUMER spectral range from 660~{\AA} to 1610~{\AA} more than 1000 spectral lines are present
that cover the vast temperature range from 0.005~MK (molecular hydrogen) to 28 MK (\ion{Fe}{xxiv}),
including the entire hydrogen Lyman series and a significant part of the Lyman continuum.
Centroiding techniques allow to detect Doppler flows down to 1--2~\kms.

UVCS supplied detailed spectroscopic observations  and diagnostics of jets in the outer corona from 1.4--10 $\mathrm{R_{\odot}}$. The instrument observes in two wavelength channels: the Ly-$\alpha$ channel covering  the range 1160--1350~{\AA} and the \ion{O}{vi} channel covering the range 940--1123~{\AA} (and 580--635~{\AA}  in the second order).  The spectrometer slit had a length of 40 arcmin.  The main lines of the two channels were the \ion{H}{i} Ly-$\alpha$ line at 1215.7~\AA, and the \ion{O}{vi} doublet at 1031.9 and 1037.6~\AA. In addition, UVCS observed lines including \ion{H}{i} Ly-$\beta$ 1025.7 \AA, \ion{C}{iii} 977.02 \AA, \ion{Mg}{x} 609.7 and 624.9~\AA, \ion{Fe}{xii} 1242~\AA,  and \ion{Si}{xii} 499.5~\AA. These lines probe plasmas with temperatures in the range 0.03--2 MK. Given the weak signals from the outer corona, the analysis of UVCS observations frequently employs some binning along the slit leading  to an effective spatial resolution $\ge20$\as. Finally, UVCS is equipped with a pinhole camera taking observations of the polarized radiance  of the outer corona in the WL in the wavelength range 4500--6000 \AA.

The {\it{Hinode}}/EIS observes the Sun observes the Sun in two narrow wavelength bands of 170--212~{\AA}  and 246--292~{\AA} that are dominated by coronal emission lines from iron but also contain some cooler lines, in particular \ion{He}{ii} \lam256.32 \citep{young07-eis}. The key advance over CDS is the use of multilayer coatings on the optical surfaces that give enhanced sensitivity and enable higher quality imaging with a spatial resolution of 3--4\as.

The Interface Region Imaging Spectrograph ({\it{IRIS}}) was launched in 2013 and supplies high-resolution spectroscopic observations of the chromosphere, transition region, and corona in the UV with a sub-arcsecond ($\sim 0.3-0.4$\as) spatial resolution and a two-second temporal resolution. It employs four pass-bands containing strong chomospheric (\ion{Mg}{ii}~H and K at 2803 and 2796~{\AA}, respectively) and transition region (\ion{C}{ii}   and \ion{Si}{iv} at 1334, 1335 and 1394, 1403~{\AA}, respectively) lines. Moreover, {\it{IRIS}} takes slit-jaw images of a maximum FOV of 130\as$\times$175\as\ in four different narrow band-passes. 

\subsection{{\it{SOHO}}/CDS Results}

A key discovery from CDS was the identification of twisting structures in macro-spicules, mostly observed in PCHs. \citet{pike97} presented the first event, which was observed just inside the solar limb at the south CH on 1996 April 11. The macrospicule was best seen in lines of \ion{He}{i} and \ion{O}{v}, but it had a weak signature in \ion{Mg}{ix} (1~MK) and so we consider it to be a coronal jet. Although Fig.~2 of this work shows \ion{He}{i} and \ion{O}{v} velocity maps with a red- and a blue-shift on opposite sides of the jet, it was only highlighted in the later paper of \citet{pike98} who found a similar signature in six other events. They coined the term ``solar tornado'' to describe this feature. Of further importance was the finding of an increasing velocity with height in the 1996 April 11 event, demonstrating that plasma continues to be accelerated along the body of the jet. Note that the velocity signatures from CDS only applied to the cool \ion{He}{i} and \ion{O}{v} lines as the signal was not strong enough in \ion{Mg}{ix} to derive accurate Doppler shifts.

The \citet{pike97} and \citet{pike98} results were derived from individual rasters. A time sequence of the evolution of a hot macrospicule was presented by \citet{banerjee00}, and this again showed weak \ion{Mg}{ix} emission, evidence for a twisted structure in \ion{O}{v}, and an increase in the \ion{O}{v} velocity with height.

\subsection{{\it{SOHO}}/SUMER Results}\label{sect.sumer}

\citet{wilhelm02} reported on the observation of a coronal jet on Mar. 8, 1999, in a single raster scan. The \ion{Ne}{viii} \lam770.41 dopplergram showed a jet-like structure extending to 35~Mm, with LOS speeds of $\sim40$~\kms.

A macrospicule at the limb of the south CH was reported by \citet{popescu07}. It was observed on 1997 Feb. 25 with a sit-and-stare study, and showed a clear signature in \ion{Ne}{viii} \lam770.41 and so we consider it to be a coronal jet. The jet extended about 36~Mm above the limb and was present for 5~minutes. The jet emission was identified by a red-shifted component at 135~\kms.

A unique observation was presented by \citet{2010A&A...510L...1K} who observed an X-ray jet at the solar limb in a CH with {\it{Hinode}}/XRT. The {\it{STEREO}}/EUVI instruments observed a co-spatial macrospicule in the 304~{\AA} filters, and SUMER and EIS rastered over the event, revealing hot emission in the \ion{Ne}{viii} \lam770.1 and \ion{Fe}{xii} \lam195.12 emission lines. The Doppler patterns in the cool lines observed by SUMER (\ion{O}{iv} \lam790.20) and EIS (\ion{He}{ii} \lam256.32) suggested a rotating motion for the broad macrospicule, and LOS speeds ranged from $+50$ to $-120$~\kms. The coronal jet was visible in \ion{Ne}{viii} \lam770.40 as a very narrow streak extending above the limb with a LOS speed of $-25$~\kms.

Another example of a CH jet observed jointly by SUMER and EIS was presented by \citet{madjarska11}, who observed a jet in an ECH on 2007 Nov. 14 using the SUMER sit-and-stare mode. The transition region lines (\ion{O}{v} \lam629.70 and \ion{N}{v} \lam1238.82) showed a strong velocity signature from the jet, but no signature was seen in the coronal \ion{Mg}{x} \lam624.94, in stark contrast to the coronal observations from EIS (Sect.~\ref{sect.eis}). The jet was demonstrated to be correlated with X-ray bursts in the BP, and magnetic reconnection was suggested as the driver for the jet.

\subsection{{\it{SOHO}}/UVCS Results}

\begin{figure*}[!ht]
\begin{center}
\includegraphics[height=0.3\textwidth]{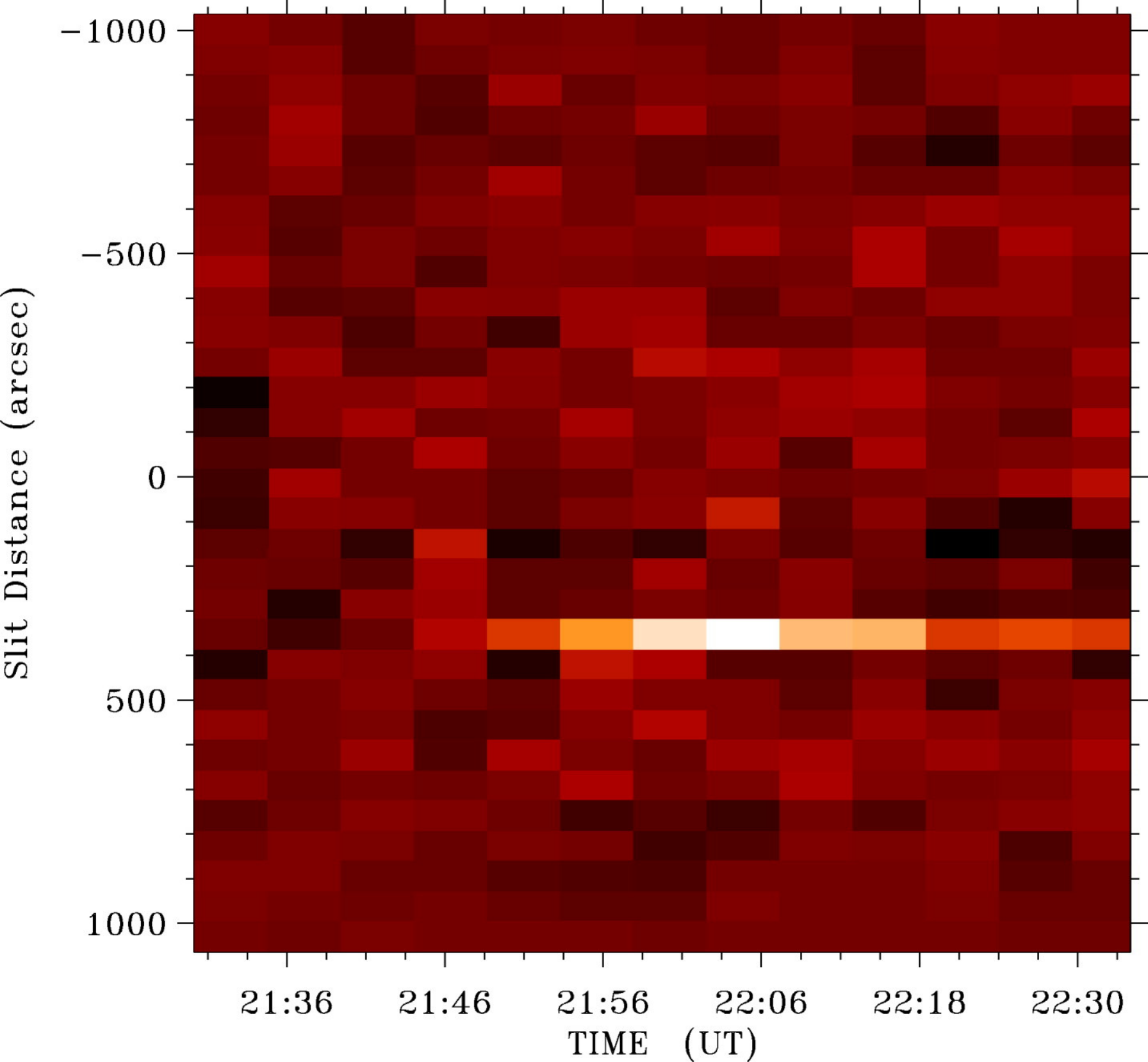}
 \includegraphics[height=0.3\textwidth]{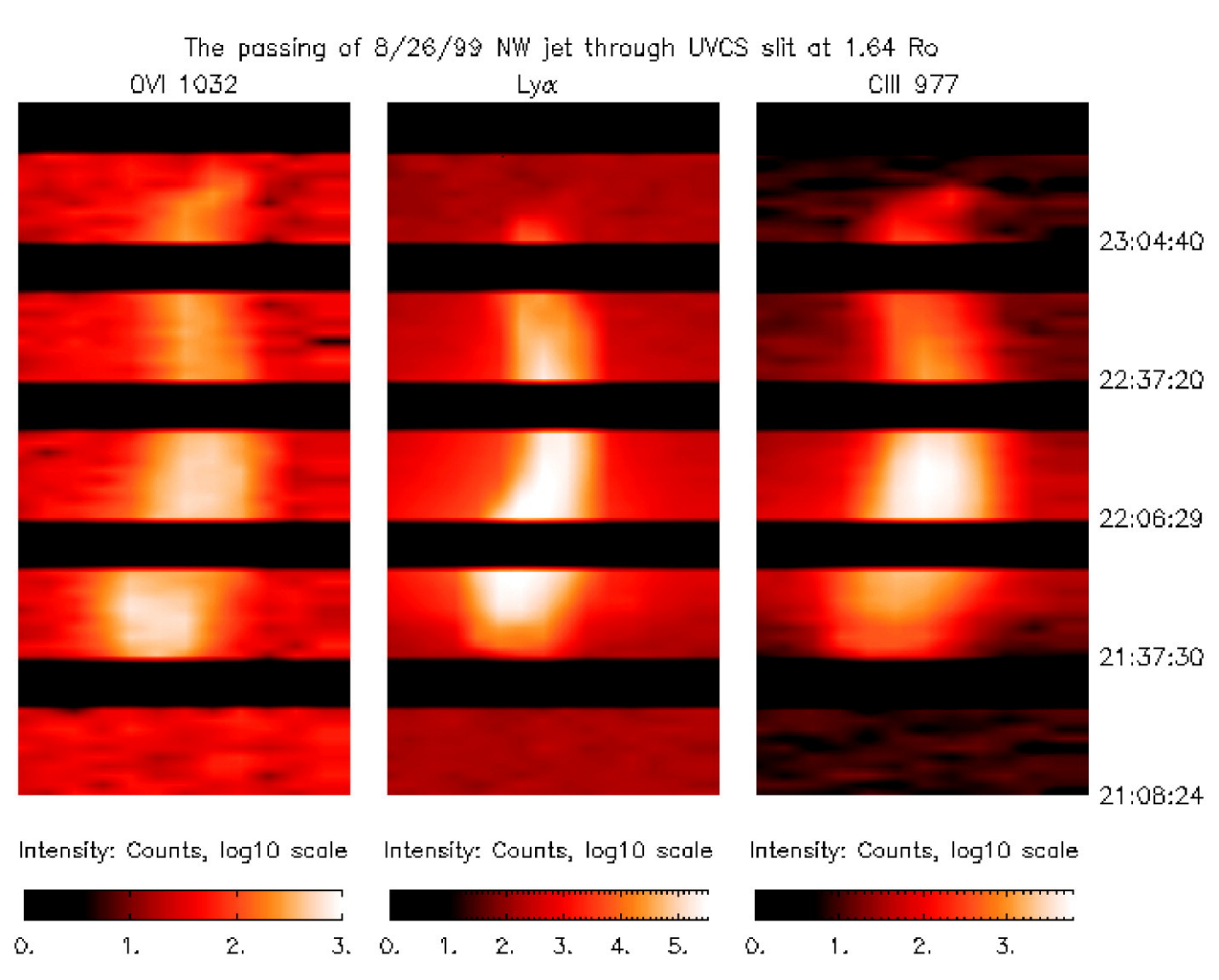}
\caption{UVCS observations of jets. {\bf{(Left)}}  \ion{H}{i}~Ly-$\alpha$ intensity a function of time and location along the UVCS slit. The jet passage corresponds to the bright stripe seen around slit distance 400\as \citep[from][]{2000ApJ...538..922D}. {\bf{(Right)}} Intensity images  formed by stacking subsequent exposures at 1.64 $\mathrm{R_{\odot}}$ in \ion{O}{vi}, \ion{H}{i}~Ly-$\alpha$ and \ion{C}{iii} at different times during a jet \citep[from][]{2005ApJ...623..519K}. }
\label{fig:1_UVCS}
\end{center}
\end{figure*}

\citet{2000ApJ...538..922D} presented the first detailed observations of coronal jets with UVCS. They analyzed a set of five polar jets, also tracked by EIT and LASCO, at radial distances 2.06-2.4 $\mathrm{R_{\odot}}$. The passage of these polar jets through the UVCS slit was manifested as increases in the intensity of the \ion{H}{i}~Ly-$\alpha$ (30--75 $\%$; see Fig.~\ref{fig:1_UVCS}) and  the \ion{O}{vi} doublet  (50--150 $\%$) lines. The increase took place either simultaneously in both  \ion{H}{i}~Ly-$\alpha$  and \ion{O}{vi} or  \ion{H}{i}~Ly-$\alpha$ had a delay of about 20 minutes with respect to \ion{O}{vi}. Interestingly the observed spectral lines became narrower during the observed jets, suggesting that the jets contained cooler plasmas than the background corona. \citet{2000ApJ...538..922D}  applied two different models to one of the observed jets. The first model is a temperature-independent line-synthesis model and supplied estimates on plasma parameters. It was found out that during the  jet passage through the UVCS slit the electron temperature decreased from $\approx0.75~\mathrm{MK}$ to $\approx0.15~\mathrm{MK}$ while the density decreased by a factor of $\approx$ 1.2 from its initial value of 4.5 $\times {10}^{6} \mathrm{{cm}^{-3}}$. The outflow speed was estimated from the Doppler dimming effect  to  be $>280~\mathrm{{km~s}^{-1}}$; the observed jets exhibited small Doppler-shifts which suggests a quasi-radial flow. The second model was a time-dependent temperature and density non-ionization  prescription of an expanding plasma parcel which showed that an initial electron temperature $<2.5$~MK and  heating rate commensurate to that of a "standard" CH were required at the coronal base. This suggests that the heating requirements of coronal jets observed in CHs in the EUV and WL can be different and lower than those for the more energetic SXR jets.

In a follow-up study, \citet{2002ApJ...565..621D} analyzed UVCS observations of a set of 6 polar jets  observed in \ion{H}{i}~Ly-$\alpha$  and \ion{O}{vi}   between  1.5-2.5 $\mathrm{R_{\odot}}$. This study extended the basic results of the \citet{2000ApJ...538..922D}  study. A heating flux of  $\approx3\times {10}^{5}$~erg~cm$^{-2}$~s$^{-1}$, based on the \citet{1994ApJ...435L.153W} plume model  and non-equilibrium ionization calculations of a moving plasma parcel, at the coronal base was required to reproduce the jet emissions as observed by UVCS. The postulated heating flux had to be concentrated into a narrow region below 1.1 $\mathrm{R_{\odot}}$ and corresponded to an electron temperature of around 2 MK. They also analyzed LASCO-C2 data of the observed jets and found densities comparable to plume values and 1.5 times higher than interplume densities at the same heights.

\citet{2005ApJ...623..519K} presented a comprehensive study of an AR coronal jet. The jet was traced from the Sun to the outer corona via an array  of instruments including UVCS. The jet was associated with huge increases of several hundred  with respect to the  background corona in  \ion{H}{i}~Ly-$\alpha$ and $\beta$, a factor 30 in \ion{C}{iii}, and a factor 8 in \ion{O}{vi} (see Fig.~\ref{fig:1_UVCS}). This suggests that the jet contained significant amounts of cool material at around ${10}^{5}$ K. Significant Doppler-shifts first towards the blue ($150~\mathrm{{km~s}^{-1}}$) and then towards the red ($100~\mathrm{{km~s}^{-1}}$) were observed by UVCS; similar Doppler-shift evolution but this time  first from the red and then to the blue was observed during the early stages of the jet at the limb by CDS and MLSO/CHIP (i.e., the Mauna Loa Solar Observatory/Chromospheric Helium-I Imaging Photometer).  The UVCS Doppler-shift pattern correlated with the corresponding outflow velocities deduced via the Doppler-dimming effect. The changing signs of the jet's Doppler-shifts both near the limb and in the outer corona may be consistent with rotation of the structure during its ascent.

\citet{2003ApJ...588..586D} analyzed five narrow CMEs (eruptions with angular width below $15^\circ$) with the aim to study the possible connection between such eruptions and jets. The deduced plasma parameters of the narrow CMEs yielded similar speeds and somewhat higher densities and temperatures by a maximum factor of 2 compared to coronal jets. Taken altogether these findings did not suggest a clear dividing line between narrow CMEs and jets, which is consistent with the blow-out jets \citep{2010ApJ...720..757M} which represent scaled-down versions of CMEs.

\citet{2007ApJ...659.1702C} analyzed observations of several cool jets during a {\it{SOHO}}-Ulysses quadrature. The jets were  first observed by EIT in its 304~\AA \, channel. Once they intercepted the UVCS slit at 1.7 $\mathrm{R_{\odot}}$ strong emissions in the cool lines \ion{H}{i}~Ly-$\alpha$ and $\beta$, \ion{O}{vi}, and \ion{C}{iii} were recorded (e.g., $>10$ times the background values in in some lines). The jets were not observed in any of the hot lines available by UVCS. Empirical modeling of the spectral line intensities resulted in jet densities in the range $(8.3-13)\times{10}^{6}$~cm$^{-3}$  and temperatures of up to $\approx1.7\times{10}^{5}$~K. The jets' average mass, gravitational, kinetic and thermal energies were estimated to  $10^{13}$~g and $1.9 \times {10}^{28}$, $2.1 \times {10}^{27}$ and  $1.5\times {10}^{26}$ erg, respectively. Finally, no conclusive evidence for an in-situ detection of these jets by Ulysses was found.

\subsection{{\it{Hinode}}/EIS Results}\label{sect.eis}

We divide the EIS jets into CH and AR categories.

\subsubsection{CH Jets}

The key advance of the EIS spectrometer relevant to coronal jets (particularly in CHs) is the high instrument sensitivity for the \ion{Fe}{xii} \lam195.12 emission line. The previously faint coronal signals of CDS and SUMER meant that, even if jets were detected, it was not possible to study velocities and line broadening. The advances of EIS were demonstrated with an impressive \ion{Fe}{xii} Doppler map in \citet{kamio07}, showing LOS velocities of up to 30~\kms\ for a jet on Jan. 09, 2007, extending about 60~Mm above a CH BP. A jet occurring in an ECH on Mar. 10, 2007, was studied by \citet{2008ApJ...673L.211M}. EIS was operating in sit-and-stare mode with the slit positioned about 11~Mm above the BP. The \ion{Fe}{xii} line showed a two component structure, with a blue-shifted component at 240~\kms. The second component was simply due to the CH background and was found at the rest wavelength of the line. This observation illustrates an important point when analyzing the \ion{Fe}{xii} line: the corona is everywhere emitting at 1.5~MK, and so the line profile of a jet is always a mixture of jet plasma and background plasma. In cases such as \citet{2008ApJ...673L.211M} the two components are clearly separated, but the \citet{kamio07} jet is an example where the two are blended. By fitting only a single Gaussian to the \ion{Fe}{xii} line, the jet velocity is underestimated. A two Gaussian fit would have led to a larger velocity for the \citet{kamio07} jet component.

The CH jet studied by \citet{2010A&A...510L...1K} (discussed in Sect.~\ref{sect.sumer})  was captured in a single EIS raster scan and observed as a very narrow streak in the \ion{Fe}{xii} \lam195.12 line, extending 75~Mm above the limb. The LOS velocity was $-20$~\kms.

Another jet observed simultaneously with SUMER was the ECH jet presented by \citet{madjarska11} and discussed in Sect.~\ref{sect.sumer}. The jet was seen in a single raster scan, and the outflowing plasma was emitting in a wide  range of lines from \ion{He}{ii} \lam256.32 to \ion{Fe}{xv} \lam284.16, and the LOS speed was up to 279~\kms. A density measurement in the jet was not possible, however.

\citet{2014SoPh..289.3313Y,2014PASJ...66S..12Y} and \citet{2015ApJ...801..124Y} presented CH observations obtained from an EIS data-set that spanned almost two days during 2011 February 8--10 and captured a number of jets. \citet{2014PASJ...66S..12Y} studied a jet on the CH boundary for which the jet took the form of an expanding loop reaching  heights of 30~Mm. The LOS speeds reached 250~\kms, and evidence was found for twisting motions based on the variation of Doppler shift in the transverse direction of the jet. The density of the jet plasma, measured with a \ion{Fe}{xii} density diagnostic, was $(0.9-1.7)\times 10^8$~cm$^{-3}$ and the temperature was 1.6~MK.

The largest and most dynamic jet from the data-set was presented by \citet{2014SoPh..289.3313Y}. It extended to 87~Mm from the BP and LOS speeds reached 250~\kms. The density of the jet plasma was $2.8\times 10^8$~cm$^{-3}$ and the temperature was 1.4~MK. A feature in common with the \citet{2014PASJ...66S..12Y} jet was the increase in LOS speed with height above the BP showing that plasma continued to be accelerated along the body of the jet. The jet BP showed a number of small, intense kernels as the jet began, reminiscent of flare kernels, and cool plasma was ejected as seen through an absorption feature in AIA 304~{\AA} images.

\citet{2015ApJ...801..124Y} demonstrated that almost half of the 24 jet events seen in the 2011 Feb. data-set showed no signature in AIA 193~{\AA} image sequences, and so referred to them as ``dark jets''. One dark jet was studied in detail, and was found to have a \ion{Fe}{xii} \lam195.12 intensity only 15--44\%\ of the CH background. The LOS speed of the jet plasma reached 107~\kms\ at a height of 30~Mm from the BP, and the temperature was 1.2--1.3~MK.

The work described above made use of narrow slit EIS data, which enables a full range of diagnostics to be applied. EIS also has a 40\as\ wide slit, enabling monochromatic imaging at high cadence. This was used by \citet{culhane07-jet} who studied how two PCH jets observed on 2007 Jan. 20 evolved with time. Ejection speeds of 360 and 150~\kms\ were measured, and the jet was found to emit in multiple ions, the hottest of which was \ion{Fe}{xv} \lam284.16 (2.2~MK). \citet{2014A&A...561A.104C} studied another PCH jet observed on 2007 Apr. 15 and derived a propagation velocity of 172~\kms\ from images in the \ion{Fe}{xii} \lam195.12.

\subsubsection{AR Jets}

\citet{2008A&A...491..279C,chifor08-jet1} presented observations of a recurring jet on the west side of AR10938 during the period 15-16 Jan. 2007. One of the jets was captured in a single raster \citep{chifor08-jet1} in emission lines formed over the temperature range $\log\,T=5.4$ to 6.4. The signature of the jets was an extended short wavelength wing to the coronal emission lines with LOS speeds of 150~\kms and very high densities (i.e., $\log\,N_{\rm e}\ge11$) as shown from diagnostics of \ion{Fe}{xii} and \ion{Fe}{xiii} lines.

Other examples of AR jets were observed in AR10960 on Jun. 05, 2007 \citep{matsui12} and at the limb in AR11082 on Jun. 27, 2010 \citep{2013ApJ...766....1L}. The ejected jet plasma in the Jun. 05, 2007 event could be identified in coronal lines up to $\log\,T=6.4$ (\ion{Fe}{xvi} \lam262.98). Taking into account the viewing geometries of the twin {\it{STEREO}} spacecraft, accurate estimates of outward jet speed were inferred through analysis of lines formed over the temperature range $\log\,T=4.9$ to 6.4. The speed was found to increase with temperature in the corona from $\approx160$ to $\approx430$~\kms, which is consistent with predictions for chromospheric evaporation during a reconnection process \citep{matsui12}. The Jun. 27, 2010 jet occurred on a large, closed loop and was very prominent in 304~{\AA} images from AIA. EIS was running in sit-and-stare mode and the slit crossed through the jet about 40~Mm above the solar surface. A signal in \ion{Fe}{xv} \lam284.16 is seen at the same time as X-ray emission is seen from XRT, confirming a jet temperature of around 2.2~MK. Evidence for twisting motions in the jet are found from simultaneous red and blue-shifts in \ion{Si}{vii} \lam275.36 (0.63~MK). For temperatures of 1--2~MK the jet appeared as a dimming region that traveled along the loop. The density of the loop was estimated at $3\times 10^8$~cm$^{-3}$ from the \ion{Fe}{xiv} \lam264.79/\lam274.20 ratio.

In summary, the AR jets observed by EIS are generally hotter than those seen in CHs, with the ejected plasma emitting in \ion{Fe}{xv} and \ion{Fe}{xvi}, suggesting temperatures of 2-3~MK. Further observations are needed to determine how common are jets with very high coronal density found by \citet{chifor08-jet1}.

\subsection{{\it{IRIS}} Results}\label{sect.iris}
\citet{2015ApJ...801...83C} reported {\it{IRIS}} observations of recurrent coronal jets in an AR over a pore within a supergranule cell. The four observed homologous jets were observed in AIA coronal channels and the TR \ion{Si}{iv} lines at 1394 and 1403~{\AA}. They were characterized by relatively well-separated red- and blue-shifts of  magnitudes of up to 50~km~s$^{-1}$ across the jets' axis. This line-shift pattern is consistent with helical motions. \citet{2014Sci...346A.315T} also reported {\it{IRIS}} observations of prevailing jet activity  in the network at spatial scales of few hundred km.

\section{Jet Dynamics: Statistics, CHs boundaries}

\subsection{Regionality of Coronal Jets}
The comparatively modest-quality observations from {\it{Yohkoh}}/SXT unveiled the most energetic jets that often occur around ARs \citep{1996PASJ...48..123S}. The recent much improved quality observations show that a higher number of jets occur in CHs \citep{2007PASJ...59S.771S}. Although CH and QS jets are smaller than AR jets \citep{2013ApJ...775...22S}, averages of the apparent speeds are comparable and are about 200~\kms. Plasma parameters such as temperature are characterized by larger error bars and are often model dependent. For details, see \S\ref{SpectroscopySect}.

\citet{2010A&A...516A..50S} investigated transient brightenings, including jets, within CHs and quiet regions. They found that CH boundaries are particularly prolific in terms of brightenings occurrence and about 70\% of these events within CHs and their boundaries show expanding loop structures and/or collimated outflows, while only 30\% of the brightenings in QS show flows. \citet{2013ApJ...775...22S} analyzed over a thousand PCH (northern) and QS jets. They found that jet occurrence rate in CH boundaries is twice as large as that in PCHs. Flux emergence/cancellation rates cannot explain this difference \citep{2013ApJ...775...22S}. \citet{yang11} reported on westward shifts in the boundaries of CHs so that their rigid rotation is maintained. It can be easily imagined that a coronal jet is produced by magnetic reconnection between the open field in a CH and the closed loop in a quiet region (i.e., interchange reconnection), which suggest that the coronal (global) magnetic topology need to be considered for understanding this phenomenon.

\subsection{Dynamics of Coronal Jets}
Except for the rare occurrence of large jets \citep{1994ApJ...431L..51S}, {\it{Yohkoh}}/SXT observations did not allow investigating the inner structure and evolution of jets. \citet{1999SoPh..190..167A} used {\it{Yohkoh}}-{\it{TRACE}} joint observations to obtain some insight into jets' fine structure. The recent high-quality X-ray/EUV data from {\it{Hinode}}, {\it{STEREO}}, and {\it{SDO}} reveal the complex structure and dynamics of these coronal events. In the following sections, we discuss the dynamics of coronal jets from the morphology and statistics point of view.

\subsubsection{Transverse Motion}

\citet{1992PASJ...44L.173S} reported on a coronal jet moving sideways with velocity of 20--30~\kms. \citet{1996ApJ...464.1016C} used jet reconnection model to interpret whip-like motions and footpoint blue-shifts of coronal jet-associated H-$\alpha$ surges. Nevertheless, detailed observations of jet transverse motions were uncovered using on-disk and mainly off-limb {\it{Hinode}} observations. A statistical study by \citet{2007PASJ...59S.771S} of 104 PCH events showed that more than 50\% of jets display transverse motions with $\sim\!35$~\kms. \citet{2014A&A...561A.104C} found that the transverse motion speed decreases with increasing height. Higher velocity ($>100$~\kms) transverse motions have been reported by \citet{2007PASJ...59S.745S} in several coronal jets, and that one event showed whip-like motion presumably following the opening of reconnected closed field lines. It can be easily speculated that whip-like motions result from the relaxation of the reconnected guide magnetic field.

There are two flavors of jet transverse motions: expanding motions and oscillations. \citet{2010ApJ...720..757M} interpreted expanding jets as ``curtain-like spires". Theoretically, the speed of the reconnected flux is $100-1000$~\kms\  assuming an Alfv\'en speed of 1000~\kms\ \citep{1996ApJ...464.1016C}. The observed speeds (i.e., $\sim\!35$~\kms) are, however, significantly smaller than the theoretical prediction. This may hint at expansion of the reconnection region rather than motion of reconnected magnetic flux. On the other hand, \citet{2007Sci...318.1580C} reported the first detection of the transverse oscillations in a coronal jet, which can be used to infer a number of physical parameters (e.g., temperature, magnetic field) in the corona using magneto-seismology. \citet{2012A&A...542A..70M} studied oscillations of a jet dark thread (i.e., the jet's inner structure) and found a period of 360 seconds. They inferred a temperature of $<3\times10^4$~K from kink mode oscillations. \citet{2014A&A...562A..98C} analyzed oscillations of the bright thread of a CH-boundary jet. They inverted the oscillation's 220 second period into magnetic field strength of 1.2 Gauss. They also reported strong damping of the transverse oscillations that are characterized by a velocity amplitude of 20~\kms.

In view of the recent results by \cite{sterling.et15}, which suggest that coronal jets may be the result of eruption of small-scale filaments (see \S\ref{subsubsec-general_observations}), a distinguishing characteristic of the emerging-flux and the minifilament ideas is the expected drift with time of the jet spire with respect to the JBP location. \cite{sterling.et15} predict that the spire should tend to drift {\it{away}} from the BP, while the emerging-flux model should result in the spire drifting {\it{toward}} the BP. \cite{Savcheva09} find that the drift is often away from the JBP (supporting the \citeauthor{sterling.et15} \citeyear{sterling.et15} interpretation), although this question should be investigated more systematically.

\subsubsection{Untwisting Motions of Jet Helical Structure}
\citet{2008ApJ...680L..73P} successfully reconstructed a coronal jet in 3D using near-simultaneous
observations from the {\it{STEREO}} twin spacecraft. They unambiguously showed the helical structure of the jet,
which was later confirmed through 3D-MHD simulations by \citet{2009ApJ...691...61P}.
The morphological analysis of \citet{2009SoPh..259...87N}
described in Sect.~2.3 confirm that helical jets are common and that untwisting motions
may also be an important property of a significant class of jets.


\begin{figure}[!th]
\begin{center}
\parbox{0.35\textwidth}{\includegraphics[width=0.35\textwidth]{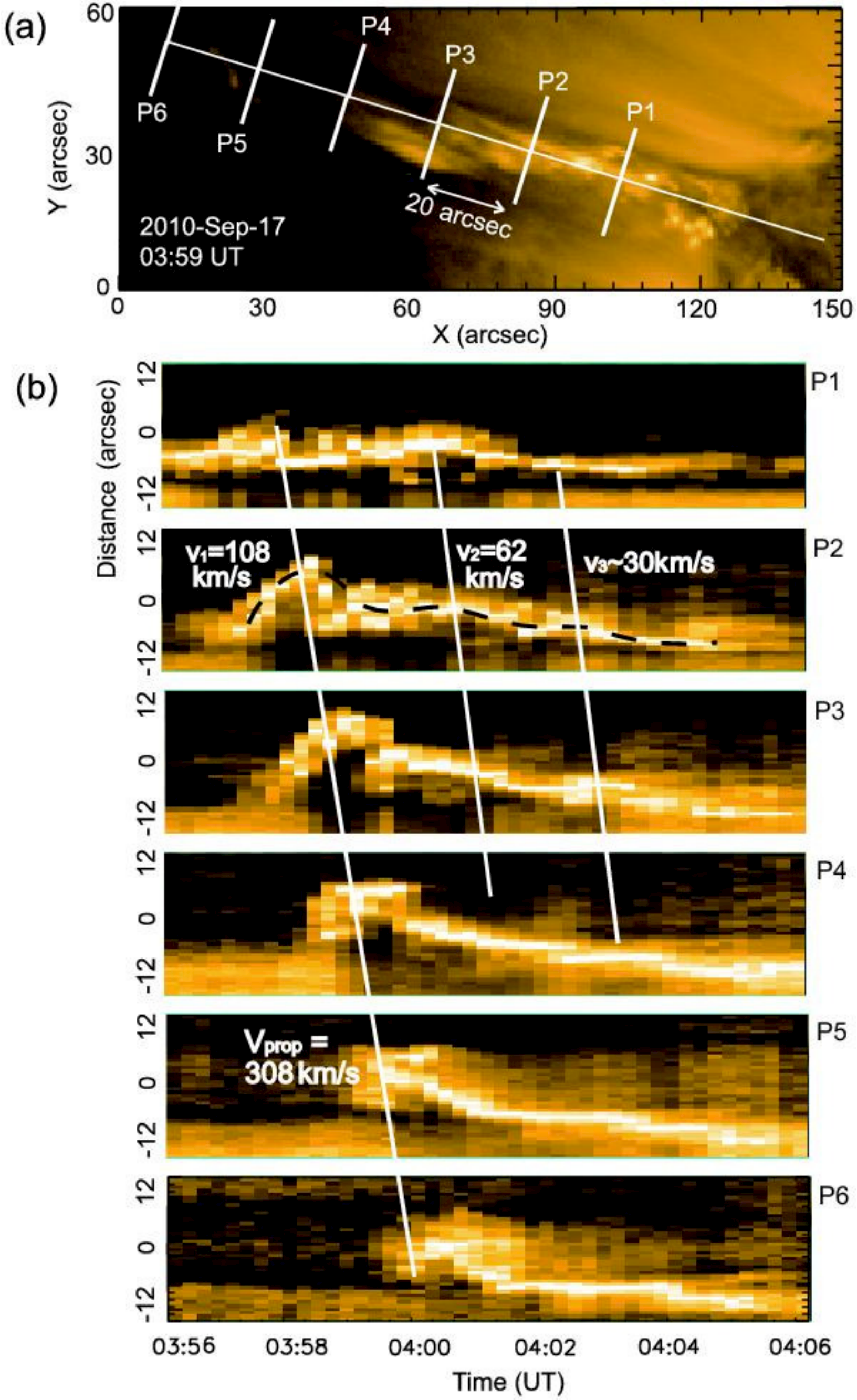}}
\parbox{0.35\textwidth}{\caption{AIA 171~{\AA} images of undulating jet, from 2010 September~17. Panel (a) shows six different heights for which distance-time plots are shown in (b).  In (b), the dashed black line highlights the undulating pattern, and the white lines show features tracked though the different panels at the indicated velocities. From \cite{2013A&A...559A...1S}.
\label{fig_schmieder_XXX} }}
\end{center}
\end{figure}

Similar to the earlier-mentioned \cite{2012ApJ...745..164S} paper, several other works have discussed winding or twisting motions of EUV jets observed in AIA  data (see Fig.~\ref{fig_schmieder_XXX}).  In this Section we highlight several of these papers, although some papers noted elsewhere also discuss twists in jets.  Such twists could be important with regards to the driving of jets, and perhaps may even be important for the energization and mass balance of the upper atmosphere.

\begin{figure}[!th]
\begin{center}
\parbox{0.4\textwidth}{\includegraphics[width=0.39\textwidth]{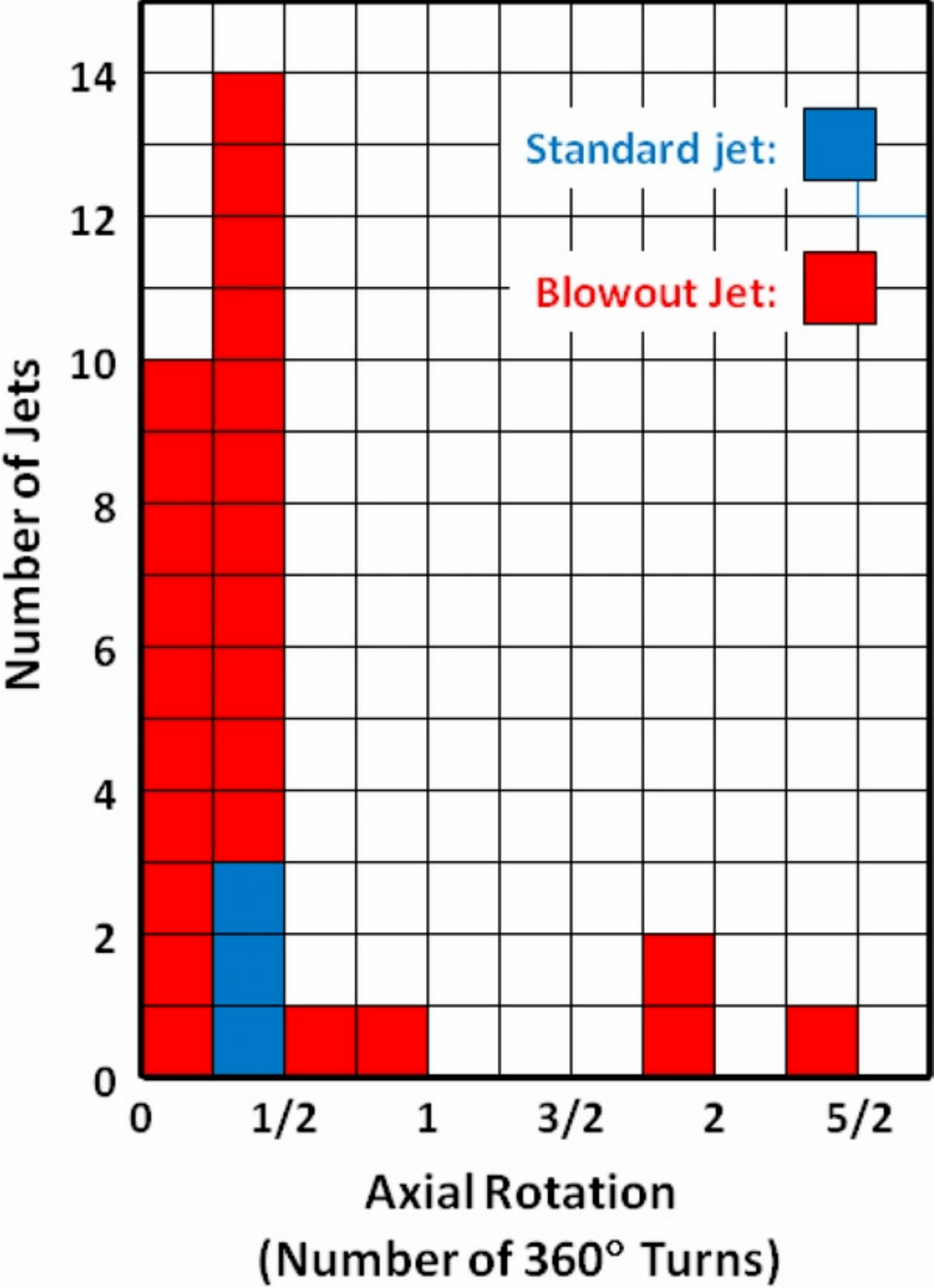}}
\parbox{0.45\textwidth}{\caption{Histogram showing amount of twist observed in AIA 304~{\AA} movies of X-ray coronal jets, giving the number of jets and the amount of rotation (twisting or unwinding) during the observed lifetime of the 304~{\AA} jets. From \cite{2013ApJ...769..134M}.   \label{fig:moore.et13_fig11_jet_twist_histogram}  }}
\end{center}
\end{figure}

Various studies reported twisting motions in coronal jets occurring in different regions (i.e., CHs, QS, ARs). While it has not been rigorously established that coronal jets in ARs are the same as those occurring in CH regions, the natural expectation is that most coronal jets would share similar or identical driving processes. Jets with $\ltsim\frac{1}{4}-2.5$ turns have been observed. Specifically, \cite{2011ApJ...735L..43S} showed a near-limb polar jet unwinding as it erupted, showing $\sim\!1-2.5$ turns. \cite{2013ApJ...769..134M} analyzed a large sample of 32 jets and found different degrees of jet twisting in 29 of their events: ten of these had modest twists of $\ltsim$ $\frac{1}{4}$ turns, 14 had twists of between $\frac{1}{4}$ and $\frac{1}{2}$ turns, while 5 jets had turns ranging from $\frac{1}{2}$ to $\frac{5}{2}$ turns (Fig.~\ref{fig:moore.et13_fig11_jet_twist_histogram}).

Velocities corresponding to the twisting motions of jets are found to range from a few km~s$^{-1}$ to $>100$~km~s$^{-1}$. \cite{2013ApJ...766....1L} observed an AR jet with a wide array of instruments and helical motions were observed, primarily in AIA 304~{\AA} images, with plane-of-sky velocities of $\sim$30--60~km~s$^{-1}$, that were decreasing with increasing height. Spectroscopic observations from {\it{Hinode}}/EIS show Doppler velocities blueshift- and redshift-velocities of $\sim$70~km~s$^{-1}$\ and 8~km~s$^{-1}$, respectively, and the authors point out that these shifts could be due to the helical motions. \cite{2014A&A...561A.134Z} also found twisting motions within a jet/surge event with an average $\sim$120~km~s$^{-1}$ that subsequently slowed down to $\sim$80~km~s$^{-1}$. An AR jet observed by \cite{2014ApJ...782...94L} showed twisting motions with velocities in the range $30-110$~km~s$^{-1}$. Other works \citep[e.g.,][]{2012RAA....12..573C,2013RAA....13..253H} also provide quantitative values for twisting motions in AIA-observed jets.

Higher temporal resolution and spectroscopic observations
could supply additional and important pieces of evidence for the
existence of twisting motions in jets. The distinctive signature of these motions
in Dopplergrams is positive and negative LOS-velocities side-by-side along the flow direction,
which is a common feature in cool jets (i.e., H-$\alpha$ Surge: \citeauthor{Ohman1968}
\citeyear{Ohman1968}, \citeauthor{1984ChA&A...8..294X} \citeyear{1984ChA&A...8..294X}, \citeauthor{1994A&A...282..240G} \citeyear{1994A&A...282..240G}, \citeauthor{1996ApJ...464.1016C} \citeyear{1996ApJ...464.1016C};
spray: \citeauthor{1987SoPh..108..251K} \citeyear{1987SoPh..108..251K}; macrospicule: \citeauthor{pike98} \citeyear{pike98},  \citeauthor{2010A&A...510L...1K} \citeyear{2010A&A...510L...1K};
 AR jet: \citeauthor{2012SoPh..280.417C} \citeyear{2012SoPh..280.417C}).
EUV observations, on the other hand, seems to point that untwisting motions may not be a
common property of EUV jets \citep{kamio07,matsui12}. This inconsistency between stereoscopic and
spectroscopic observations maybe due to the fact that spectrometers
can detect subresolution flows.

One other possibility is that the sensitivity, spatial resolution and time cadence
of current spectroscopic observations may not be sufficient for the measurement of untwisting motions.
If this is correct, untwisting motions should be significantly slower than the flow.
To understand the driving mechanism of coronal jets, especially to evaluate the contribution
of ${\bf{J}}\times{\bf{B}}$ force for accelerating a coronal jet \citep{1986SoPh..103..299S}, it is essential
to know the properties of untwisting motion. New instruments with better capabilities are also needed.
Recent {\it{IRIS}} spectroscopic  observations with their superior spatial and temporal resolution show
ubiquitous untwisting motions  at chromospheric level \citep{2014Sci...346D.315D},
and twisting motions in AR jets at transition region temperatures have been reported by \citet{2015ApJ...801...83C}.
However, {\it{IRIS}} may not be able to detect coronal signatures of twisting as the only
available coronal line (\ion{Fe}{xii} 1349~{\AA}) is too weak.

In addition to twisting motions, periodic dynamics, ranging from $\sim50$~s \citep{2012A&A...542A..70M} to $\sim20$~min \citep{2014ApJ...782...94L,2014A&A...561A.134Z},  have been also found in connection with these motions.

\subsubsection{Apparent Speed of Coronal Jets}
Since the flow speed is a key parameter for understanding the acceleration mechanism(s) of jets, different approaches have been adopted for the measurement of apparent speeds. Since polar jets are nearly-radial, the difference between the apparent and real speeds may be small. For AR jets, additional information, such as spectroscopic measurements, is needed. \citet{matsui12} and \citet{Sako2014} measured speed for relatively long jets ($<$ few $\times10^4$~km) using {\it{STEREO}} data. They found that apparent speeds are $10-20$\% smaller than real velocities. This indicated that using apparent speeds might be adequate in most cases keeping in mind that these velocities are the lower limits.

\begin{figure}[!ht]
\begin{center}
\includegraphics[width=.75\textwidth]{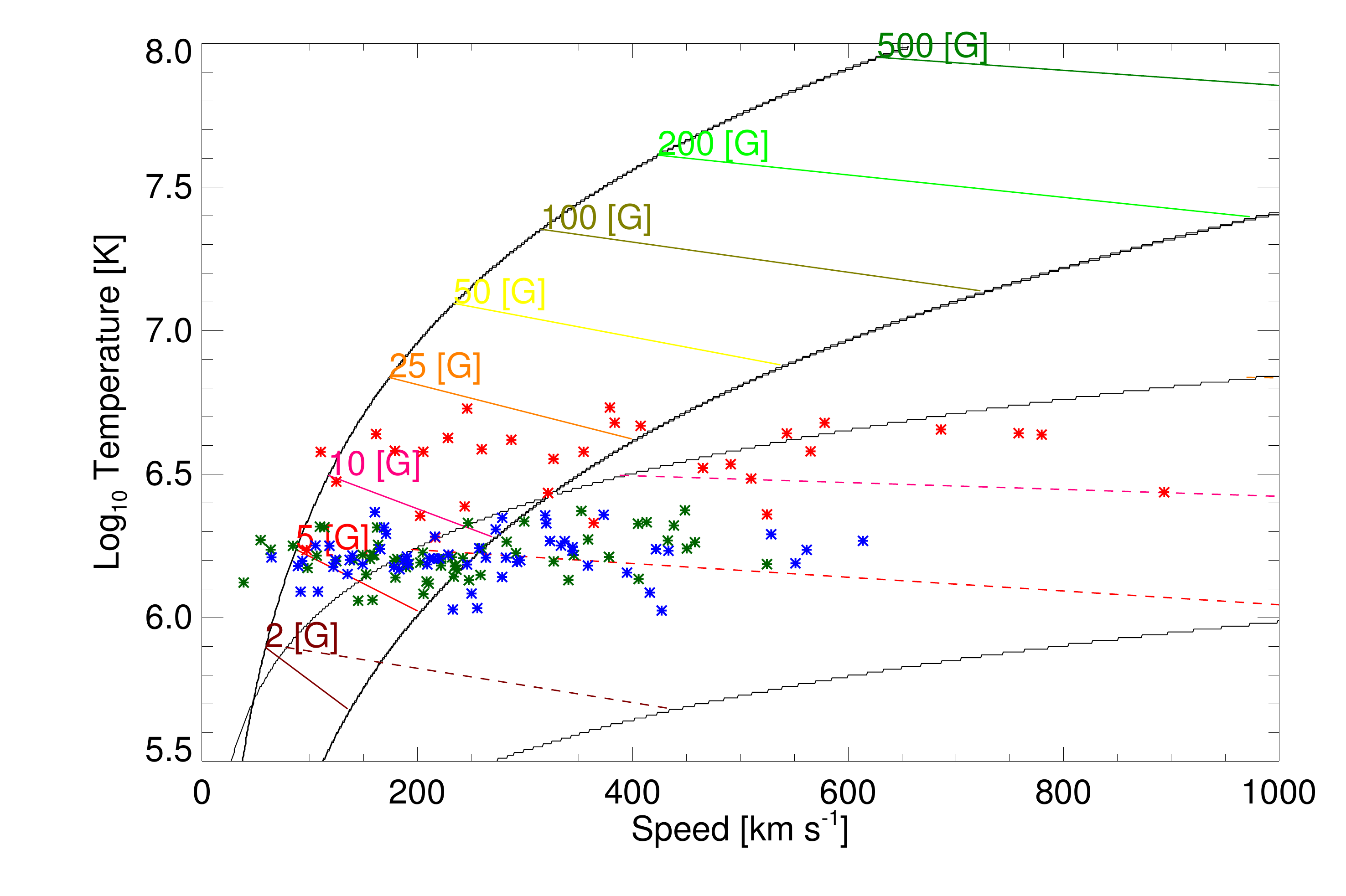}
\caption{Temperature-speed map for thermally- (evaporation) and magnetically-driven jets. The regions between the thin and solid curves correspond to the temperature-speed domains where magnetically- and thermally-driven jets occur, respectively. The intersection between the two regions indicates where both jet classes coexist. The red, green and blue symbols correspond to jets occurred in ARs, QS, and CHs, respectively. The colored solid/dashed lines indicate theoretical predictions of the magnetic field strength. From \citet{Sako2014}.}
\label{Sako2014_fg8h}
\end{center}
\end{figure}

Most measurements of jet apparent speeds are around 200~\kms, which is smaller than jet's sound speed. The 200~\kms\ is also comparable to the sound speed at coronal temperatures \citep[e.g.,][]{1996PASJ...48..123S,2007PASJ...59S.771S}. This led \citet{2000ApJ...542.1100S} to conclude that jets' high temperatures and dense flows are the result of chromospheric evaporation, which can also explain the jet mass. Chromospheric evaporation is therefore considered a strong candidate for jet acceleration. Numerical simulations of jets show the presence of flows with Alfv\'en speed \citep[$\sim\!1000$~\kms; e.g.,][]{1994ApJ...436L.197Y}. This has been confirmed by \citet{2007Sci...318.1580C} using {\it{Hinode}}/XRT observations. The high-speed component may not be fundamental as it appears intermittently during the jet flow evolution at the much lower apparent speed. The two flows are produced simultaneously in a jet, which has been predicted by \citet{1992PASJ...44L.173S}. It remains, however, unclear what component dominates and whether the fast component is magnetically driven. \citet{Sako2014} studied a number of long jets and classified them into thermal or magnetic dominated events based on comparison of the observed temperature -- speed relation with theoretical predictions (Fig.~\ref{Sako2014_fg8h}). He found that both classes exist all over the solar disk and that in ARs the number of thermally-driven jets is larger than the magnetically-driven ones. In CHs and QS, however, jets are distributed in the overlapping region of the temperature-speed map (Fig.~\ref{Sako2014_fg8h}). From their result, it is not clear what physical parameter is most important for the selection of the jet dominant component. The magnetic field strength is an important parameter but not critical since both jet classes occur in all regions. Magnetic topology and free energy may be necessary to address this question.

Six AIA~304~{\AA} coronal jets (it is not clear whether these jets had hotter counterparts) analyzed by \citet{2013SoPh..284..427M} are found to accelerate at a fraction of the value of solar gravity, i.e., their dynamics are determined by forces other than gravity \citep[see also][for a similar study]{2014A&A...561A.134Z}. Two of these jets displayed helical patterns during ejection. A PCH cool jet (with a lifetime of $\sim$21~min, and a height of $\sim$72~Mm) was studied by \cite{2011A&A...534A..62S}. They were able to reproduce quantitatively the jet dynamics with a 2-D MHD simulation by launching a velocity pulse in the low atmosphere. 

\section{Relationship to other coronal structures}

\subsection{Jet -- Plume Connection}

Coronal jets are characterized by their transient nature and often appear as collimated beams presumably
guided by open magnetic fields. In contrast, coronal plumes, which are most prominent and pervasive within CHs,
are hazy without sharp edges as seen in EUV 171~{\AA} and WL images. They are also significantly wider than
jets \citep[$\sim\!20-40$~Mm;][]{2006A&A...455..697W} , may reach up to $\sim\!30\ R_\odot$ above the
solar surface \citep[][]{1997SoPh..175..393D}, and are quasi-stationary during their lifetime spanning in the order of days.
Prior to the {\it{STEREO}}/{\it{Hinode}} era, coronal jets and plumes have been studied independently
and any eventual interrelationship has not been considered. They, however, share common characteristics.
They are both episodic in nature and are usually rooted in the chromospheric network where
magnetic reconnection is thought to play a key role in their formation \citep[][]{1996ApJ...464.1016C,1998ApJ...501L.145W}.
The connection between jets and plumes is important for understanding their formation processes and evolution
and their eventual contribution to the heating and acceleration of the SW.

\begin{figure}[!h]
\begin{center}
\parbox{0.6\textwidth}{\includegraphics[width=0.6\textwidth]{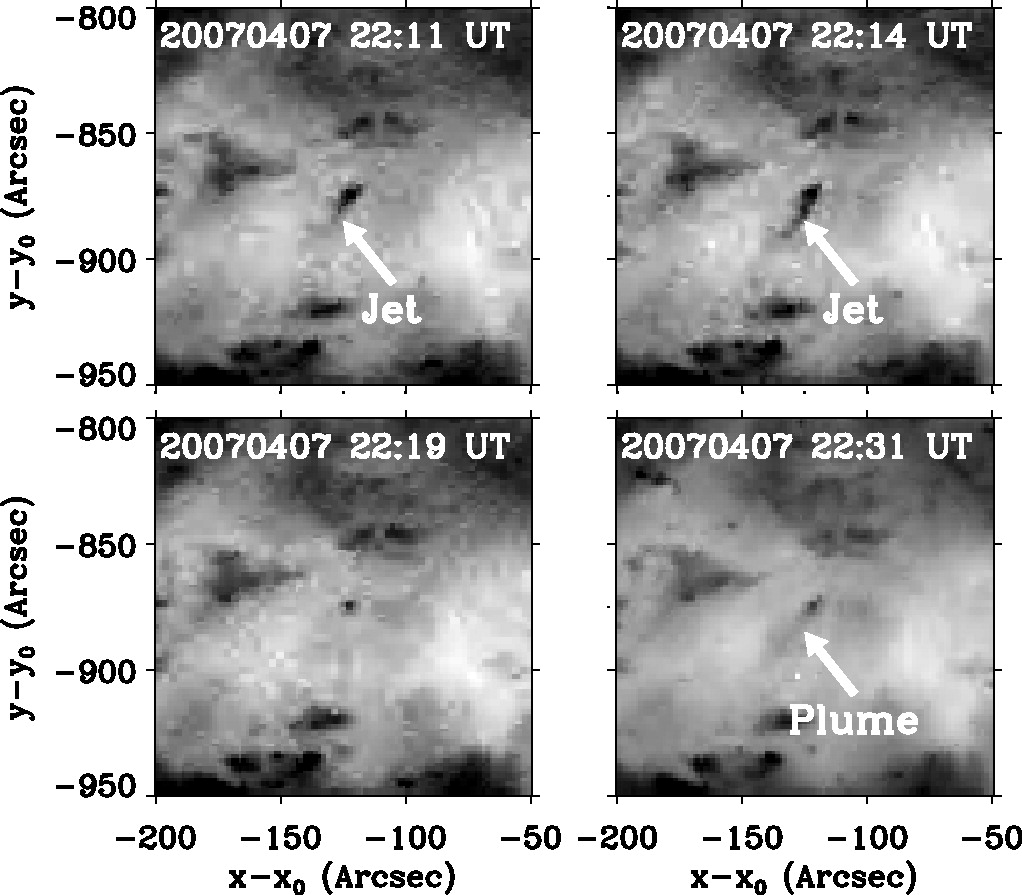}  }
\parbox{0.33\textwidth}{
\caption{EUV 171~{\AA} images from SECCHI/EUVI on {\it{STEREO}}-A illustrating the connection between coronal jets and plumes. Jets are often observed to erupt prior to and during the lifetime of plumes. Top panels: eruption of a CBP into a jet (white arrows). Bottom-left: disappearance of the coronal jet. Bottom-right: appearance of plume haze (white arrow). A time lag ranging from minutes to several tens of minutes is typically observed between the jet disappearance and plume formation. From \citet{2008ApJ...682L.137R}.\label{Fig2_NER2008}}  }
\end{center}
\end{figure}

\cite{2008ApJ...682L.137R} investigated the jet-plume relationship using {\it{Hinode}}/XRT and {\it{STEREO}} EUV observations that were recorded during the deep solar minimum (2007 Apr. 7-8; see Fig.~\ref{Fig2_NER2008}). A total of 28 X-ray jets were analyzed. \cite{2008ApJ...682L.137R} discovered that $>90\%$ of the jets were associated with plume haze and $\sim\!70\%$ of these jets are followed by polar plumes with a time delay ranging from minutes to tens of minutes. \cite{2008ApJ...682L.137R} and \cite{2009ASPC..415..144R} argue that coronal jets are precursors of plumes. They also noted that jet eruptions within the footpoints of preexisting plumes cause an enhancement of the plumes' brightness. In addition, short-lived, jet-like events occur within the footpoints of plumes. \cite{2008ApJ...682L.137R} argue that these jet-like events ensure the continuous rise of haze and may contribute to the change in plume brightness \citep[see][]{1997SoPh..175..393D}. Their interpretation of the observations is that jets result from impulsive magnetic reconnection of emerging flux \citep[e.g.,][]{1995Natur.375...42Y}, where plumes may be the result of lower-rate magnetic reconnection as shown by the short-lived, small-scale BPs and jet-like events observed within their footpoints.

More recent findings \citep[e.g.,][]{2010A&A...510L...1K,wilhelm11} provide evidence
that polar jets occur in the {\it interplume} region next to a plume and that
the jet plasma probably feeds the plume.
SUMER observations \citep{{2002A&A...382..328W}} revealed an increased line width in plumes, but no net flow in plumes.
\citet{2015JApA...36..185D} presented a model assuming a plume as quasi-closed volume,
where plasma of the polar jet is trapped and continuously moves up and down, thus operating the strong FIP effect.

Recently, \cite{2014ApJ...787..118R} took advantage of the high-quality data from the {\it{SDO}}/AIA and HMI to analyze the coronal conditions leading to the formation and evolution of coronal plumes. Prior to the {\it{SDO}} era, the spatial and temporal resolution of the imaging instruments were limited and therefore coronal plumes and their fine structure could not be fully resolved and their temporal evolution could not be analyzed in detail. This might be the reason that coronal plumes are often referred to in the literature as ``hazy" structures and heuristic assumptions on the plasma heating and acceleration at their source regions are typically the norm in numerical models.

\cite{2014ApJ...787..118R} focused particularly on the fine structure within plume footpoints and on the role of transient magnetic activity in sustaining these structures for relatively lengthy periods of time (several days). In addition to nominal jets occurring prior to and during the development of plumes, the data show that a large number of small jets (``jetlets") and plume transient BPs (PTBPs) occur on timescales of tens of seconds to a few minutes. These features are the result of quasi-random cancellations of fragmented and diffuse minority magnetic polarity with the dominant unipolar magnetic field concentration over an extended period of time. They unambiguously reflect a highly dynamical evolution at the footpoints and are seemingly the main energy source for plumes. This suggests a tendency for plumes to be dependent on the occurrence of transients (i.e., jetlets, and PTBPs) resulting from low-rate magnetic reconnection. The present findings may be of great importance for the interpretation of future measurements by different instruments on board the Solar Probe Plus and Solar Orbiter. These future measurements may provide further details on the plasma thermodynamics, such as heating and acceleration of SW particles within coronal plumes.

\subsection{Jet-Sigmoid Connection}
Historically, the distinctive collimated structure of coronal jets inspired the ``anemone" model, in which the jet is the result of a dipolar magnetic structure reconnecting with a slanted background field \citep{2000ApJ...542.1100S} and that the previously trapped plasma is channeled into a collimated beam on open magnetic field lines \citep{1995Natur.375...42Y,1998ApJ...495..491K}. Although this model has been shown to reproduce many morphological features of coronal jets, it also exhibits shortcomings in explaining other properties such as helical structures \citep[see][]{1996PASJ...48..123S,1996ApJ...464.1016C,2008ApJ...680L..73P,2009SoPh..259...87N} and apparent transverse motions \citep[see][]{Savcheva09} of numerous jet events.

\begin{figure*}[!ht]
\begin{center}
\parbox{0.55\textwidth}{\includegraphics[width=0.55\textwidth]{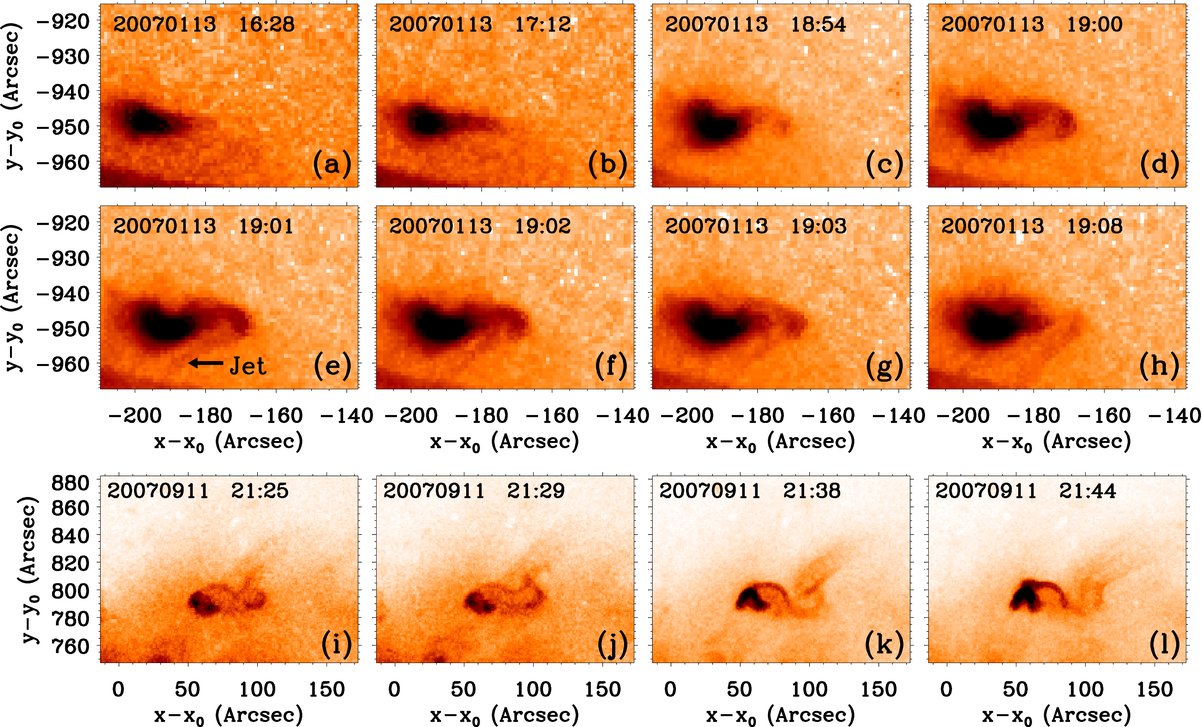}}
\parbox{0.32\textwidth}{\caption{{\it{Hinode}}/XRT observations of XBPs evolving into micro-sigmoid and then erupting into jets. For details see \citet{2010ApJ...718..981R}.\label{Fig2_NER2010}  }}
\end{center}
\end{figure*}

\cite{2010ApJ...718..981R} used observations from XRT, EIS, and EUVI to study the morphology and fine structure of XBPs leading to coronal jets in conjunction with their characteristics (i.e., untwisting motions, transverse apparent motions, etc.). The resolved structure of some BPs is more complex than previously assumed. Several jet events in the CHs are found to erupt from small-scale, S-shaped bright regions (see Fig.~\ref{Fig2_NER2010}). This finding suggests that coronal small-scale (micro-) sigmoids \citep{1996ApJ...464L.199R} may well be progenitors of coronal jets. Moreover, the presence of these structures may explain numerous observed characteristics of jets such as helical structures, apparent transverse motions, and shapes.

Patsourakos \& Raouafi (2016, in preparation) investigated the case of a small sigmoidal BP within an ECH. The underlying magnetic fields show significant emergence as well as cancellation that lead to the formation of the sigmoid, which lasted for several hours. The full eruption of this sigmoid was the result of the cancellation of large magnetic flux that resulted in a large jet. This analysis shows that the evolution of the small-scale coronal features have similarities with large magnetic regions.

\section{Jets, Solar Wind, and Solar Energetic Particles}

\subsection{Association Between Jets and Impulsive Solar Energetic Particle Events}
Solar energetic particle (SEP) events are typically put into classes: ``gradual" and ``impulsive". The former are intense, long-lasting, and are generally associated with large, fast CMEs. They are less correlated with impulsive flares and are characterized by the abundances and charge states of the SW. The continuous injection and acceleration of the particles are believed to originate at the CME-driven shock wave \cite[e.g.,][]{1984JGR....89.9683K}. Impulsive SEP events, on the other hand, are smaller, less intense, last less than a day, and are characterized by high charge states \citep[Fe $Q=20.5\pm1.2$, see][]{1987ApJ...317..951L,1988ApJ...327..998R}. The particles seem to originate from source regions with temperatures $\gg1$~MK (e.g., flare sites with $T>10^7$~K). They are typically rich in heavy ion species relative to coronal abundances (e.g., ${\rm{Fe/O}}\!\sim\!10$) and are often referred to as ``$^3$He-rich events" because of their high $^3{\rm{He}}/^4{\rm{He}}$ ratio of $\approx\!10^3$ \citep[see][]{1970ApJ...162L.191H}. Impulsive SEP events are also well correlated with metric and kilometric type III radio bursts \citep[see][]{1987SoPh..107..385K}. They are particularly found to coincide also with X-ray jets \citep{1995ApJ...447L.135K,1996A&A...306..299R,2012ApJ...754....9G}. A review by \cite{1999SSRv...90..413R} provides details on the characteristics of these two classes of SEP events.

Historically, the lack of understanding of the origin of impulsive SEP events may be due to the lack of high-resolution, both temporal and spatial, solar observations that are necessary for the identification of small-scale activity. SEPs during the latest solar maximum were observed with unprecedented precision and breadth by a new generation of instruments on ACE, Wind, SAMPEX, {\it{SOHO}}, {\it{TRACE}}, {\it{Hinode}}, and {\it{RHESSI}}.

\begin{figure*}[!h]
\begin{center}
\includegraphics[width=0.9\textwidth]{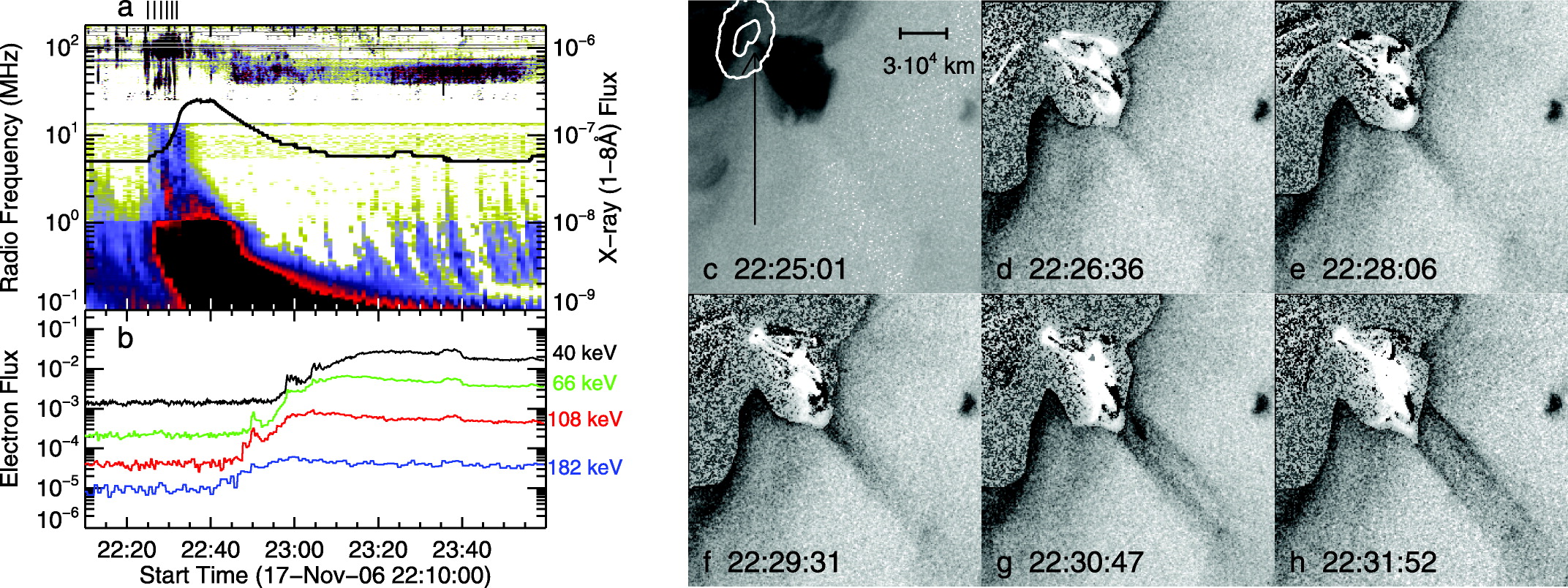}
\caption{Evolution of a coronal jet coinciding with a type III burst and an electron event. (a) Radio dynamic spectra and X-ray light curve. (b) The electron flux at 1 AU. (c) XRT negative intensity image of the jet source region. (d-h) Running-difference images illustrating the evolution of the flare and the jet. From \citet{2008ApJ...675L.125N}.\label{Fig2_NittaEtAl2008}}
\end{center}
\end{figure*}

\cite{2006ApJ...639..495W} investigated the solar sources of 25 $^3$He-rich events through the analysis of EUV and WL coronagraph observations from {\it{SOHO}}/EIT and LASCO, respectively. They also used a potential field source surface (PFSS) magnetic field model to determine the magnetic connectivity of Earth to the solar surface, as well as \ion{He}{i} 10830~{\AA} images to locate CHs. They suggested that $^3$He-rich events originated from coronal jets erupting at the interface between ARs and adjacent open field regions.

\cite{2006ApJ...650..438N} studied 117 impulsive SEP events between Dec. 1994 and Dec. 2002. They used particle measurements from the WIND/EPACT/LEMT \citep{1995SSRv...71..155V}, solar EUV and X-ray images from {\it{SOHO}}/EIT and {\it{Yohkoh}}/SXT, Type III radio burst observations, as well as PFSS models. They argue that most of these events have clear velocity dispersions, which is suggestive of new injection at the Sun. They found solar sources in 69 events where solar images show jets erupting shortly after the type III bursts. Open field lines are found in around 80\% of the source regions, but only in 40\% of the cases open field lines are found close to both the source regions and the Parker spiral coordinates at the source surface. Other events are found associated with CMEs and type II radio bursts \citep{2004ASPC..325..401Y} and the enhancement of ultra-heavy elements \citep{2001ApJ...562..558K,2004ApJ...610..510R,2006ApJ...648.1247P}. \cite{2006ApJ...639..495W} argue that in some cases these CMEs are faint, narrow, and resemble WL jets \citep{2002ApJ...575..542W}. The most convincing case of connection between coronal jets and $^3$He-rich SEP events is found by \cite[][Fig.~\ref{Fig2_NittaEtAl2008}]{2008ApJ...675L.125N} through the use of {\it{Hinode}}/XRT images. The close temporal correlation of the jet with both the type III bursts at metric to kilometric ranges and the electron event strengthens the link between the jet and the $^3$He-rich SEP event.

Similarly, \cite{2011A&A...531L..13I} found the source of quasi-periodic type~III radio bursts seen in WIND/WAVES dynamic spectra to be EUV jets observed in {\it{SDO}}/AIA 211~{\AA} images. \cite{2013ApJ...769...96C} also studied a jet with AIA and HMI that occurred during a \goes\ C-class flare and that generated a {\it{RHESSI}}-observed HXR event, and a type~III radio burst observed by WIND/WAVES and two ground-based systems. From a differential emission measure (DEM)  analysis, they found that very high temperatures ($\sim$10$^7$~K) were produced at the base of the jet, which was also the location of the HXR source. The HXRs, EUV emissions, and radio bursts occurred nearly simultaneously.  For our purposes here perhaps the key finding of this work is that in many ways emissions from the base of this jet seem to mimic emissions and characteristics found in ``normal'' flares. The implication is that these jets gave rise to interplanetary electron streams. 

\subsection{Contribution of Jets to the Solar Wind}

Although we have learnt a great deal about coronal jets during the last two or three decades through improved quality X-ray and EUV observations, this knowledge remained confined to very low latitudes in the solar corona. \cite{2007Sci...318.1580C} analyzed the dynamics of a number of prominent X-ray jets and found two speed components. The higher speed component is comparable to the local Alfv\'en velocity in the low corona, suggesting that Alfv\'en waves may be an energy source for the acceleration of jets. The acceleration mechanisms of the jet plasma at high altitudes and their mass and energy inputs into the SW remain unclear. Only a handful of studies of coronal jets at high coronal latitudes are available in the literature. 

\cite{1992PASJ...44L.173S} found through analysis of time series of {\it{Yohkoh}}/SXT images that the typical jet size ranges from $5\times10^3$~km to $4\times10^5$~km and their velocity in the range $30$ to $300~{\rm{km~s}}^{-1}$, which correspond to kinetic energy $10^{25}-10^{28}$~erg. A more recent study by \cite{2007ApJ...659.1702C} of cool jets using {\it{SOHO}}/UVCS spectral observations shows that the mass estimate of these jets at $1.7~R_\odot$ is of the order of $10^{13}$~g. \cite{2012ApJ...754....9G} reported on the first observation of several hard X-ray coronal sources in an interchange reconnection type jet event. The event occurred on the solar limb with flare footpoints occulted. They found that plasma temperature during the impulsive phase as high as 13 MK and that early electrons were accelerated to tens of keV. The jet velocity is $417\pm73~{\rm{km~s}}^{-1}$ and the non-thermal electron energy as calculated with the thin-target model is estimated to be $10^{28}~{\rm{erg}}$.

\cite{1998ApJ...508..899W} discovered narrow, radially-moving jet-like features using coronagraph images recorded by LASCO-C2 coronagraph above the solar poles. Through analysis of EUV observations, they found that these WL features are the manifestations of EUV jets observed by EIT. They also found that the leading edges of the WL jets propagate outward at speeds ranging from 400 to 1100 km~s$^{-1}$ (median $\sim\!590$ km~s$^{-1}$), whereas their centroidal velocities are much lower, ranging from 140 to 360 km~s$^{-1}$ (median $\sim\!260$ km~s$^{-1}$) in the region $2.9$ to $3.7~{\rm{R}}_\odot$. They argue that the large difference in velocities between the leading edge and the bulk of the jet material is the main cause of the jet elongation as it propagates away from the solar surface. It is clear that the velocity of the bulk of the jet material is significantly less than the escape speed, and added to the fact that there is no evidence for jet material falling back to the solar surface suggests that in situ acceleration prevents this material from falling back to the solar surface. The relatively narrow range of centroidal velocities ($v_{cen}\simeq250\pm110~{\rm{km~s}}^{-1}$) measured around 3.3 $R_\odot$ and the absence of any correlation between $v_{cen}$ and $v_{lead}$ suggest that the bulk of the jet material moves through the C2 FOV at the speed of the background SW. The last conclusion is potentially the most important of this study, since it raises the possibility that the jets can be used to determine the dynamical properties of the polar wind itself in the immediate vicinity of the Sun.

\cite{2014ApJ...784..166Y} used {\it{Hinode}}/XRT, {\it{SOHO}}/LASCO-C2, {\it{STEREO}}/COR2 and the Solar Mass Ejection Imager \cite[SMEI;][]{2003SoPh..217..319E,2004SoPh..225..177J} observations to trace the evolution of three large X-ray jets in Sep. 2007. \cite{2014ApJ...784..166Y} argue that they are in fact tracing the fast component of the jet through the solar corona \cite[see][]{1999ApJ...523..444W,2007Sci...318.1580C}. The analysis of the SMEI WL data shows that all three jets have similar mass and energy, averaging $1.3\times10^{14}$~g and $1.8\times10^{29}$~erg, respectively. \cite{2014ApJ...784..166Y} found that the jets contribute $6\times10^{15}$~g to the SW mass within three weeks. They argue that the jets contribute $\sim\!3.2\%$ of the mass of SW and $\sim\!1.6\%$ of the SW energy.

It has been reported that coronal jets cause disturbances in the SW \citep{2012ApJ...750...50N,2014ApJ...784..166Y}. However, according to the frequency distribution of coronal jets occurring in PCHs, the total visible energy (total kinetic and thermal energy of coronal plasma) of coronal jets is not sufficient to accelerate the fast SW \citep{2013ApJ...775...22S}. Furthermore, plasma of a coronal jet falls down to the solar surface after cooling because the speed of most coronal jets does not exceed the escape velocity \citep{culhane07-jet,2014A&A...561A.104C}. Additional, and possibly important contributions to the SW energetics, could arise from the consideration of other jet-related sources of energy such as waves, as well of faint and small-scale jets  \citep[i.e., ``jetlets",][]{2014ApJ...787..118R}.

The results by \citet{culhane07-jet} and \citet{2014A&A...561A.104C} seem to contradict those by \cite{1998ApJ...508..899W} concerning the existence or not of jet material falling back to the solar surface. It is, however, noteworthy that \citet{culhane07-jet} and \citet{2014A&A...561A.104C} studied jets very close to the solar surface using {\it{Hinode}}/EIS and XRT whereas \cite{1998ApJ...508..899W} analyzed coronal jets a significantly higher coronal altitudes using mainly WL images from LASCO-C2.

\begin{figure*}[!h]
\begin{center}
\parbox{.67\textwidth}{\includegraphics[width=.65\textwidth]{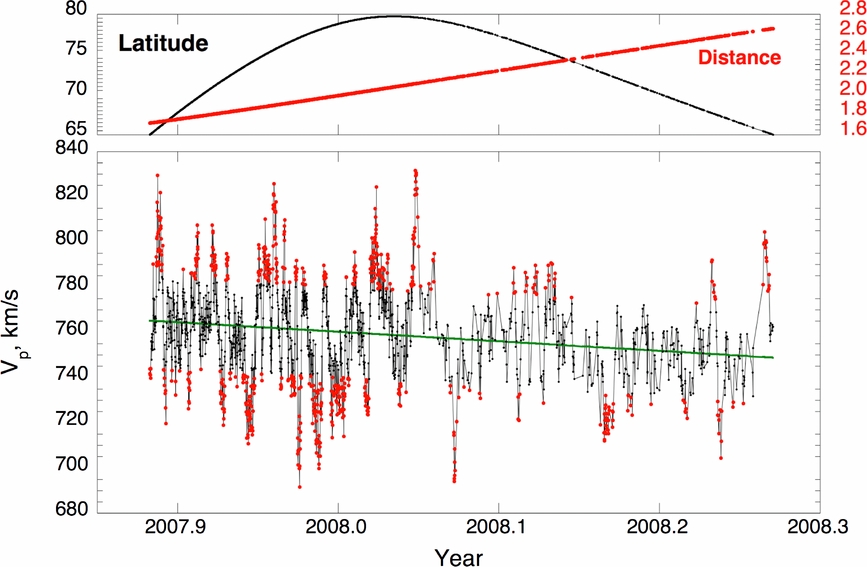}}
\parbox{.32\textwidth}{\caption{Top: heliographic latitude (black, in degrees) and distance (red, in AU) of the Ulysses spacecraft {\it{vs.}} year. Bottom: hourly averages of proton speed {\it{vs.}} year. The green line is the average SW speed. Red symbols mark hours for which the speeds were more than $\pm20~{\rm{km~s}}^{-1}$ above or below the average. From \citet{2012ApJ...750...50N}.\label{Fig_Neug2012}
}}
\end{center}
\end{figure*}

In-situ measurements at $>1$~AU show that the fast polar SW has much less structure than the slow wind \citep{1977JGR....82.1487B}. It is, however, not totally uniform, but exhibits some structures that are probably of solar origin \citep[see Fig.~\ref{Fig_Neug2012};][]{2012ApJ...750...50N}. Such structures include the so-called micro-streams \citep{1995JGR...10023389N}. \cite{1995JGR...10023389N} and \cite{2012ApJ...750...50N} showed that micro-streams  exhibit velocity fluctuations of $\pm35~{\rm{km~s}}^{-1}$ and are characterized by mean half-width of 0.4 days, recurrence on timescales of two to three days, higher plasma temperatures, density and temperature pileup on positive velocity gradients and expansion otherwise, and no latitude dependence of temporal duration or recurrence rate. These properties, together with increased abundance of low-FIP elements, led \cite{2012ApJ...750...50N} to conclude that fast-SW micro-streams are related to episodic rather than quasi-stationary sources. Unlike previous analyses which were rather inconclusive about the origin(s) of these structures (e.g., polar plumes, jets, BPs), \cite{2012ApJ...750...50N} suggests that the detected structures are of solar origin and their properties can be explained if the fast micro-streams result from the magnetic reconnection of BP loops, which leads to jets. In this work jets are favored over plumes for the majority of the micro-stream peaks.

\section{Models of Jet Formation Mechanisms}\label{ModelSection9}

The explosively dynamic nature, morphology, and magnetic environment of coronal jets has led to a broad consensus that they result from magnetic reconnection occurring in the solar corona. The common factor in the numerous models that have been explored is the existence of a null point (or null line, in systems with axial symmetry) in the coronal magnetic field configuration that gives rise to the jet. A coronal null is a region that is susceptible to the build-up of thin, strong current sheets where reconnection can occur in an explosive manner \citep[e.g.,][]{1990MmSAI..61..369A,1990ApJ...350..672L}. Such a region forms naturally whenever a bipole of one (parasitic) polarity is embedded within a larger-scale domain of the opposite polarity; this background flux consists of magnetic field lines that may be open to the heliosphere (in a CH) or closed back to solar surface (in a larger-scale loop). The presence of a null prior to a jet has been inferred from the specific shapes of the jet emission in X-rays \citep{1992PASJ...44L.173S} and the circular flare ribbons observed in the chromosphere at the base of some jets \citep{2012ApJ...760..101W} and confirmed from magnetic field extrapolations \citep{2008ApJ...673L.211M,2011ApJ...735L..18L,2012ApJ...746...19Z,2013A&A...559A...1S}.

Two principal scenarios have been investigated for the occurrence of coronal jets. The first is {\it flux emergence} from below the photosphere, in which the newly emerging field collides with the ambient coronal field and, where favorably oriented, the two flux systems reconnect. In this scenario, the reconnection and the resulting jet are driven directly by the flows (vertical and horizontal) associated with flux emergence, which is accompanied by an increase in the amount of unsigned vertical magnetic flux in or adjacent to the jet source. The second is {\it onset of instability or loss of equilibrium}, in which the stressed, nonpotential, closed flux beneath the null point begins to reconnect with the ambient, quasi-potential flux exterior to the fan surface. The reconnection occurs explosively, after some critical threshold is reached in response to quasi-static footpoint motions. In this scenario, the reconnection is initiated spontaneously by an internal rearrangement of the coronal field in the closed flux system, and there need be no change whatsoever in the amount of unsigned vertical magnetic flux in or near the jet source. In both scenarios, the jet is driven by magnetic energy released via reconnection between the magnetic fields of the jet-source region and the surrounding corona. Three basic processes that occur, alone or in combination, during the reconnection episodes are the {\it slingshot}, {\it untwisting}, and {\it evaporation}.

In the {\it slingshot} mechanism, the plasma in the immediate vicinity of the reconnection site is accelerated to Alfv\'enic velocities by the rapid motion of the newly reconnected field lines as they straighten under magnetic tension. The outflowing plasma also can be heated due to the energy released at the reconnection site. Multiple models have been studied that simulate the physics of the {\it slingshot} mechanism \citep[e.g.,][]{1996PASJ...48..353Y,2008ApJ...673L.211M,2008ApJ...683L..83N,2013ApJ...769L..21A,2013ApJ...777...16Y}.

The {\it untwisting} mechanism relies on the presence of shear and/or twist in the closed region beneath the null point prior to reconnection onset. When reconnection occurs, the newly reconnected open field lines are twisted at the low-atmosphere end and untwisted at the heliosphere end. This creates conditions for the propagation along the open (or large-scale closed) field lines of a nonlinear Alfv\'en wave, whose magnetic pressure accelerates and compresses the plasma upstream of the wave as it travels through it. Depending upon the dimensionality of the system, the Alfv\'en wave that develops along the untwisting field lines can be a shear wave \citep[e.g.,][]{1995ApJ...450..422K,1998ApJ...495..491K} or a torsional wave \citep[e.g.,][]{2009ApJ...691...61P,2010ApJ...714.1762P,2015A&A...573A.130P,2009ApJ...704..485T,2013ApJ...769L..21A,2013ApJ...771...20M,2014ApJ...789L..19F,2015ApJ...798L..10L}.

The {\it evaporation} mechanism results indirectly from the reconnection process, which deposits energy into the surrounding plasma in the form of heat, accelerated energetic particles, and compressive flows. The energy and particles from the reconnection site travel down to and are deposited in the chromosphere, increasing the pressure and temperature. This creates an evaporation flow, which supplies plasma to the closed and open field lines. The jet is then accelerated by the pressure gradient along the magnetic field. Relatively few existing numerical studies include thermal conduction, which is necessary for producing the evaporation jet \citep[e.g.,][]{2004ApJ...614.1042M,2014ApJ...789L..19F}.

It is possible that all three of these processes play a role during jets, but to different extents in different magnetic configurations and/or during different phases of a single jet event. For example, it is plausible that jets exhibiting helical motions primarily reflect the occurrence of the {\it untwisting} mechanism \citep{2010ApJ...714.1762P,2013ApJ...771...20M, 2013ApJ...769L..21A}, whereas straight jets may instead be due to the {\it slingshot} mechanism \citep{2015A&A...573A.130P}; both may have some contribution from the {\it evaporation} mechanism. The observed rotations and wavy motions in some jets seem to be well explained by the untwisting mechanism \citep{1996ApJ...464.1016C,2004ApJ...610.1129J,2007Sci...318.1580C,2008ApJ...680L..73P,2009ApJ...707L..37L,2010A&A...510L...1K,2012RAA....12..573C,2013RAA....13..253H}.

Since the 1990s, multiple modeling strategies aimed at reproducing the mechanism and characteristics of coronal jets have been proposed. The seminal Shibata schematic \citep[][]{1992PASJ...44L.173S} of a coronal jet due to flux emergence identified the importance of reconnection in the jet phenomenon. Developments in the intervening years include one-dimensional hydrodynamic (1D HD) and multidimensional MHD (2D, 2.5D, and 3D) simulations. Different MHD simulations have included various effects, such as a range of plasma beta values, viscosity, resistivity, gravity, plasma heat conduction, and radiation. More recently, magnetofrictional simulations have been used to study the equilibrium magnetic field structure and topology associated with jets.

\subsection{Hydrodynamic (HD) Simulations}

Most MHD simulations have been performed neglecting heat conduction, mainly due to the rather high computational cost. On the other hand, heat conduction is essential for comparing simulations with observations, because the spatial distribution of X-ray and EUV brightness depends on the distribution of temperature and emission measure that are controlled by heat conduction and radiation. \citet{2001ApJ...550.1051S} assumed that a coronal jet is a hot plasma flow that is created by chromospheric evaporation in a flux tube, whose shape does not change, and performed 1D HD simulations including heat conduction and radiation. The result of their simulations shows that the intensity distribution along an X-ray jet cannot be reproduced by injecting energy into a single tube. The inconsistency is caused by heat conduction that carries energy to the transition region and upper chromosphere quickly. Based on the reconnection picture of coronal jets, they constructed a pseudo-2D model in which different flux loops are heated successively. This simulation could reproduce the observed intensity distribution along an X-ray jet. The HD approach is, however, superseded by more sophisticated approaches (i.e., MHD models) as described in the following sections.

\subsection{MHD Simulations: Flux Emergence Scenario}

\subsubsection{2D Simulations}

One mechanism that has been suggested to lead to jets is the emergence of flux into a preexisting open or closed field. \citet{1992PASJ...44L.173S} showed that a new small loop system appeared by the side of a larger emerging loop system during an X-ray jet observed by {\it{Yohkoh}}/SXT. The observations suggest that magnetic reconnection occurred between the emerging loop system and the ambient vertical coronal fields. The configuration is similar to the emerging flux model of a confined flare, which was proposed by \citet{1977ApJ...216..123H}. \citet{1995Natur.375...42Y, 1996PASJ...48..353Y} performed 2D MHD simulations using two magnetic initial configurations: one is an {\it anemone} type, the Shibata model with vertical ambient fields, and the other one is a {\it two-sided} type that occurs during the reconnection between emerging flux and horizontal ambient fields. They succeeded in reproducing not only a hot jet but also a cool jet simultaneously. The result is consistent with the observations that show a H-$\alpha$ surge occurring simultaneously with a coronal jet. The hot jet is not a reconnection outflow directly. At first, the reconnection outflow collides with coronal fields, and then produces a fast-mode MHD shock. The outflow is deflected by the shock, and becomes a hot jet along the large-scale coronal magnetic field. The reconnection outflow is not only diverted but additionally accelerated by pressure gradients. The plasma of the cool jet is provided by emerging flux, which carries chromospheric plasma to the corona without heating in the process. When the magnetic reconnection occurs, chromospheric plasma is ejected from the emerging flux along the coronal fields. Another important result from their simulations is the plasmoid ejection from the current sheet. During the evolution of the current sheet between the emerging flux and the coronal fields, the magnetic island (plasmoid) is formed by tearing and coalescence instabilities. When the plasmoid ejects from the current sheet, the reconnection rate suddenly increases and the main energy release phase, which includes the formation of a coronal jet, starts. Based on these results, \citet{1994ApJ...436L.197Y} proposed that the plasmoid ejection is a key process for producing fast reconnection. Since their simulation is two dimensional, the plasmoid completely disappears during the reconnection with the coronal fields. Nevertheless, the plasmoid ejection in their simulations may account for at least some blow-out jets.

\citet{2001ApJ...549.1160Y} developed a 2D MHD simulation code including the effects of heat conduction and radiation, and compared the results of the simulations with the standard model of flares. \citet{2004ApJ...614.1042M} used the code for simulating a coronal jet. They performed the simulations of an emerging flux with horizontal coronal fields (two-sided type). Their result shows that magnetic reconnection produces two different types of jets simultaneously. One is a low-density jet, which properties are similar to that shown in \citet{1995Natur.375...42Y}, and the other one is a high-density jet produced by chromospheric evaporation. Based on the results of their simulation, they suggested that the mass of a coronal jet that is produced by chromospheric evaporation could be estimated from the magnetic field strength and temperature of the corona, the size of the emerging flux and the duration of the jet.

\subsubsection{3D Simulations}

The first 3D MHD simulation of flux emergence producing a hot jet in a CH was performed by \citet{2008ApJ...673L.211M}, including a comparison with {\it{Hinode}}/XRT observations. A follow-up study was published in \cite{2013ApJ...771...20M}. The experiment was carried out for a domain that contained the top $3.7$~Mm of the solar interior, the low atmosphere and the corona. To understand the jet behavior in a CH, in the simulation the corona was uniformly magnetized at time $t=0$ with $B=10$\,G and field lines subtending an angle of $25^{\circ}$ to the vertical. These simulations belong to the category of the so-called {\it idealized} models in which the gas was assumed to behave like a simple ideal gas, radiation transfer was not included, and the only entropy sources considered were those associated with ohmic dissipation and viscosity. On the other hand, the values for temperature, density and Alfv\'en speed used for the corona were close to those expected for the Sun.

\begin{figure}[!ht]
\begin{center}
\parbox{.35\textwidth}{\includegraphics[width=.35\textwidth]{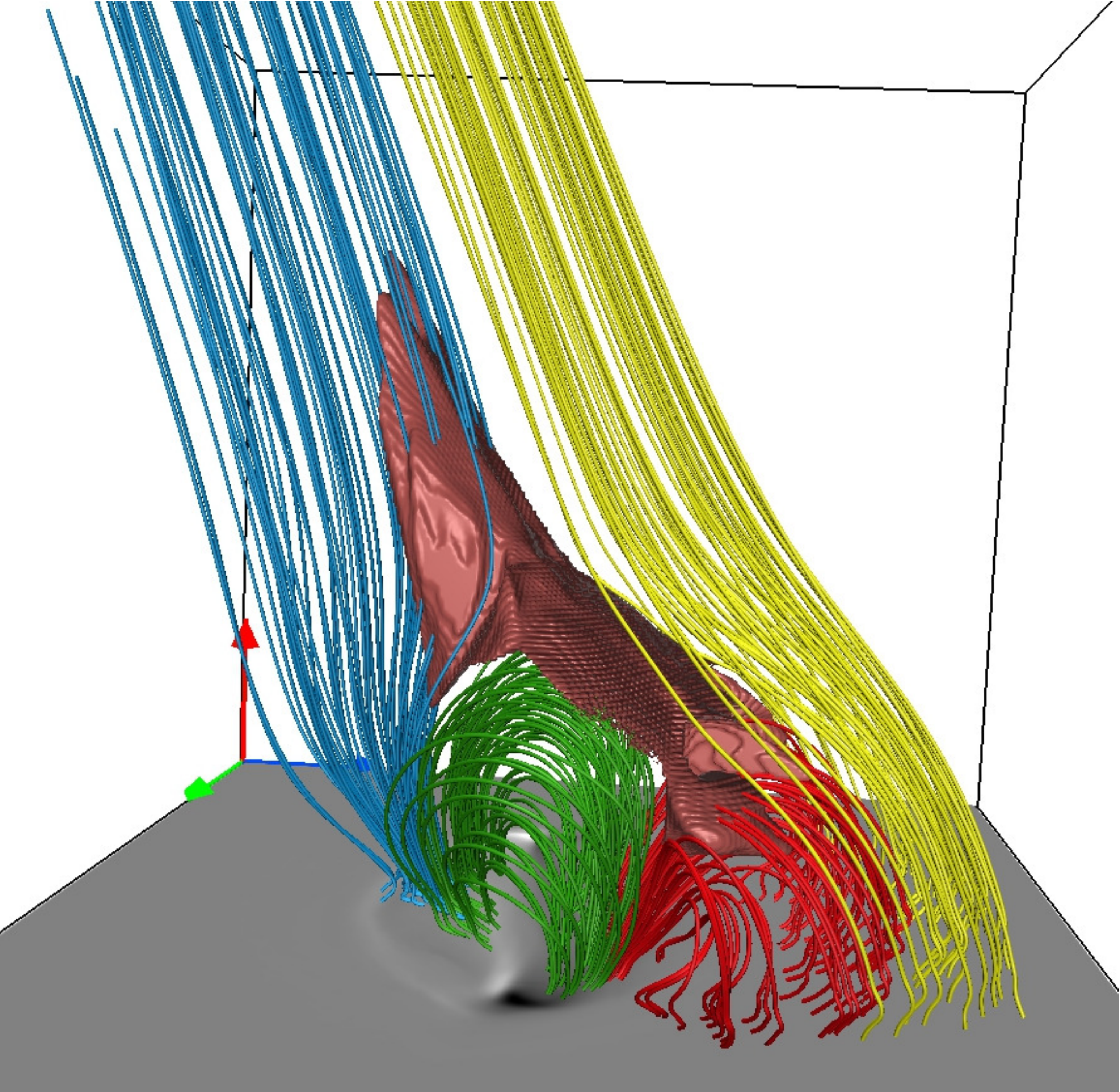}}
\parbox{.5\textwidth}{\caption{Field lines (blue/green/red/yellow) in selected regions of the jet experiment by \citet{2013ApJ...771...20M} clearly delineate the four basic connectivity domains in the model. A temperature isosurface (brown) at $T=7$~MK that encompasses the jet, the reconnection region, and the top of the hot loops is also shown.
\label{Fig_MorInsGal2013}}}
\end{center}
\end{figure}

In the simulation by \citet{2008ApJ...673L.211M} the initial condition was a magnetic tube inserted near the bottom of the domain and endowed with a density distribution leading to the formation of a buoyant $\Omega$-loop. Given the sign and direction of the tube's magnetic field in the experiment, when the rising magnetized plasma meets the coronal field, a concentrated, curved, blanket-like current sheet is formed covering the emerging plasma, and reconnection between the two magnetic systems starts. One of the outflow regions from the reconnection site leads to the emission of a thin jet up along the slanted field lines of the model CH (Fig.~\ref{Fig_MorInsGal2013}). Below the reconnection site, a double-chambered vault structure can be discerned consisting of closed magnetic loops: one of them contains the field lines of the emerged system (green in the figure), which have not been reconnected yet. The other contains a new set of closed loops (in red in the figure) resulting from the reconnection, and the plasma at their top has high temperature (several MK) because of the ohmic heating it experienced when going through the reconnection site. The high-temperature regions in the model, i.e., the jet, the reconnection site and the hot closed reconnected loops, taken together, have the shape of an inverted Y (i.e., Eiffel tower), very much as observed in X-ray by {\it{Yohkoh}}/SXT and {\it{Hinode}}/XRT.

Various quantitative features in those models are amenable to comparison with observations. The jet phenomenon lasts in the model for about $20$ to $30$\,min, but the high-temperature phase is shorter, some $10$ to $20$ min, which is well within the observed range. The jet velocities are also within the observed values, namely $100-300$~km~s$^{-1}$. Of particular interest is that the jet in the model suffers a horizontal drift due to the gradual change of connectivity of the emerged loops which turn into reconnected loops: the resulting sideways velocity in the experiment, about $10$~km~s$^{-1}$, is also compatible with observed values for the horizontal drifts.

The jet structure and emission process were analyzed in depth in the study by \citet{2013ApJ...771...20M}. The jet has the shape of an inclined hollow cane, or, more precisely: the plasma flows preferentially in a surface with the shape of a hollow semi-cylinder (Fig.~\ref{Fig_MorInsGal2013}), and has fast and slow streams, with the fast ones reaching $200-300$ km~s$^{-1}$. Typical temperatures in the jet are around $5-6$\,MK. A number of null points and plasmoids were identified in the reconnection site, and the topological changes as the emergence and jet emission process advances was described in some detail. A further aspect of the model is the appearance of a dense wall-like structure extending to heights of several megameters and surrounding the domain constituted by the emerged field region and the hot reconnected loops. The density of the plasma in the wall can be up to a few orders of magnitude above the values in the standard corona at the same height. The dense wall has at most transition region temperatures and velocities typically below $50$ km sec$^{-1}$. Whether this cold wall can be assimilated to the phenomenon of cold jets introduced by \citet{1996PASJ...48..353Y} or \citet{2008ApJ...683L..83N} is still debated.

Many EUV/X-ray jets have been seen to be followed by a phase of violent eruptions. This transition from a quiescent jet to highly dynamical eruptions was shown to be a natural process in the model by \citet{2013ApJ...771...20M}. In fact, it was seen to occur in two successive steps. When the quiescent jet was already well into the decaying phase, a first eruption took place: by that time, the emerged domain directly below the jet had adopted the shape of a comparatively thin wedge containing a highly sheared magnetic arcade. The opposite polarities across the wedge got increasingly close to each other and reconnection started. This unleashes an unstable process of the classical tether-cutting type: most of the wedge is violently ejected upward as a flux rope and impinges upon the overlying coronal structure. This sort of process had already been described to occur in the late stages of an emergence episode by \citet{2004ApJ...610..588M} and \citet{2008ApJ...674L.113A, 2012A&A...537A..62A}. However, when this first eruption was decaying, a collection of several repeated eruptions of a different physical nature took place in the experiment. Unlike the first one, they occurred around the location of the opposite polarities at the surface resulting from the initial dipole emergence. Also, the instability process was of a different nature: in one of them, for instance, a twisted loop of semi-toroidal shape was expelled upward maintaining its roots in the photosphere. The level of twist was slightly above the threshold for the kink instability; also, twist was seen to convert into writhe as time advanced, a process reminiscent of the idealized case described by \citet{2005ApJ...630L..97T}. While in the lower $\approx$$10$\,Mm of the corona, the rope being ejected was dense ($n \sim 10^{11}$\,cm$^{-3}$) and cool ($T < 4\;10^5$\,K) compared with the surroundings. Later on, when the rope collides with the overlying magnetized corona, acceleration and heating of the plasma takes place which leads to high velocities and temperatures. The kinetic energy involved in this eruption was close to $10^{27}$\,erg. It remains to study whether the simulated observational signatures one can obtain from this kind of eruptive process share common features with the actual observations of blow-out jets.

\begin{figure}
\centering
\includegraphics[width=0.85\columnwidth]{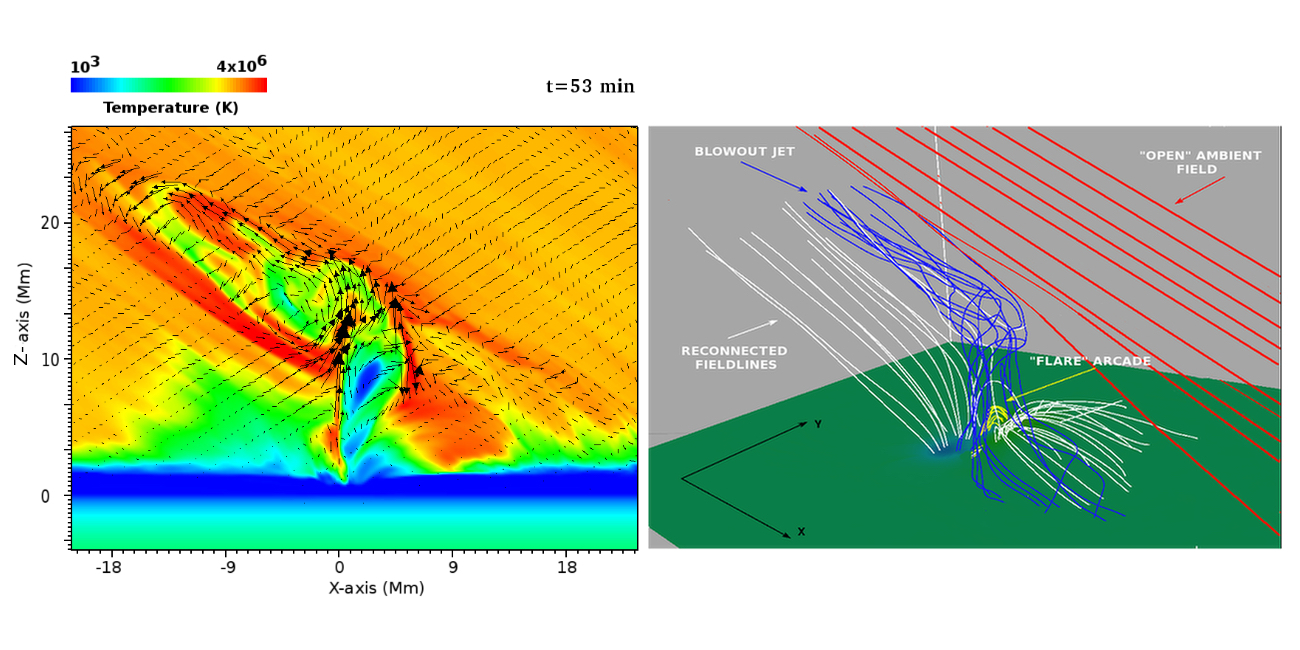}
\caption{\textbf{(Left):} Temperature distribution at the vertical midplane during the ejection of the blow-out jet. The arrows indicate the projected velocity field on the plane. \textbf{(Right):} 3D magnetic field topology during the ejection of the blow-out jet. See text for details. Adapted from \citet{2013ApJ...769L..21A}.  }
\label{fig:archontis_blow-out_jet}
\end{figure}

The onset and evolution of blow-out jets was studied in another flux emergence simulation by \citet{2013ApJ...769L..21A}. They modeled the interaction of an emerging twisted flux tube with the ambient coronal magnetic field. Initially, the emerging field interacted with the ambient field creating bi-directional jet outflows. The upward outflow was directed (as expected) along the channel of the ``open'' reconnected field lines. The downward reconnection flow collided with the magnetic field underneath it, heating the plasma locally to 10\,MK. The overall hot plasma emission formed the ``standard'' Y-shaped jets. Eventually, a {\it new} magnetic flux rope was formed due to reconnection of sheared field lines along the polarity inversion line of the emerging region. The flux rope became eruptive, blowing out the envelope field lines. During this blow-out eruption, both hot ($\approx$$10^7$\,K) and cool (5-15$\times10^4$\,K) plasma is emitted into the corona (left panel of Fig. \ref{fig:archontis_blow-out_jet}). The reconnection of the twisted field lines of the flux rope with the non-twisted oblique ambient field created an untwisting motion during the ejection of the blow-out jet (blue field lines in the right panel of Fig. \ref{fig:archontis_blow-out_jet}). Since the eruption occurred over a long distance within the emerging flux region, the blow-out jet appeared to be much wider than the standard jet and it consisted of many filament-like outflows along its width. The shape and the physical properties of the blow-out jet in this model are in agreement with those of the observed blow-out jets.

\citet{2009A&A...506L..45G} reported on the formation of an AR jet in a flux emergence simulation. They found that a fast (117\,\kms) and hot ($\approx$1~MK) bi-directional flow was formed after the emergence of new flux in the vicinity of the AR. To model this jet, they performed a numerical experiment where a small AR was formed by the emergence of a twisted flux tube and the nearby emergence of another (weaker) flux tube. The reconnection between the AR field and the newly emerging flux gave onset to a jet, which was comparable with the observed ones in terms of physical properties (e.g., temperature, velocity) and geometrical shape. \citet{2010A&A...512L...2A}, in the study of a similar system, revealed the recurrent emission of jets at the edge of the AR due to oscillatory reconnection between the emerging and the pre-existing AR magnetic field. The dynamical interaction of the two magnetic systems reconfigured the magnetic field connectivity, forming new magnetic regions (post-emergence arcade and envelope fields), which eventually started to reconnect as well. The recurrent jets moved along the ``closed'' field lines of the envelope field (confined jets). The overall system had a specific reservoir of magnetic flux and energy, which eventually became exhausted leading to a gradual annihilation of the jets.

\citet{2015ApJ...798L..10L} performed numerical experiments and they reported on the recurrent onset of helical ``blow-out'' jets in an emerging flux region (EFR). They found that these jets grow with velocities comparable to the local Alfv{\'e}n speed and they transfer a vast amount of heavy plasma into the outer solar atmosphere. During their emission, they undergo an untwisting motion as a result of reconnection between the twisted emerging and the non-twisted pre-existing magnetic field in the solar atmosphere. This study provides direct evidence that the untwisting motion of a blow-out jet is associated with the propagation of torsional Alfv{\'e}n waves in the corona.

The emergence of a small twisted flux tube into a large-scale, arcade-like coronal magnetic field was modeled by \citet{2009ApJ...704..485T}. The focus of their study was the topological change of the coronal magnetic field in response to flux emergence. The simulation was motivated by puzzling {\it{Hinode}}/XRT observations of a small limb event that took place next to a quiescent prominence cavity. The event first exhibited the typical morphological characteristics of a standard jet. Shortly after, however, a second closed loop system adjacent to the first one became visible, which is typically not observed in coronal jets.

\citet{2009ApJ...704..485T} employed a $\beta=0$ MHD simulation to model the magnetic evolution that may lead to such an event. They used the coronal flux rope model by \citet{titov99} to construct the magnetic field of the prominence and the surrounding arcade, and followed by the emergence of a second, much smaller flux rope in its vicinity. Note that the flux emergence was driven kinematically, i.e., the slow, bodily emergence of the small flux rope was imposed as a boundary condition at the bottom plane of the computational domain \citep[see e.g.,][]{2004ApJ...609.1123F}. Since the energy equation was neglected, the simulation does not provide information on the plasma properties of the system, but it allows for the study of topological changes of the magnetic field caused by the emergence of small-scale twisted flux into locally open magnetic field.

\begin{figure*}[!ht]
\includegraphics[width=1.\textwidth]{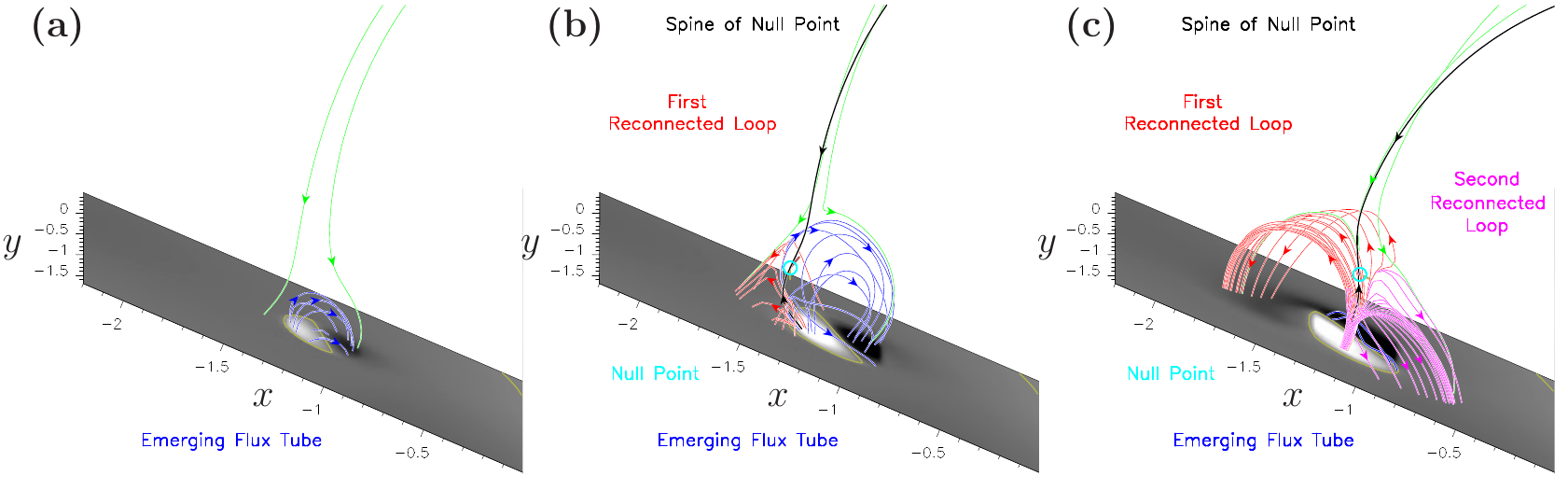}
\caption{Side view on the flux emergence region in the simulation by \cite{2009ApJ...704..485T} at three consecutive times. Green and blue magnetic field lines outline the coronal arcade and the emerging flux rope, respectively, while the red and magenta field lines show two loop systems that are successively formed by reconnection.}
\label{fig:torok09}
\end{figure*}

The simulation revealed a two-step reconnection process, which is depicted in Fig.\,\ref{fig:torok09}. Initially, the evolution corresponds exactly to the standard jet scenario -- the emerging flux rope reconnects with the semi-open background field, which leads to the formation of closed loops on the left-hand side of the emergence region (red field lines in Fig.\,\ref{fig:torok09}b). However, since the horizontal orientation of the emerging field is rotating as flux located closer to the rope axis emerges \citep[see Fig.\,6 in][]{2009ApJ...704..485T}, the reconnection site (the null point) slowly drifts towards the other side of the emerging parasitic polarity. This leads to a successive displacement of the footpoints of the reconnected field lines until, eventually, the field lines come into contact with the background field on the right-hand side of the emergence region and reconnect to form a second loop system (magenta field lines in Fig.\,\ref{fig:torok09}c). The reconnected field lines collectively form a fan-spine configuration that significantly extends over the parasitic polarity. \citet{2009ApJ...704..485T} therefore suggested that such a two-step reconnection process may play a role in the formation of anemone regions.

Extending the above simulations, \citet{2014ApJ...789L..19F} carried out the first 3D MHD simulation of the emergence of a flux rope into open field including field-aligned thermal conduction. The simulation was performed with the Block Adaptive Tree Solar-wind Roe Upwind Scheme \citep[BATS-R-US;][]{1999JCoPh.154..284P,2012JCoPh.231..870T}. In this simulation the jet also experiences two phases $-$ a standard and a blowout phase. During the blowout jet cold and dense plasma is ejected in a spinning motion along untwisting field lines, driven by Lorentz forces, in addition to the hot reconnection outflow. The authors compared their results to a run without heat conduction and constructed synthetic {\it{SDO}}/AIA emission images for both cases. They found that the run with heat conduction produces plasma emission that is in better agreement with the underlying magnetic field structure, because the heat conduction efficiently distributes the energy release from the reconnection region into the lower atmosphere and promotes the ejection of dense plasma into the corona along the field lines.

\subsection{MHD Simulations: Instability Onset Scenario}

The untwisting-jet model was first demonstrated to generate solar-like jets by \citet{2009ApJ...691...61P,2010ApJ...714.1762P}, with subsequent extensions by \citet{2012EAS....55..201D}, \citet{2015A&A...573A.130P}, and \citet{Karpen15}. The physical mechanism underlying energy release and jet initiation is kink-instability-induced interchange magnetic reconnection occurring at 3D null points.

To simulate the jets numerically, an initially potential magnetic configuration is assumed  \citep{2009ApJ...691...61P,2010ApJ...714.1762P}. A vertical magnetic dipole, positioned below the photosphere to generate a closed flux system above, is embedded within a uniform, inclined (with respect to the vertical direction), open background field. A highly conducting, low-pressure plasma with initially uniform temperature (and mass density, in these gravity-free studies) fills the corona. The nonlinear equations of ideal, single-fluid MHD are advanced in time using the Adaptively Refined Magnetohydrodynamics Solver \citep[ARMS; e.g.,][]{2008ApJ...680..740D}. Magnetic free energy and helicity are introduced into the closed-flux region by imposing photospheric twisting motions at the bottom boundary.

Analyses of the simulation results \citep{2010ApJ...714.1762P} reveal that the jet generation consists of distinct phases of energy storage and explosive energy release. During the energy-storage phase, a highly localized, thin current sheet develops gradually at the null point. Larger inclination angles (with respect to the vertical direction) of the background field introduce greater asymmetries into the  strengthening current sheet. Configurations with sufficiently inclined fields eventually begin to reconnect quasi-steadily, with an associated slow release of free energy \citep{2010ApJ...714.1762P,2012EAS....55..201D}. This process generates a straight jet of tension-driven outflows due to the retraction of newly reconnected field lines \citep{2015A&A...573A.130P}, the so-called ``slingshot effect'' \citep[e.g.,][]{2001ApJ...553..905L}. If the energy-storage rate exceeds the slow energy-release rate, which always occurs at small inclination angles where no straight jet is generated, the magnetic energy continues to accumulate. Eventually, the configuration experiences an explosive energy-release phase driven by an ideal kink-like instability \citep{2010ApJ...715.1556R}, during which a very broad, highly dynamical current sheet develops along the fan-surface field lines that separate the closed and open flux systems. Reconnection across this current sheet causes an impulsive release of both free energy and helicity that generates a helical jet. The helical jet is driven by large-amplitude, torsional Alfv\'en waves that propagate upward along newly reconnected, open field lines. These waves carry away a large fraction of the free energy and helicity initially stored in the closed-flux region. The helical jet is generated irrespective of the inclination of the open field, but its properties vary with those of the precursor straight jet: a stronger straight jet reduces the energy released during the subsequent helical jet more substantially, and also delays the triggering of the helical jet for a longer time.

\subsection{Recent Simulations and Work in Progress}

\subsubsection{MHD Simulation: Instability Onset Scenario}

\begin{figure*}[t]
\centering \includegraphics[width=0.9\textwidth]{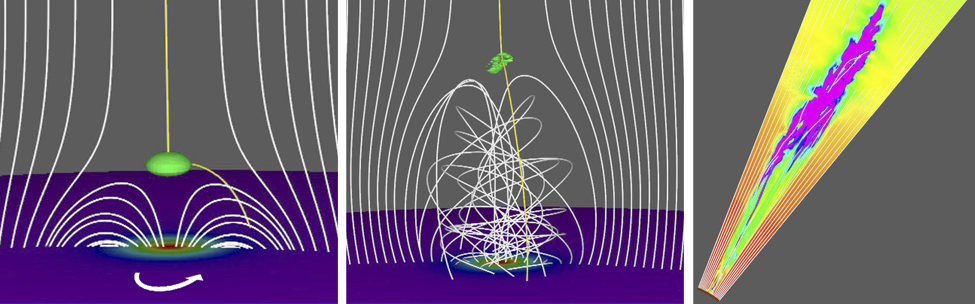}
\caption{Untwisting jet model in spherical geometry with solar gravity and wind. {\bf{(Left)}} Initial potential configuration in the low corona showing magnetic field lines (white curves), radial magnetic field component at the solar surface (color shading), an isosurface (green) of high plasma beta , and a schematic representation of the twisting motions imposed at the bottom boundary (white arrow). {\bf{(Middle)}} Strongly twisted configuration in the low corona, just prior to reconnection onset, showing magnetic field lines (white curves), radial magnetic field component at the surface (color shading), and an isosurface of high plasma beta (green). {\bf{(Right)}} Propagation of the jet into the inner heliosphere, showing magnetic field lines (white curves) and plasma velocity magnitude against the plane of the sky (color shading). From \citet{Karpen15}.}
\label{fig:UJM}
\end{figure*}

Recent work on the ARMS model has focused on its extension to spherical geometry, the inclusion of solar gravity and wind, and the predicted signatures of jets in the inner heliosphere \citep{Karpen15}. The results obtained thus far corroborate the conclusions of the cartesian-geometry, gravity-free simulations summarized above. A configuration with a strictly radial background field \citep[i.e., zero inclination angle, as in][]{2009ApJ...691...61P} is shown in Fig. \ref{fig:UJM}. The left, middle, and right panels illustrate, respectively, the initial potential state, the strongly twisted configuration just prior to reconnection onset and jet initiation, and the late state of propagation of torsional Alfv\'en waves into the inner heliosphere. Isosurfaces in the left and middle panels show regions where the thermal pressure is twice the magnetic pressure (plasma $\beta = 2$): the compact volume that surrounds the null point initially (left) fragments and spreads around the top of the fan surface near the time of reconnection onset (middle). The outermost edge of the heliospheric view (right) is at $\approx5~R_\odot$, and the time is 25 min after reconnection onset in the low corona. Thus, the Alfv\'enic jet averaged $\approx2000$~km~s$^{-1}$ as it traversed the corona.

The reconnection-driven untwisting jets that occur at stressed 3D null points above embedded bipoles reproduce key features of observed polar jets. Highly impulsive, obviously helical plasma motions are generated in untwisting jets at all inclination angles studied thus far. If the inclination angle is large enough, a precursor phase of gentler, more linearly directed plasma motions occurs. If the photospheric driving is maintained over a long interval, recurrent quasi-homologous helical jets can be generated from a single structure. The torsional Alfv\'en waves that drive the helical jets can propagate well out into the inner heliosphere, producing signatures that have been observed in the corona by {\it{STEREO}} and that may be detectable in the SW by {\it Solar Probe Plus} and {\it Solar Orbiter}. Finally, the straight/helical jets obtained in the simulations may replicate the observed standard/blow-out jet classification proposed by \citet{2010ApJ...720..757M,2013ApJ...769..134M}. The straight jets are strongly collimated, possess the classical inverse-Y shape and a narrow spire, and show little evidence of rotation, all of which are principal criteria for standard jets. The helical jets, on the other hand, exhibit strong rotational motions and possess a broad spire, matching key properties of blow-out jets.

\subsubsection{MHD Simulation: Flux Emergence Scenario}

Very recently,  \cite{torok15} employed the 3D MHD model of the solar corona by Predictive Science Inc. (PSI) to simulate jets. The model uses a spherical computational domain that extends to 20 solar radii, and it incorporates thermal conduction, radiation losses, background coronal heating, and the SW \citep[see][]{lionello09}. For simplicity, \citet{torok15} chose a purely radial magnetic field (with a field strength of $\approx 6$\,G at the surface) and a radially dependent coronal-heating function. A constant number density of $1.4 \times 10^{12}$ cm$^{-3}$ and a temperature of $2 \times 10^4$\,K are prescribed at the lower boundary of the simulation domain. After a steady-state solution of the large-scale corona and SW is obtained, the emergence of a flux rope is modeled by successively imposing a time-dependent electric field, which is calculated using magnetic fields and plasma velocities extracted from a flux-emergence simulation by \cite{leake13}, at the lower boundary \citep[see][for a detailed description of the method]{2013ApJ...777...76L}.

As the flux rope expands in the corona, a current layer is formed and reconnection across this layer triggers a standard jet, similar to the simulations described in Section\,8.2.2. If the emergence is continued for a sufficiently long time, a blow-out jet is produced by the eruption of the flux rope.

\begin{figure*}
\begin{center}
\parbox{0.68\textwidth}{\includegraphics[width=0.68\textwidth]{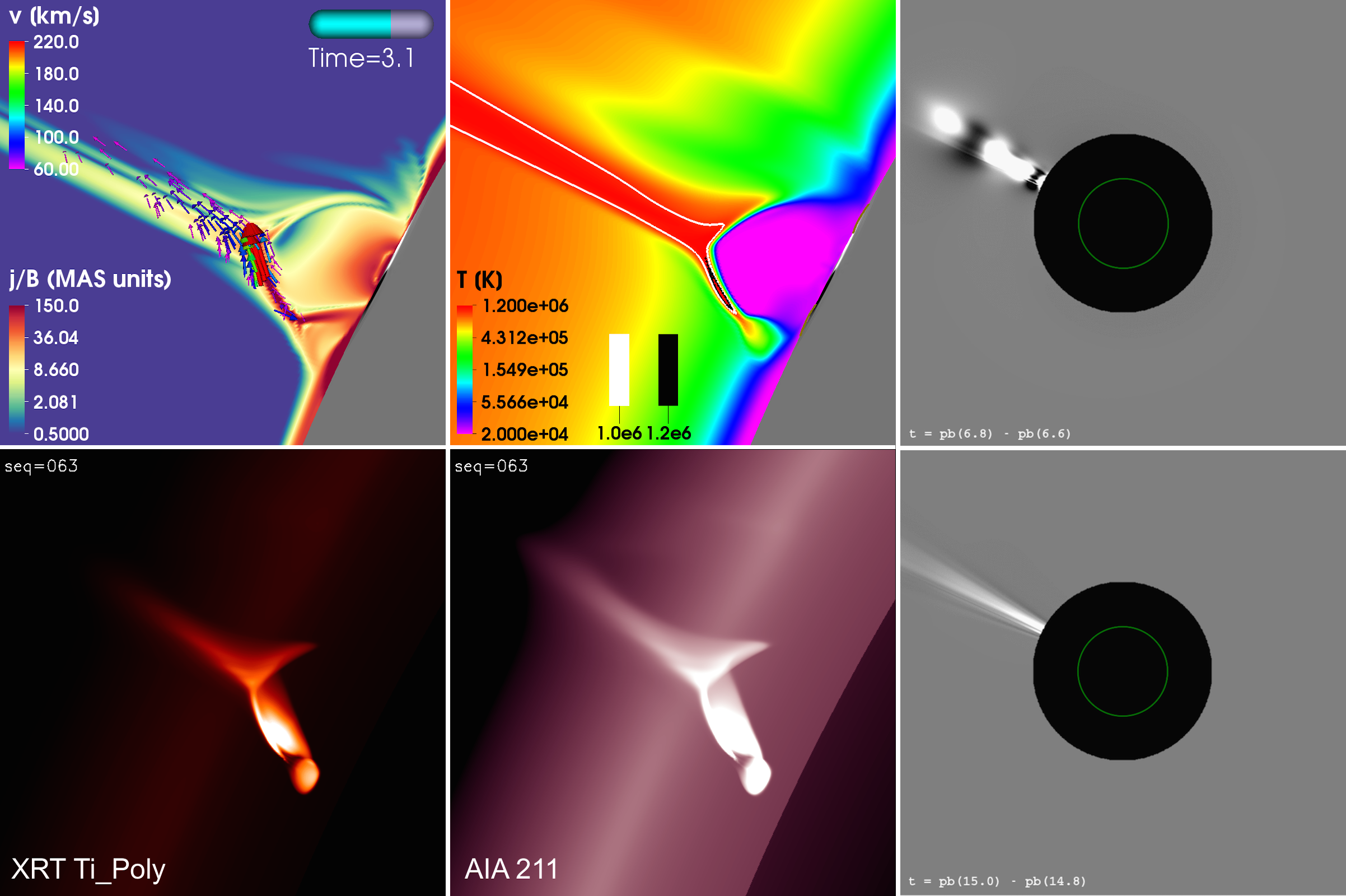}}
\parbox{0.3\textwidth}{\caption{``Thermodynamic'' MHD simulation of a standard jet. The four panels on the left show electric currents and plasma flows (arrows), plasma temperature, and two synthetic emission images, respectively. The right panels show synthetic running-difference coronagraph images at an intermediate (top) and at a later (bottom) time. The green circle outlines the solar surface. Adapted from \citet{torok15}.
\label{fig_torok_2015}
}}
\end{center}
\end{figure*}

Fig.~\ref{fig_torok_2015} shows one of the simulations. In this case ohmic heating was turned off and no blow-out jet was modeled. The temperature increase to a realistic value of $\approx\,1.1$\,MK in the jet spire is caused predominantly by compressional heating in the current layer and in the reconnection outflow regions. The synthetic emission images nicely show the jet spire, the inverted Y-shape, and the BP. The reconnection occurs in episodic bursts, which manifest as ``blobs'' in the synthetic coronagraph images (top right). At a later time, when the emergence is stopped and reconnection has ceased, the WL signature of the jet evolves into a structure reminiscent of a plume (bottom right). These simulations will allow us to investigate the plasma heating and dynamics in coronal jets, as well as their mass and energy contributions to the SW, in much more detail than before.

\subsection{Magnetofrictional Simulations}

\begin{figure*}[t]
\centering \includegraphics[width=0.7\textwidth]{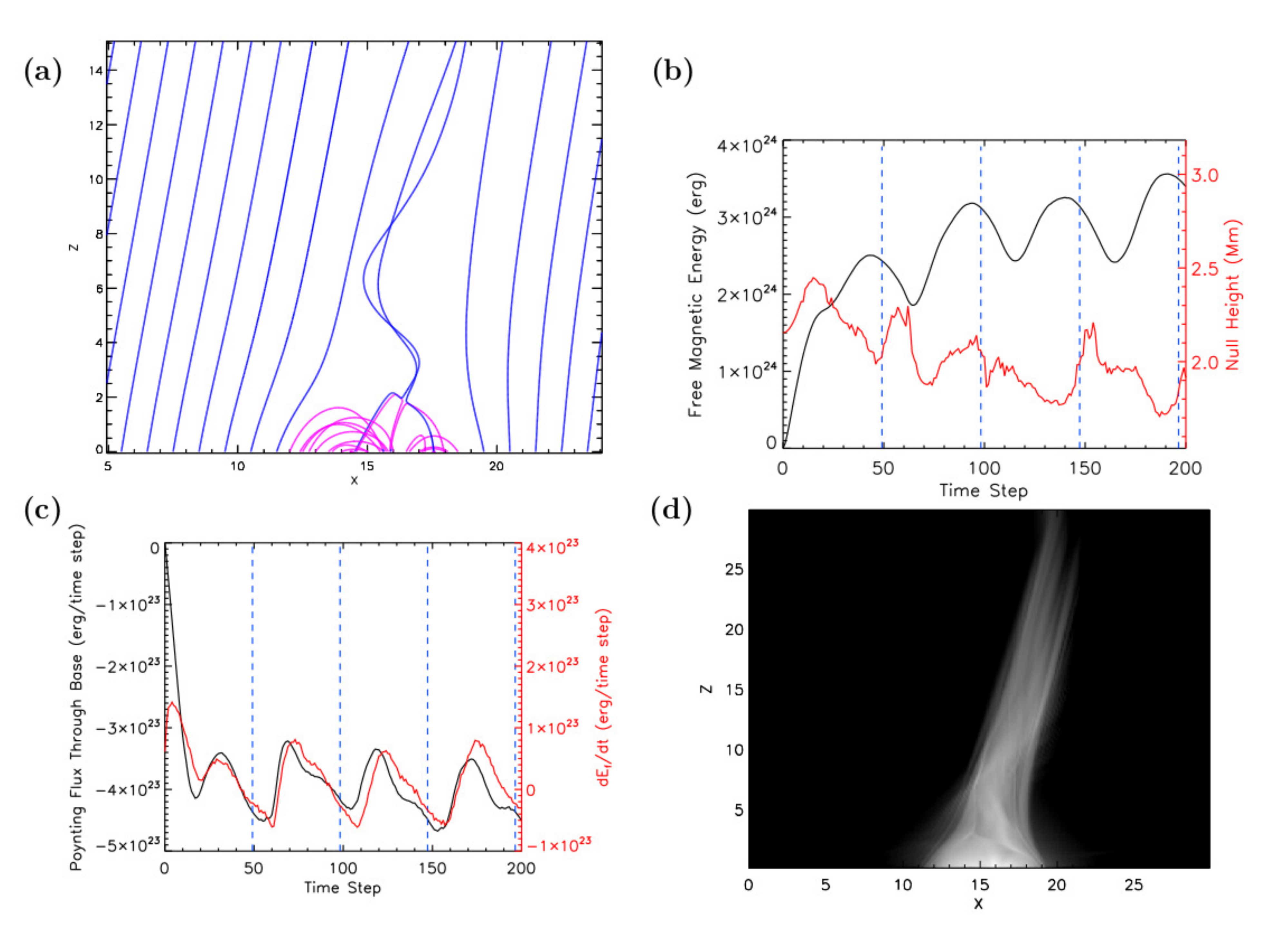}
\caption{(a) Closed (magenta) and open (blue) magnetic field lines viewed in the $xz$ plane at $y=15$\,Mm, at $t=120$ time steps. (b) Free magnetic energy (black) and height of null point (red) as a function of time. (c) Poynting flux through the photosphere (black) and time derivative of free magnetic energy $\mathrm{d}E_f/\mathrm{d}t$ as a function of time. Vertical blue dashed lines indicate times between full laps of the positive magnetic polarity about the midpoint of the box. (d) Logarithm of LOS-integrated current, viewed in the $xz$ plane. Adapted from \citet{Meyer15}.}
\label{meyer15}
\end{figure*}

\citet{Meyer15} have considered different configurations of jets using the magnetofrictional technique of \citet{2000ApJ...539..983V}. The method calculates the evolution of the magnetic field through a series of quasi-static equilibria in response to photospheric footpoint motions. Since the technique considers equilibria, it is not able to capture the eruptive stage of a jet. It is, however, very useful for considering the build-up of electric current systems and free magnetic energy in the lead-up to the eruption. It also has the advantage that it is computationally inexpensive, allowing for the modeling of a wide variety of different situations and configurations, and the exploration of the parameter space. \citet{Meyer15} consider a series of simple, theoretical situations including: a magnetic polarity in a uniform opposite-polarity background magnetic field, rotating around its axis similar to \cite{2009ApJ...691...61P}; a magnetic polarity rotating in a circle around an outside axis; a magnetic flyby, where two opposite-polarity magnetic features shear past one another; flux cancellation; and flux emergence. For all cases, they show the evolution of the free energy, helicity, and height of the null point with time, as shown in Fig.\,\ref{meyer15}. The current structure of the jet and the simulated emission based on integrating the square of the current shows a collimated standard jet in the case similar to \cite{2009ApJ...691...61P}, a broader filamentary curtain of current (see Fig.\,\ref{meyer15}d) reminiscent of a blow-out jet in the circular motion and flux cancellation cases. A follow-up statistical study is planned to compare these theoretical configurations with a catalog of observed solar jets.

\citet{Savcheva15} modeled the magnetic field structure of a standard and blow-out jet that appeared in the outskirts of an AR. The null point topology of the standard jet was obtained from a potential field extrapolation from an HMI magnetogram, and the blow-out jet was modeled using the flux rope insertion method \citep[e.g.,][]{2004ApJ...612..519V, 2012ApJ...750...15S}. The flux rope insertion method produces a nonlinear force-free field (NLFFF) containing a flux rope embedded in a potential field. In the initial configuration the flux rope is stable under a null point. This initial condition is used in a dynamic magnetofrictional simulation similar to \citet{2014ApJ...782...71G}. During the simulation the magnetic flux rope expands and pushes on the null point where reconnection takes place and twist propagates along the large-scale AR field. The simulations are used to resolve LOS effects in interpreting {\it{IRIS}} observations of these jets.

The same {\it{IRIS}} observations of a recurrent blow-out jet were considered along with the data-driven magnetofrictional simulation of \cite{2015ApJ...801...83C}. In contrast to the flux rope insertion method that uses LOS magnetograms, the quasi-static NLFFF equilibria in \cite{2015ApJ...801...83C} re-extrapolated based on vector magnetograms. The blow-out jet in this simulation is more strongly driven than the one in \cite{Savcheva15} due to the detectable rotation motion in the vector magnetograms.

\section{Conclusion and Prospects}
Imaging and spectroscopic observations over the last two decades have provided unprecedented insights into the formation and evolution of solar coronal transients, particularly coronal jets. Recent space missions, such as {\it{Hinode}}, {\it{STEREO}}, {\it{SDO}}, {\it{IRIS}} and are instrumental in advancing our understanding of this phenomenon.

Instrument improvement in terms of both spatial and temporal resolution and also temperature coverage are key in the numerous discoveries made concerning the different facets of coronal jets. Thanks to the multiple discoveries made using high quality observations (both remote sensing and in situ), it is now widely believed that coronal jets play an important role in the multi-scale solar activity, coronal heating, and the contribution to the SW. The Feature Finding Team (FFT) of {\it{SDO}} developed dedicated computer tools that allow identifying and characterizing coronal features including coronal jets. This has the potential to build massive jet statistics all over the solar disk. This would hopefully give us truly global statistics of the energy and mass contents of jets and their role in the energy and mass supply of the solar atmosphere

However, there are still many aspects of coronal jets that remain ambiguous and need further investigation. It is still unclear whether the scale size spectrum of coronal jets extends to scales much smaller than the spatial resolution available now. Do they extend to the nano-flare scales where it is believed that the contribution to the coronal heating could be significant? The contribution of jets to the SW and to the population of energetic particles is still unclear as well as their role in driving the formation and evolution of other coronal structures such as plumes and chromospheric features such as spicules. An area of improvement in the study of coronal jets that need to be deepened is spectroscopy, which provides insights into the plasma properties of jets. This aspect is still in its infancy compared to imaging. The latter needs further improvements in terms of spatial resolution.

The extensive jet modeling in the past decade or so has achieved many successes including the detailed representation of the jet morphology and dynamics that match observations of jets well. In addition, we have achieved the successful modeling of the transition between standard and blow-out jets as well as the ability to create recurrent jets and jet/plume structures. Jets have been produced both in the open field of CHs and the large-scale closed field of solar ARs. The scenario for producing jets that has been most widely explored in the simulations is flux emergence \citep{2008ApJ...673L.211M, 2009ApJ...704..485T, 2013ApJ...771...20M,2013ApJ...769L..21A}, but increasing attention has been given to the instability-onset scenario \citep{2009ApJ...691...61P,2010ApJ...714.1762P,2015A&A...573A.130P}. The mechanism of flux cancellation has not been explored with MHD simulations, while a small study of flux cancellation has been performed in a magnetofrictional simulation. Recent observations \citep[e.g.,][]{2014SoPh..289.3313Y} demonstrate that the flux cancellation mechanism may be as important as the emergence.

Most current simulations lack a full thermodynamic treatment and do not include thermal conduction or radiation effects. These ingredients are clearly important for reproducing the observed plasma properties and understanding the emission and spectral observations of jets. As a consequence the MHD simulations so far either under- or overestimate the temperature of jets. As more plasma diagnostics of coronal jets become available through analysis of {\it{Hinode}}/EIS and {\it{IRIS}} data, we need to create increasingly realistic MHD simulations of jets. Current MHD simulations only deal with idealized boundary and initial conditions. Hence, the ultimate goal is to develop data-constrained, and eventually, data-driven MHD simulations with useful energy equations to model observed events. Other investigations, both observational and theoretical, should clarify whether mini-filament eruptions play a larger role than previously recognized in jet and jet-bright-point formation \citep{sterling.et15}.

We believe that future missions such as NASA's Solar Probe Plus and ESA's Solar Orbiter will provide further insights into the physics of coronal jets and the related phenomena. For instance, Solar Probe Plus will fly through coronal structures including jets, which would provide close by imaging observations as well as in situ measurements of these features. Solar Orbiter's above the ecliptic observations will provide unprecedented view of the solar poles where coronal jets are prominent. This includes magnetic field measurements, spectroscopy, imaging, and in situ measurements.

\begin{acknowledgements}
The ``Solar Jets" team members are grateful for the International Space Science Institute (ISSI, Bern, Switzerland) that hosted two meetings on March 2013 and March 2014 within the frame of the international team on the ``Solar Coronal Jets (http://www.issibern.ch/teams/solarjets)". This work benefited greatly from discussions held at these meetings. S. P. acknowledges support from an FP7 Marie Curie Grant (FP7-PEOPLE-2010-RG/268288) as well as European Union (European Social Fund $-$ ESF) and Greek national funds through the Operational Program ``Education and Lifelong Learning" of the National Strategic Reference Framework (NSRF) - Research Funding Program: Thales. Investing in knowledge society through the European Social Fund. ACS thanks R. L. Moore for useful discussions. ACS was supported by funding from the Heliophysics Division  of NASA's Science Mission Directorate through the Living With a Star Targeted Research and Technology Program, and by funding from the {\it{Hinode}} Project Office at NASA/MSFC. PRY acknowledges funding from National Science Foundation grant AGS-1159353. TT was supported by NASA's HSR and LWS programs. KD acknowledges support from the Computational and Information Systems Laboratory and from the HAO, as well as support from the AFOSR under award FA9550-15-1-0030.

The {\it{SOHO}} is a mission of international cooperation between ESA and NASA. {\it{Hinode}} is a Japanese mission
developed and launched by ISAS/JAXA, with NAOJ as a domestic partner and NASA and STFC (UK) as international partners. It is operated by these agencies in cooperation with the ESA and NSC (Norway). The {\it{STEREO}}/SECCHI data used here are produced by an international consortium of the NRL (USA), LMSAL (USA), NASA GSFC (USA), RAL (UK), Univ. Birmingham (UK), MPS (Germany), CSL (Belgium), IOTA (France), and IAS (France). {\it{SDO}} is the first mission to be launched for NASA's Living With a Star (LWS) Program. {\it{IRIS}}  is  a  NASA  small  explorer  mission developed  and  operated  by  LMSAL  with  mission  operations executed  at  NASA  Ames  Research  center  and major contributions to downlink communications funded by the Norwegian Space Center (NSC, Norway) through an ESA PRODEX contract.
\end{acknowledgements}



\begin{thebibliography}{248}
\ifx \bisbn   \undefined \def \bisbn  #1{ISBN #1}\fi
\ifx \binits  \undefined \def \binits#1{#1} \fi
\ifx \bauthor  \undefined \def \bauthor#1{#1} \fi
\ifx \bjtitle  \undefined \def \bjtitle#1{\textrm{#1}}\fi
\ifx \batitle  \undefined \def \batitle#1{#1} \fi
\ifx \bctitle  \undefined \def \bctitle#1{#1} \fi
\ifx \bvolume  \undefined \def \bvolume#1{\textbf{#1}}\fi
\ifx \byear  \undefined \def \byear#1{#1} \fi
\ifx \bissue  \undefined \def \bissue#1{#1} \fi
\ifx \bfpage  \undefined \def \bfpage#1{#1} \fi
\ifx \blpage  \undefined \def \blpage #1{#1} \fi
\ifx \burl  \undefined \def \burl#1{#1} \fi
\ifx \doiurl  \undefined \def \doiurl#1{#1} \fi
\ifx \betal  \undefined \def \betal{et al.} \fi
\ifx \binstitute  \undefined \def \binstitute#1{#1} \fi
\ifx \beditor  \undefined \def \beditor#1{#1} \fi
\ifx \bpublisher  \undefined \def \bpublisher#1{#1} \fi
\ifx \bbtitle  \undefined \def \bbtitle#1{\textit{#1}} \fi
\ifx \bedition  \undefined \def \bedition#1{#1} \fi
\ifx \bseriesno  \undefined \def \bseriesno#1{#1} \fi
\ifx \blocation  \undefined \def \blocation#1{#1} \fi
\ifx \bsertitle  \undefined \def \bsertitle#1{#1} \fi
\ifx \bsnm \undefined \def \bsnm#1{#1} \fi
\ifx \bsuffix \undefined \def \bsuffix#1{#1} \fi
\ifx \bparticle \undefined \def \bparticle#1{#1} \fi
\ifx \barticle \undefined \def \barticle#1{#1} \fi
\ifx \botherref \undefined \def \botherref #1{#1} \fi
\ifx \url \undefined \def \url#1{#1} \fi
\ifx \bchapter \undefined \def \bchapter#1{#1} \fi
\ifx \bbook \undefined \def \bbook#1{#1} \fi
\ifx \bcomment \undefined \def \bcomment#1{#1} \fi
\ifx \oauthor \undefined \def \oauthor#1{#1} \fi
\ifx \citeauthoryear \undefined \def \citeauthoryear#1{#1} \fi
\ifx \texttildelow  \undefined \def \texttildelow{\symbol{126}} \fi
\def \endbibitem {}
\ifx \bconflocation  \undefined \def \bconflocation#1{#1} \fi

\bibitem[\protect\citeauthoryear{{Adams} et~al.}{2014}]{2014ApJ...783...11A}
\begin{barticle}
\bauthor{\binits{M.} \bsnm{{Adams}}},
\bauthor{\binits{A.C.} \bsnm{{Sterling}}},
\bauthor{\binits{R.L.} \bsnm{{Moore}}},
\bauthor{\binits{G.A.} \bsnm{{Gary}}},
\batitle{{A Small-scale Eruption Leading to a Blowout Macrospicule Jet in an
  On-disk Coronal Hole}}.
\bjtitle{\apj}
\bvolume{783},
\bfpage{11}
(\byear{2014}).
doi:\doiurl{10.1088/0004-637X/783/1/11}
\end{barticle}
\endbibitem

\bibitem[\protect\citeauthoryear{{Alexander} \&
  {Fletcher}}{1999}]{1999SoPh..190..167A}
\begin{barticle}
\bauthor{\binits{D.} \bsnm{{Alexander}}},
\bauthor{\binits{L.} \bsnm{{Fletcher}}},
\batitle{{High-resolution Observations of Plasma Jets in the Solar Corona}}.
\bjtitle{\solphys}
\bvolume{190},
\bfpage{167}
(\byear{1999}).
doi:\doiurl{10.1023/A:1005213826793}
\end{barticle}
\endbibitem

\bibitem[\protect\citeauthoryear{{Antiochos}}{1990}]{1990MmSAI..61..369A}
\begin{barticle}
\bauthor{\binits{S.K.} \bsnm{{Antiochos}}},
\batitle{{Heating of the corona by magnetic singularities}}.
\bjtitle{\memsai}
\bvolume{61},
\bfpage{369}
(\byear{1990}).
\end{barticle}
\endbibitem

%
%

\bibitem[\protect\citeauthoryear{{Archontis} \& {Hood}}{2008}]{2008ApJ...674L.113A}
\begin{barticle}
\bauthor{\binits{V.} \bsnm{{Archontis}}},
\bauthor{\binits{A.W.} \bsnm{{Hood}}},
\batitle{{A Flux Emergence Model for Solar Eruptions}}.
\bjtitle{\apjl}
\bvolume{674},
\bfpage{113}
(\byear{2008}).
doi:\doiurl{10.1086/529377}
\end{barticle}
\endbibitem

\bibitem[\protect\citeauthoryear{{Archontis}
  et~al.}{2010}]{2010A&A...512L...2A}
\begin{barticle}
\bauthor{\binits{V.} \bsnm{{Archontis}}},
\bauthor{\binits{K.} \bsnm{{Tsinganos}}},
\bauthor{\binits{C.} \bsnm{{Gontikakis}}},
\batitle{{Recurrent solar jets in active regions}}.
\bjtitle{\aap}
\bvolume{512},
\bfpage{2}
(\byear{2010}).
doi:\doiurl{10.1051/0004-6361/200913752}
\end{barticle}
\endbibitem

\bibitem[\protect\citeauthoryear{{Archontis} \& {Hood}}{2012}]{2012A&A...537A..62A}
\begin{barticle}
\bauthor{\binits{V.} \bsnm{{Archontis}}},
\bauthor{\binits{A.W.} \bsnm{{Hood}}},
\batitle{{Magnetic flux emergence: a precursor of solar plasma expulsion}}.
\bjtitle{\aap}
\bvolume{537},
\bfpage{62}
(\byear{2012}).
doi:\doiurl{10.1051/0004-6361/201116956}
\end{barticle}
\endbibitem

\bibitem[\protect\citeauthoryear{{Archontis} \& {Hood}}{2013}]{2013ApJ...769L..21A}
\begin{barticle}
\bauthor{\binits{V.} \bsnm{{Archontis}}},
\bauthor{\binits{A.W.} \bsnm{{Hood}}},
\batitle{{A Numerical Model of Standard to Blowout Jets}}.
\bjtitle{\apjl}
\bvolume{769},
\bfpage{21}
(\byear{2013}).
doi:\doiurl{10.1088/2041-8205/769/2/L21}
\end{barticle}
\endbibitem

\bibitem[\protect\citeauthoryear{{Bain} \& {Fletcher}}{2009}]{2009A&A...508.1443B}
\begin{barticle}
\bauthor{\binits{H.M.} \bsnm{{Bain}}},
\bauthor{\binits{L.} \bsnm{{Fletcher}}},
\batitle{{Hard X-ray emission from a flare-related jet}}.
\bjtitle{\aap}
\bvolume{508},
\bfpage{1443}
(\byear{2009}).
doi:\doiurl{10.1051/0004-6361/200911876}
\end{barticle}
\endbibitem

\bibitem[\protect\citeauthoryear{{Bame} et~al.}{1977}]{1977JGR....82.1487B}
\begin{barticle}
\bauthor{\binits{S.J.} \bsnm{{Bame}}},
\bauthor{\binits{J.R.} \bsnm{{Asbridge}}},
\bauthor{\binits{W.C.} \bsnm{{Feldman}}},
\bauthor{\binits{J.T.} \bsnm{{Gosling}}},
\batitle{{Evidence for a structure-free state at high solar wind speeds}}.
\bjtitle{\jgr}
\bvolume{82},
\bfpage{1487}
(\byear{1977}).
doi:\doiurl{10.1029/JA082i010p01487}
\end{barticle}
\endbibitem

\bibitem[\protect\citeauthoryear{{Banerjee} et~al.}{2000}]{banerjee00}
\begin{barticle}
\bauthor{\binits{D.} \bsnm{{Banerjee}}},
\bauthor{\binits{E.} \bsnm{{O'Shea}}},
\bauthor{\binits{J.G.} \bsnm{{Doyle}}},
\batitle{{Giant macro-spicule as observed by CDS on SOHO}}.
\bjtitle{\aap}
\bvolume{355},
\bfpage{1152}
(\byear{2000}).
\end{barticle}
\endbibitem

\bibitem[\protect\citeauthoryear{{Beckers}}{1968}]{1968SoPh....3..367B}
\begin{barticle}
\bauthor{\binits{J.M.} \bsnm{{Beckers}}},
\batitle{{Solar Spicules (Invited Review Paper)}}.
\bjtitle{\solphys}
\bvolume{3},
\bfpage{367}
(\byear{1968}).
doi:\doiurl{10.1007/BF00171614}
\end{barticle}
\endbibitem

\bibitem[\protect\citeauthoryear{{Beckers}}{1972}]{1972ARA&A..10...73B}
\begin{barticle}
\bauthor{\binits{J.M.} \bsnm{{Beckers}}},
\batitle{{Solar Spicules}}.
\bjtitle{\araa}
\bvolume{10},
\bfpage{73}
(\byear{1972}).
doi:\doiurl{10.1146/annurev.aa.10.090172.000445}
\end{barticle}
\endbibitem


\bibitem[\protect\citeauthoryear{{Berkebile-Stoiser} et~al.}{2009}]{2009A&A...505..811B}
\begin{barticle}
\bauthor{\binits{S.} \bsnm{{Berkebile-Stoiser}}},
\bauthor{\binits{P.} \bsnm{{G{\"o}m{\"o}ry}}},
\bauthor{\binits{A.M.} \bsnm{{Veronig}}}, \betal,
\batitle{{Multi-wavelength fine structure and mass flows in solar
  microflares}}.
\bjtitle{\aap}
\bvolume{505},
\bfpage{811}
(\byear{2009}).
doi:\doiurl{10.1051/0004-6361/200912100}
\end{barticle}
\endbibitem

\bibitem[\protect\citeauthoryear{{Bohlin} et~al.}{1975}]{1975ApJ...197L.133B}
\begin{barticle}
\bauthor{\binits{J.D.} \bsnm{{Bohlin}}},
\bauthor{\binits{S.N.} \bsnm{{Vogel}}},
\bauthor{\binits{J.D.} \bsnm{{Purcell}}}, \betal,
\batitle{{A newly observed solar feature - Macrospicules in He II 304 A}}.
\bjtitle{\apjl}
\bvolume{197},
\bfpage{133}
(\byear{1975}).
doi:\doiurl{10.1086/181794}
\end{barticle}
\endbibitem

\bibitem[\protect\citeauthoryear{{Bout} et~al.}{2002}]{2002ESASP.508..379B}
\begin{bchapter}
\bauthor{\binits{M.} \bsnm{{Bout}}},
\bauthor{\binits{P.} \bsnm{{Lamy}}},
\bauthor{\binits{A.} \bsnm{{Llebaria}}},
\bctitle{{Electron density in coronal jets}},
in \bbtitle{From Solar Min to Max: Half a Solar Cycle with SOHO},
ed. by \beditor{\binits{A.} \bsnm{{Wilson}}},
\bsertitle{ESA SP-508},
pp. \bfpage{379}--\blpage{382}
(\byear{2002}).
\end{bchapter}
\endbibitem

\bibitem[\protect\citeauthoryear{{Bray} \& {Loughhead}}{1974}]{1974soch.book.....B}
\begin{bbook}
\bauthor{\binits{R.J.} \bsnm{{Bray}}},
\bauthor{\binits{R.E.} \bsnm{{Loughhead}}},
\bbtitle{{The solar chromosphere}},
\bsertitle{The International Astrophysics Series, London: Chapman and Hall}
(\byear{1974}).
\end{bbook}
\endbibitem

\bibitem[\protect\citeauthoryear{{Brueckner} \& {Bartoe}}{1978}]{1978BAAS...10R.416B}
\begin{bchapter}
\bauthor{\binits{G.E.} \bsnm{{Brueckner}}},
\bauthor{\binits{J.-D.F.} \bsnm{{Bartoe}}},
\bctitle{{High Velocity Jets in the ``Quiet'' Sun as a Possible Source of the
  Solar Wind and the Heating of the Corona}},
\bjtitle{\baas}
\bvolume{10},
\bfpage{416}
(\byear{1978}).
\end{bchapter}
\endbibitem

\bibitem[\protect\citeauthoryear{{Brueckner}}{1980}]{1980HiA.....5..557B}
\begin{barticle}
\bauthor{\binits{G.E.} \bsnm{{Brueckner}}},
\batitle{{A high-resolution view of the solar chromosphere and corona}}.
\bjtitle{Highlights of Astronomy}
\bvolume{5},
\bfpage{557}--\blpage{569}
(\byear{1980}).
\end{barticle}
\endbibitem

\bibitem[\protect\citeauthoryear{{Brueckner} \& {Bartoe}}{1983}]{1983ApJ...272..329B}
\begin{barticle}
\bauthor{\binits{G.E.} \bsnm{{Brueckner}}},
\bauthor{\binits{J.-D.F.} \bsnm{{Bartoe}}},
\batitle{{Observations of high-energy jets in the corona above the quiet sun,
  the heating of the corona, and the acceleration of the solar wind}}.
\bjtitle{\apj}
\bvolume{272},
\bfpage{329}
(\byear{1983}).
doi:\doiurl{10.1086/161297}
\end{barticle}
\endbibitem

\bibitem[\protect\citeauthoryear{{Brueckner} et~al.}{1995}]{1995SoPh..162..357B}
\begin{barticle}
\bauthor{\binits{G.E.} \bsnm{{Brueckner}}},
\bauthor{\binits{R.A.} \bsnm{{Howard}}},
\bauthor{\binits{M.J.} \bsnm{{Koomen}}}, \betal,
\batitle{{The Large Angle Spectroscopic Coronagraph (LASCO)}}.
\bjtitle{\solphys}
\bvolume{162},
\bfpage{357}
(\byear{1995}).
doi:\doiurl{10.1007/BF00733434}
\end{barticle}
\endbibitem

\bibitem[\protect\citeauthoryear{{Canfield} et~al.}{1996}]{1996ApJ...464.1016C}
\begin{barticle}
\bauthor{\binits{R.C.} \bsnm{{Canfield}}},
\bauthor{\binits{K.P.} \bsnm{{Reardon}}},
\bauthor{\binits{K.D.} \bsnm{{Leka}}}, \betal,
\batitle{{H alpha Surges and X-Ray Jets in AR 7260}}.
\bjtitle{\apj}
\bvolume{464},
\bfpage{1016}
(\byear{1996}).
doi:\doiurl{10.1086/177389}
\end{barticle}
\endbibitem

\bibitem[\protect\citeauthoryear{{Chae}}{2003}]{2003ApJ...584.1084C}
\begin{barticle}
\bauthor{\binits{J.} \bsnm{{Chae}}},
\batitle{{The Formation of a Prominence in NOAA Active Region 8668. II. Trace
  Observations of Jets and Eruptions Associated with Canceling Magnetic
  Features}}.
\bjtitle{\apj}
\bvolume{584},
\bfpage{1084}
(\byear{2003}).
doi:\doiurl{10.1086/345739}
\end{barticle}
\endbibitem

\bibitem[\protect\citeauthoryear{{Chandrashekhar}
  et~al.}{2014a}]{2014A&A...561A.104C}
\begin{barticle}
\bauthor{\binits{K.} \bsnm{{Chandrashekhar}}},
\bauthor{\binits{A.} \bsnm{{Bemporad}}},
\bauthor{\binits{D.} \bsnm{{Banerjee}}}, \betal,
\batitle{{Characteristics of polar coronal hole jets}}.
\bjtitle{\aap}
\bvolume{561},
\bfpage{104}
(\byear{2014}a).
doi:\doiurl{10.1051/0004-6361/201321213}
\end{barticle}
\endbibitem

\bibitem[\protect\citeauthoryear{{Chandrashekhar}
  et~al.}{2014b}]{2014A&A...562A..98C}
\begin{barticle}
\bauthor{\binits{K.} \bsnm{{Chandrashekhar}}},
\bauthor{\binits{R.J.} \bsnm{{Morton}}},
\bauthor{\binits{D.} \bsnm{{Banerjee}}},
\bauthor{\binits{G.R.} \bsnm{{Gupta}}},
\batitle{{The dynamical behaviour of a jet in an on-disk coronal hole observed
  with AIA/SDO}}.
\bjtitle{\aap}
\bvolume{562},
\bfpage{98}
(\byear{2014}b).
doi:\doiurl{10.1051/0004-6361/201322408}
\end{barticle}
\endbibitem

\bibitem[\protect\citeauthoryear{{Chen} et~al.}{2008}]{2008A&A...478..907C}
\begin{barticle}
\bauthor{\binits{H.D.} \bsnm{{Chen}}},
\bauthor{\binits{Y.C.} \bsnm{{Jiang}}},
\bauthor{\binits{S.L.} \bsnm{{Ma}}},
\batitle{{Observations of H{$\alpha$} surges and ultraviolet jets above
  satellite sunspots}}.
\bjtitle{\aap}
\bvolume{478},
\bfpage{907}
(\byear{2008}).
doi:\doiurl{10.1051/0004-6361:20078641}
\end{barticle}
\endbibitem

\bibitem[\protect\citeauthoryear{{Chen} et~al.}{2009}]{2009SoPh..255...79C}
\begin{barticle}
\bauthor{\binits{H.} \bsnm{{Chen}}},
\bauthor{\binits{Y.} \bsnm{{Jiang}}},
\bauthor{\binits{S.} \bsnm{{Ma}}},
\batitle{{An EUV Jet and H{$\alpha$} Filament Eruption Associated with Flux
  Cancelation in a Decaying Active Region}}.
\bjtitle{\solphys}
\bvolume{255},
\bfpage{79}
(\byear{2009}).
doi:\doiurl{10.1007/s11207-008-9298-1}
\end{barticle}
\endbibitem

\bibitem[\protect\citeauthoryear{{Chen} et~al.}{2012}]{2012RAA....12..573C}
\begin{barticle}
\bauthor{\binits{H.-D.} \bsnm{{Chen}}},
\bauthor{\binits{J.} \bsnm{{Zhang}}},
\bauthor{\binits{S.-L.} \bsnm{{Ma}}},
\batitle{{The kinematics of an untwisting solar jet in a polar coronal hole
  observed by SDO/AIA}}.
\bjtitle{\raa}
\bvolume{12},
\bfpage{573}
(\byear{2012}).
doi:\doiurl{10.1088/1674-4527/12/5/009}
\end{barticle}
\endbibitem

\bibitem[\protect\citeauthoryear{{Chen} et~al.}{2013}]{2013ApJ...769...96C}
\begin{barticle}
\bauthor{\binits{N.} \bsnm{{Chen}}},
\bauthor{\binits{W.-H.} \bsnm{{Ip}}},
\bauthor{\binits{D.} \bsnm{{Innes}}},
\batitle{{Flare-Associated Type III Radio Bursts and Dynamics of the EUV Jet
  from SDO/AIA and RHESSI Observations}}.
\bjtitle{\apj}
\bvolume{769},
\bfpage{96}
(\byear{2013}).
doi:\doiurl{10.1088/0004-637X/769/2/96}
\end{barticle}
\endbibitem

\bibitem[\protect\citeauthoryear{{Cheung} et~al.}{2015}]{2015ApJ...801...83C}
\begin{barticle}
\bauthor{\binits{M.C.M.} \bsnm{{Cheung}}},
\bauthor{\binits{B.} \bsnm{{de Pontieu}}},
\bauthor{\binits{T.D.} \bsnm{{Tarbell}}}, \betal,
\batitle{{Homologous Helical Jets: Observations By IRIS, SDO, and Hinode and
  Magnetic Modeling With Data-Driven Simulations}}.
\bjtitle{\apj}
\bvolume{801},
\bfpage{83}
(\byear{2015}).
doi:\doiurl{10.1088/0004-637X/801/2/83}
\end{barticle}
\endbibitem

\bibitem[\protect\citeauthoryear{{Chifor} et~al.}{2008a}]{2008A&A...491..279C}
\begin{barticle}
\bauthor{\binits{C.} \bsnm{{Chifor}}},
\bauthor{\binits{H.} \bsnm{{Isobe}}},
\bauthor{\binits{H.E.} \bsnm{{Mason}}}, \betal,
\batitle{{Magnetic flux cancellation associated with a recurring solar jet
  observed with Hinode, RHESSI, and STEREO/EUVI}}.
\bjtitle{\aap}
\bvolume{491},
\bfpage{279}--\blpage{288}
(\byear{2008}a).
doi:\doiurl{10.1051/0004-6361:200810265}
\end{barticle}
\endbibitem

\bibitem[\protect\citeauthoryear{{Chifor} et~al.}{2008b}]{chifor08-jet1}
\begin{barticle}
\bauthor{\binits{C.} \bsnm{{Chifor}}},
\bauthor{\binits{P.R.} \bsnm{{Young}}},
\bauthor{\binits{H.} \bsnm{{Isobe}}}, \betal,
\batitle{{An active region jet observed with Hinode}}.
\bjtitle{\aap}
\bvolume{481},
\bfpage{57}
(\byear{2008}b).
doi:\doiurl{10.1051/0004-6361:20079081}
\end{barticle}
\endbibitem

\bibitem[\protect\citeauthoryear{{Christe} et~al.}{2008}]{2008ApJ...680L.149C}
\begin{barticle}
\bauthor{\binits{S.} \bsnm{{Christe}}},
\bauthor{\binits{S.} \bsnm{{Krucker}}},
\bauthor{\binits{R.P.} \bsnm{{Lin}}},
\batitle{{Hard X-Rays Associated with Type III Radio Bursts}}.
\bjtitle{\apjl}
\bvolume{680},
\bfpage{149}--\blpage{152}
(\byear{2008}).
doi:\doiurl{10.1086/589971}
\end{barticle}
\endbibitem

\bibitem[\protect\citeauthoryear{{Cirtain} et~al.}{2007}]{2007Sci...318.1580C}
\begin{barticle}
\bauthor{\binits{J.W.} \bsnm{{Cirtain}}},
\bauthor{\binits{L.} \bsnm{{Golub}}},
\bauthor{\binits{L.} \bsnm{{Lundquist}}}, \betal,
\batitle{{Evidence for Alfv{\'e}n Waves in Solar X-ray Jets}}.
\bjtitle{Science}
\bvolume{318},
\bfpage{1580}
(\byear{2007}).
doi:\doiurl{10.1126/science.1147050}
\end{barticle}
\endbibitem

\bibitem[\protect\citeauthoryear{{Corti} et~al.}{2007}]{2007ApJ...659.1702C}
\begin{barticle}
\bauthor{\binits{G.} \bsnm{{Corti}}},
\bauthor{\binits{G.} \bsnm{{Poletto}}},
\bauthor{\binits{S.T.} \bsnm{{Suess}}}, \betal,
\batitle{{Cool-Plasma Jets that Escape into the Outer Corona}}.
\bjtitle{\apj}
\bvolume{659},
\bfpage{1702}
(\byear{2007}).
doi:\doiurl{10.1086/512233}
\end{barticle}
\endbibitem

\bibitem[\protect\citeauthoryear{{Culhane} et~al.}{2007a}]{culhane07}
\begin{barticle}
\bauthor{\binits{J.L.} \bsnm{{Culhane}}},
\bauthor{\binits{L.K.} \bsnm{{Harra}}},
\bauthor{\binits{A.M.} \bsnm{{James}}}, \betal,
\batitle{{The EUV Imaging Spectrometer for Hinode}}.
\bjtitle{\solphys}
\bvolume{243},
\bfpage{19}
(\byear{2007}a).
doi:\doiurl{10.1007/s01007-007-0293-1}
\end{barticle}
\endbibitem

\bibitem[\protect\citeauthoryear{{Culhane} et~al.}{2007b}]{culhane07-jet}
\begin{barticle}
\bauthor{\binits{L.} \bsnm{{Culhane}}},
\bauthor{\binits{L.K.} \bsnm{{Harra}}},
\bauthor{\binits{D.} \bsnm{{Baker}}}, \betal,
\batitle{{Hinode EUV Study of Jets in the Sun's South Polar Corona}}.
\bjtitle{\pasj}
\bvolume{59},
\bfpage{751}
(\byear{2007}b).
doi:\doiurl{10.1093/pasj/59.sp3.S751}
\end{barticle}
\endbibitem

\bibitem[\protect\citeauthoryear{{Curdt} et~al.}{2012}]{2012SoPh..280.417C}
\begin{barticle}
\bauthor{\binits{W.} \bsnm{{Curdt}}},
\bauthor{\binits{H.} \bsnm{{Tian}}},
\bauthor{\binits{S.} \bsnm{{Kamio}}},
\batitle{{Explosive Events: Swirling Transition Region Jets}}.
\bjtitle{\solphys}
\bvolume{280},
\bfpage{417}
(\byear{2012}).
doi:\doiurl{10.1007/s11207-012-9940-9}
\end{barticle}
\endbibitem

\bibitem[\protect\citeauthoryear{{Dalmasse} et~al.}{2012}]{2012EAS....55..201D}
\begin{bchapter}
\bauthor{\binits{K.} \bsnm{{Dalmasse}}},
\bauthor{\binits{E.} \bsnm{{Pariat}}},
\bauthor{\binits{S.K.} \bsnm{{Antiochos}}},
\bauthor{\binits{C.R.} \bsnm{{DeVore}}},
\bctitle{{Coronal jets in an inclined coronal magnetic field : a parametric 3D MHD study}},
in \bbtitle{EAS Publications Series},
vol. \bseriesno{55},
pp. \bfpage{201}--\blpage{205}
(\byear{2012}).
doi:\doiurl{10.1051/eas/1255028}
\end{bchapter}
\endbibitem

%

%

%

\bibitem[\protect\citeauthoryear{{de Pontieu}
  et~al.}{2014a}]{2014SoPh..289.2733D}
\begin{barticle}
\bauthor{\binits{B.} \bsnm{{de Pontieu}}},
\bauthor{\binits{A.M.} \bsnm{{Title}}},
\bauthor{\binits{J.R.} \bsnm{{Lemen}}}, \betal,
\batitle{{The Interface Region Imaging Spectrograph (IRIS)}}.
\bjtitle{\solphys}
\bvolume{289},
\bfpage{2733}
(\byear{2014}a).
doi:\doiurl{10.1007/s11207-014-0485-y}
\end{barticle}
\endbibitem

\bibitem[\protect\citeauthoryear{{de Pontieu}
  et~al.}{2014b}]{2014Sci...346D.315D}
\begin{barticle}
\bauthor{\binits{B.} \bsnm{{de Pontieu}}},
\bauthor{\binits{L.} \bsnm{{Rouppe van der Voort}}},
\bauthor{\binits{S.W.} \bsnm{{McIntosh}}}, \betal,
\batitle{{On the prevalence of small-scale twist in the solar chromosphere and
  transition region}}.
\bjtitle{Science}
\bvolume{346},
\bfpage{1255732}
(\byear{2014}b).
doi:\doiurl{10.1126/science.1255732}
\end{barticle}
\endbibitem

\bibitem[\protect\citeauthoryear{{Deforest} et~al.}{1997}]{1997SoPh..175..393D}
\begin{barticle}
\bauthor{\binits{C.E.} \bsnm{{Deforest}}},
\bauthor{\binits{J.T.} \bsnm{{Hoeksema}}},
\bauthor{\binits{J.B.} \bsnm{{Gurman}}}, \betal,
\batitle{{Polar Plume Anatomy: Results of a Coordinated Observation}}.
\bjtitle{\solphys}
\bvolume{175},
\bfpage{393}
(\byear{1997}).
doi:\doiurl{10.1023/A:1004955223306}
\end{barticle}
\endbibitem

\bibitem[\protect\citeauthoryear{{Delaboudini{\`e}re}
  et~al.}{1995}]{1995SoPh..162..291D}
\begin{barticle}
\bauthor{\binits{J.-P.} \bsnm{{Delaboudini{\`e}re}}},
\bauthor{\binits{G.E.} \bsnm{{Artzner}}},
\bauthor{\binits{J.} \bsnm{{Brunaud}}}, \betal,
\batitle{{EIT: Extreme-Ultraviolet Imaging Telescope for the SOHO Mission}}.
\bjtitle{\solphys}
\bvolume{162},
\bfpage{291}
(\byear{1995}).
doi:\doiurl{10.1007/BF00733432}
\end{barticle}
\endbibitem

\bibitem[\protect\citeauthoryear{{Demastus} et~al.}{1973}]{1973SoPh...31..449D}
\begin{barticle}
\bauthor{\binits{H.L.} \bsnm{{Demastus}}},
\bauthor{\binits{W.J.} \bsnm{{Wagner}}},
\bauthor{\binits{R.D.} \bsnm{{Robinson}}},
\batitle{{Coronal Disturbances. I: Fast Transient Events Observed in the Green
  Coronal Emission Line During the Last Solar Cycle}}.
\bjtitle{\solphys}
\bvolume{31},
\bfpage{449}
(\byear{1973}).
doi:\doiurl{10.1007/BF00152820}
\end{barticle}
\endbibitem

\bibitem[\protect\citeauthoryear{{DeVore} \& {Antiochos}}{2008}]{2008ApJ...680..740D}
\begin{barticle}
\bauthor{\binits{C.R.} \bsnm{{DeVore}}},
\bauthor{\binits{S.K.} \bsnm{{Antiochos}}},
\batitle{{Homologous Confined Filament Eruptions via Magnetic Breakout}}.
\bjtitle{\apj}
\bvolume{680},
\bfpage{740}
(\byear{2008}).
doi:\doiurl{10.1086/588011}
\end{barticle}
\endbibitem

\bibitem[\protect\citeauthoryear{{Dobrzycka}
  et~al.}{2000}]{2000ApJ...538..922D}
\begin{barticle}
\bauthor{\binits{D.} \bsnm{{Dobrzycka}}},
\bauthor{\binits{J.C.} \bsnm{{Raymond}}},
\bauthor{\binits{S.R.} \bsnm{{Cranmer}}},
\batitle{{Ultraviolet Spectroscopy of Polar Coronal Jets}}.
\bjtitle{\apj}
\bvolume{538},
\bfpage{922}
(\byear{2000}).
doi:\doiurl{10.1086/309173}
\end{barticle}
\endbibitem

\bibitem[\protect\citeauthoryear{{Dobrzycka}
  et~al.}{2002}]{2002ApJ...565..621D}
\begin{barticle}
\bauthor{\binits{D.} \bsnm{{Dobrzycka}}},
\bauthor{\binits{S.R.} \bsnm{{Cranmer}}},
\bauthor{\binits{J.C.} \bsnm{{Raymond}}}, \betal,
\batitle{{Polar Coronal Jets at Solar Minimum}}.
\bjtitle{\apj}
\bvolume{565},
\bfpage{621}
(\byear{2002}).
doi:\doiurl{10.1086/324431}
\end{barticle}
\endbibitem

\bibitem[\protect\citeauthoryear{{Dobrzycka}
  et~al.}{2003}]{2003ApJ...588..586D}
\begin{barticle}
\bauthor{\binits{D.} \bsnm{{Dobrzycka}}},
\bauthor{\binits{J.C.} \bsnm{{Raymond}}},
\bauthor{\binits{D.A.} \bsnm{{Biesecker}}}, \betal,
\batitle{{Ultraviolet Spectroscopy of Narrow Coronal Mass Ejections}}.
\bjtitle{\apj}
\bvolume{588},
\bfpage{586}
(\byear{2003}).
doi:\doiurl{10.1086/374047}
\end{barticle}
\endbibitem

\bibitem[\protect\citeauthoryear{{Domingo} et~al.}{1995}]{1995SoPh..162....1D}
\begin{barticle}
\bauthor{\binits{V.} \bsnm{{Domingo}}},
\bauthor{\binits{B.} \bsnm{{Fleck}}},
\bauthor{\binits{A.I.} \bsnm{{Poland}}},
\batitle{{The SOHO Mission: an Overview}}.
\bjtitle{\solphys}
\bvolume{162},
\bfpage{1}
(\byear{1995}).
doi:\doiurl{10.1007/BF00733425}
\end{barticle}
\endbibitem

\bibitem[\protect\citeauthoryear{{Dwivedi} \& {Wilhelm}}{2015}]{2015JApA...36..185D}
\begin{barticle}
\bauthor{\binits{B.N.} \bsnm{{Dwivedi}}},
\bauthor{\binits{K.} \bsnm{{Wilhelm}}},
\batitle{{Solar Coronal Plumes and the Fast Solar Wind}}.
\bjtitle{\jaa}
\bvolume{36},
\bfpage{185}
(\byear{2015}).
doi:\doiurl{10.1007/s12036-015-9326-0}
\end{barticle}
\endbibitem

\bibitem[\protect\citeauthoryear{{Eyles} et~al.}{2003}]{2003SoPh..217..319E}
\begin{barticle}
\bauthor{\binits{C.J.} \bsnm{{Eyles}}},
\bauthor{\binits{G.M.} \bsnm{{Simnett}}},
\bauthor{\binits{M.P.} \bsnm{{Cooke}}}, \betal,
\batitle{{The Solar Mass Ejection Imager (SMEI)}}.
\bjtitle{\solphys}
\bvolume{217},
\bfpage{319}
(\byear{2003}).
doi:\doiurl{10.1023/B:SOLA.0000006903.75671.49}
\end{barticle}
\endbibitem

\bibitem[\protect\citeauthoryear{{Fan} \& {Gibson}}{2004}]{2004ApJ...609.1123F}
\begin{barticle}
\bauthor{\binits{Y.} \bsnm{{Fan}}},
\bauthor{\binits{S.E.} \bsnm{{Gibson}}},
\batitle{{Numerical Simulations of Three-dimensional Coronal Magnetic Fields
  Resulting from the Emergence of Twisted Magnetic Flux Tubes}}.
\bjtitle{\apj}
\bvolume{609},
\bfpage{1123}
(\byear{2004}).
doi:\doiurl{10.1086/421238}
\end{barticle}
\endbibitem

\bibitem[\protect\citeauthoryear{{Fang} et~al.}{2014}]{2014ApJ...789L..19F}
\begin{barticle}
\bauthor{\binits{F.} \bsnm{{Fang}}},
\bauthor{\binits{Y.} \bsnm{{Fan}}},
\bauthor{\binits{S.W.} \bsnm{{McIntosh}}},
\batitle{{Rotating Solar Jets in Simulations of Flux Emergence with Thermal
  Conduction}}.
\bjtitle{\apjl}
\bvolume{789},
\bfpage{19}
(\byear{2014}).
doi:\doiurl{10.1088/2041-8205/789/1/L19}
\end{barticle}
\endbibitem

\bibitem[\protect\citeauthoryear{{Feng} et~al.}{2012}]{2012A&A...538A..34F}
\begin{barticle}
\bauthor{\binits{L.} \bsnm{{Feng}}},
\bauthor{\binits{B.} \bsnm{{Inhester}}},
\bauthor{\binits{J.} \bsnm{{de Patoul}}}, \betal,
\batitle{{Particle kinetic analysis of a polar jet from SECCHI COR data}}.
\bjtitle{\aap}
\bvolume{538},
\bfpage{34}
(\byear{2012}).
doi:\doiurl{10.1051/0004-6361/201117071}
\end{barticle}
\endbibitem

\bibitem[\protect\citeauthoryear{{Filippov} et~al.}{2013}]{2013SoPh..286..143F}
\begin{barticle}
\bauthor{\binits{B.} \bsnm{{Filippov}}},
\bauthor{\binits{S.} \bsnm{{Koutchmy}}},
\bauthor{\binits{E.} \bsnm{{Tavabi}}},
\batitle{{Formation of a White-Light Jet Within a Quadrupolar Magnetic
  Configuration}}.
\bjtitle{\solphys}
\bvolume{286},
\bfpage{143}
(\byear{2013}).
doi:\doiurl{10.1007/s11207-011-9911-6}
\end{barticle}
\endbibitem

\bibitem[\protect\citeauthoryear{{Gibb} et~al.}{2014}]{2014ApJ...782...71G}
\begin{barticle}
\bauthor{\binits{G.P.S.} \bsnm{{Gibb}}},
\bauthor{\binits{D.H.} \bsnm{{Mackay}}},
\bauthor{\binits{L.M.} \bsnm{{Green}}},
\bauthor{\binits{K.A.} \bsnm{{Meyer}}},
\batitle{{Simulating the Formation of a Sigmoidal Flux Rope in AR10977 from
  SOHO/MDI Magnetograms}}.
\bjtitle{\apj}
\bvolume{782},
\bfpage{71}
(\byear{2014}).
doi:\doiurl{10.1088/0004-637X/782/2/71}
\end{barticle}
\endbibitem

\bibitem[\protect\citeauthoryear{{Glesener} et~al.}{2012}]{2012ApJ...754....9G}
\begin{barticle}
\bauthor{\binits{L.} \bsnm{{Glesener}}},
\bauthor{\binits{S.} \bsnm{{Krucker}}},
\bauthor{\binits{R.P.} \bsnm{{Lin}}},
\batitle{{Hard X-Ray Observations of a Jet and Accelerated Electrons in the
  Corona}}.
\bjtitle{\apj}
\bvolume{754},
\bfpage{9}
(\byear{2012}).
doi:\doiurl{10.1088/0004-637X/754/1/9}
\end{barticle}
\endbibitem

\bibitem[\protect\citeauthoryear{{Golub} et~al.}{2007}]{2007SoPh..243...63G}
\begin{barticle}
\bauthor{\binits{L.} \bsnm{{Golub}}},
\bauthor{\binits{E.} \bsnm{{Deluca}}},
\bauthor{\binits{G.} \bsnm{{Austin}}}, \betal,
\batitle{{The X-Ray Telescope (XRT) for the Hinode Mission}}.
\bjtitle{\solphys}
\bvolume{243},
\bfpage{63}
(\byear{2007}).
doi:\doiurl{10.1007/s11207-007-0182-1}
\end{barticle}
\endbibitem

\bibitem[\protect\citeauthoryear{{Gontikakis}
  et~al.}{2009}]{2009A&A...506L..45G}
\begin{barticle}
\bauthor{\binits{C.} \bsnm{{Gontikakis}}},
\bauthor{\binits{V.} \bsnm{{Archontis}}},
\bauthor{\binits{K.} \bsnm{{Tsinganos}}},
\batitle{{Observations and 3D MHD simulations of a solar active region jet}}.
\bjtitle{\aap}
\bvolume{506},
\bfpage{45}
(\byear{2009}).
doi:\doiurl{10.1051/0004-6361/200913026}
\end{barticle}
\endbibitem

\bibitem[\protect\citeauthoryear{{Gu} et~al.}{1994}]{1994A&A...282..240G}
\begin{barticle}
\bauthor{\binits{X.M.} \bsnm{{Gu}}},
\bauthor{\binits{J.} \bsnm{{Lin}}},
\bauthor{\binits{K.J.} \bsnm{{Li}}}, \betal,
\batitle{{Kinematic characteristics of the surge on March 19, 1989}}.
\bjtitle{\aap}
\bvolume{282},
\bfpage{240}
(\byear{1994}).
\end{barticle}
\endbibitem

\bibitem[\protect\citeauthoryear{{Handy} et~al.}{1999}]{1999SoPh..187..229H}
\begin{barticle}
\bauthor{\binits{B.N.} \bsnm{{Handy}}},
\bauthor{\binits{L.W.} \bsnm{{Acton}}},
\bauthor{\binits{C.C.} \bsnm{{Kankelborg}}}, \betal,
\batitle{{The transition region and coronal explorer}}.
\bjtitle{\solphys}
\bvolume{187},
\bfpage{229}
(\byear{1999}).
doi:\doiurl{10.1023/A:1005166902804}
\end{barticle}
\endbibitem

\bibitem[\protect\citeauthoryear{{Harrison} et~al.}{1995}]{harrison95}
\begin{barticle}
\bauthor{\binits{R.A.} \bsnm{{Harrison}}},
\bauthor{\binits{E.C.} \bsnm{{Sawyer}}},
\bauthor{\binits{M.K.} \bsnm{{Carter}}}, \betal,
\batitle{{The Coronal Diagnostic Spectrometer for the Solar and Heliospheric
  Observatory}}.
\bjtitle{\solphys}
\bvolume{162},
\bfpage{233}
(\byear{1995}).
doi:\doiurl{10.1007/BF00733431}
\end{barticle}
\endbibitem

\bibitem[\protect\citeauthoryear{{Heyvaerts}
  et~al.}{1977}]{1977ApJ...216..123H}
\begin{barticle}
\bauthor{\binits{J.} \bsnm{{Heyvaerts}}},
\bauthor{\binits{E.R.} \bsnm{{Priest}}},
\bauthor{\binits{D.M.} \bsnm{{Rust}}},
\batitle{{An emerging flux model for the solar flare phenomenon}}.
\bjtitle{\apj}
\bvolume{216},
\bfpage{123}
(\byear{1977}).
doi:\doiurl{10.1086/155453}
\end{barticle}
\endbibitem

\bibitem[\protect\citeauthoryear{{Hong} et~al.}{2011}]{2011ApJ...738L..20H}
\begin{barticle}
\bauthor{\binits{J.} \bsnm{{Hong}}},
\bauthor{\binits{Y.} \bsnm{{Jiang}}},
\bauthor{\binits{R.} \bsnm{{Zheng}}}, \betal,
\batitle{{A Micro Coronal Mass Ejection Associated Blowout Extreme-ultraviolet Jet}}.
\bjtitle{\apjl}
\bvolume{738},
\bfpage{20}
(\byear{2011}).
doi:\doiurl{10.1088/2041-8205/738/2/L20}
\end{barticle}
\endbibitem

\bibitem[\protect\citeauthoryear{{Hong} et~al.}{2013}]{2013RAA....13..253H}
\begin{barticle}
\bauthor{\binits{J.-C.} \bsnm{{Hong}}},
\bauthor{\binits{Y.-C.} \bsnm{{Jiang}}},
\bauthor{\binits{J.-Y.} \bsnm{{Yang}}}, \betal,
\batitle{{Twist in a polar blowout jet}}.
\bjtitle{\raa}
\bvolume{13},
\bfpage{253}
(\byear{2013}).
doi:\doiurl{10.1088/1674-4527/13/3/001}
\end{barticle}
\endbibitem

\bibitem[\protect\citeauthoryear{{Hong} et~al.}{2014}]{2014ApJ...796...73H}
\begin{barticle}
\bauthor{\binits{J.} \bsnm{{Hong}}},
\bauthor{\binits{Y.} \bsnm{{Jiang}}},
\bauthor{\binits{J.} \bsnm{{Yang}}}, \betal,
\batitle{{Coronal Bright Points Associated with Minifilament Eruptions}}.
\bjtitle{\apj}
\bvolume{796},
\bfpage{73}
(\byear{2014}).
doi:\doiurl{10.1088/0004-637X/796/2/73}
\end{barticle}
\endbibitem

\bibitem[\protect\citeauthoryear{{Howard} et~al.}{2008}]{2008SSRv..136...67H}
\begin{barticle}
\bauthor{\binits{R.A.} \bsnm{{Howard}}},
\bauthor{\binits{J.D.} \bsnm{{Moses}}},
\bauthor{\binits{A.} \bsnm{{Vourlidas}}}, \betal,
\batitle{{Sun Earth Connection Coronal and Heliospheric Investigation
  (SECCHI)}}.
\bjtitle{\ssr}
\bvolume{136},
\bfpage{67}
(\byear{2008}).
doi:\doiurl{10.1007/s11214-008-9341-4}
\end{barticle}
\endbibitem

\bibitem[\protect\citeauthoryear{{Hsieh} \& {Simpson}}{1970}]{1970ApJ...162L.191H}
\begin{barticle}
\bauthor{\binits{K.C.} \bsnm{{Hsieh}}},
\bauthor{\binits{J.A.} \bsnm{{Simpson}}},
\batitle{{The Relative Abundances and Energy Spectra of $^3$He and $^4$He from
  Solar Flares}}.
\bjtitle{\apjl}
\bvolume{162},
\bfpage{191}
(\byear{1970}).
doi:\doiurl{10.1086/180652}
\end{barticle}
\endbibitem

\bibitem[\protect\citeauthoryear{{Innes} et~al.}{2011}]{2011A&A...531L..13I}
\begin{barticle}
\bauthor{\binits{D.E.} \bsnm{{Innes}}},
\bauthor{\binits{R.H.} \bsnm{{Cameron}}},
\bauthor{\binits{S.K.} \bsnm{{Solanki}}},
\batitle{{EUV jets, type III radio bursts and sunspot waves investigated using
  SDO/AIA observations}}.
\bjtitle{\aap}
\bvolume{531},
\bfpage{13}
(\byear{2011}).
doi:\doiurl{10.1051/0004-6361/201117255}
\end{barticle}
\endbibitem

%

\bibitem[\protect\citeauthoryear{{Jackson} et~al.}{2004}]{2004SoPh..225..177J}
\begin{barticle}
\bauthor{\binits{B.V.} \bsnm{{Jackson}}},
\bauthor{\binits{A.} \bsnm{{Buffington}}},
\bauthor{\binits{P.P.} \bsnm{{Hick}}}, \betal,
\batitle{{The Solar Mass-Ejection Imager (SMEI) Mission}}.
\bjtitle{\solphys}
\bvolume{225},
\bfpage{177}
(\byear{2004}).
doi:\doiurl{10.1007/s11207-004-2766-3}
\end{barticle}
\endbibitem


\bibitem[\protect\citeauthoryear{{Jiang} et~al.}{2007}]{2007A&A...469..331J}
\begin{barticle}
\bauthor{\binits{Y.C.} \bsnm{{Jiang}}},
\bauthor{\binits{H.D.} \bsnm{{Chen}}},
\bauthor{\binits{K.J.} \bsnm{{Li}}}, \betal,
\batitle{{The H{$\alpha$} surges and EUV jets from magnetic flux emergences and
  cancellations}}.
\bjtitle{\aap}
\bvolume{469},
\bfpage{331}
(\byear{2007}).
doi:\doiurl{10.1051/0004-6361:20053954}
\end{barticle}
\endbibitem

\bibitem[\protect\citeauthoryear{{Jibben} \& {Canfield}}{2004}]{2004ApJ...610.1129J}
\begin{barticle}
\bauthor{\binits{P.} \bsnm{{Jibben}}},
\bauthor{\binits{R.C.} \bsnm{{Canfield}}},
\batitle{{Twist Propagation in H{$\alpha$} Surges}}.
\bjtitle{\apj}
\bvolume{610},
\bfpage{1129}
(\byear{2004}).
doi:\doiurl{10.1086/421727}
\end{barticle}
\endbibitem

\bibitem[\protect\citeauthoryear{{Kahler} et~al.}{1984}]{1984JGR....89.9683K}
\begin{barticle}
\bauthor{\binits{S.W.} \bsnm{{Kahler}}},
\bauthor{\binits{N.R.} \bsnm{{Sheeley}} \bsuffix{Jr.}},
\bauthor{\binits{R.A.} \bsnm{{Howard}}}, \betal,
\batitle{{Associations between coronal mass ejections and solar energetic
  proton events}}.
\bjtitle{\jgr}
\bvolume{89},
\bfpage{9683}
(\byear{1984}).
doi:\doiurl{10.1029/JA089iA11p09683}
\end{barticle}
\endbibitem

\bibitem[\protect\citeauthoryear{{Kahler} et~al.}{1987}]{1987SoPh..107..385K}
\begin{barticle}
\bauthor{\binits{S.W.} \bsnm{{Kahler}}},
\bauthor{\binits{R.P.} \bsnm{{Lin}}},
\bauthor{\binits{D.V.} \bsnm{{Reames}}}, \betal,
\batitle{{Characteristics of solar coronal source regions producing He-3-rich
  particle events}}.
\bjtitle{\solphys}
\bvolume{107},
\bfpage{385}
(\byear{1987}).
doi:\doiurl{10.1007/BF00152032}
\end{barticle}
\endbibitem

\bibitem[\protect\citeauthoryear{{Kahler} et~al.}{2001}]{2001ApJ...562..558K}
\begin{barticle}
\bauthor{\binits{S.W.} \bsnm{{Kahler}}},
\bauthor{\binits{D.V.} \bsnm{{Reames}}},
\bauthor{\binits{N.R.} \bsnm{{Sheeley}} \bsuffix{Jr.}},
\batitle{{Coronal Mass Ejections Associated with Impulsive Solar Energetic
  Particle Events}}.
\bjtitle{\apj}
\bvolume{562},
\bfpage{558}
(\byear{2001}).
doi:\doiurl{10.1086/323847}
\end{barticle}
\endbibitem

\bibitem[\protect\citeauthoryear{{Kaiser} et~al.}{2008}]{2008SSRv..136....5K}
\begin{barticle}
\bauthor{\binits{M.L.} \bsnm{{Kaiser}}},
\bauthor{\binits{T.A.} \bsnm{{Kucera}}},
\bauthor{\binits{J.M.} \bsnm{{Davila}}}, \betal,
\batitle{{The STEREO Mission: An Introduction}}.
\bjtitle{\ssr}
\bvolume{136},
\bfpage{5}
(\byear{2008}).
doi:\doiurl{10.1007/s11214-007-9277-0}
\end{barticle}
\endbibitem

\bibitem[\protect\citeauthoryear{{Kamio} et~al.}{2007}]{kamio07}
\begin{barticle}
\bauthor{\binits{S.} \bsnm{{Kamio}}},
\bauthor{\binits{H.} \bsnm{{Hara}}},
\bauthor{\binits{T.} \bsnm{{Watanabe}}}, \betal,
\batitle{{Velocity Structure of Jets in a Coronal Hole}}.
\bjtitle{\pasj}
\bvolume{59},
\bfpage{757}
(\byear{2007}).
\end{barticle}
\endbibitem

\bibitem[\protect\citeauthoryear{{Kamio} et~al.}{2010}]{2010A&A...510L...1K}
\begin{barticle}
\bauthor{\binits{S.} \bsnm{{Kamio}}},
\bauthor{\binits{W.} \bsnm{{Curdt}}},
\bauthor{\binits{L.} \bsnm{{Teriaca}}}, \betal,
\batitle{{Observations of a rotating macrospicule associated with an X-ray jet}}.
\bjtitle{\aap}
\bvolume{510},
\bfpage{1}
(\byear{2010}).
doi:\doiurl{10.1051/0004-6361/200913269}
\end{barticle}
\endbibitem

\bibitem[\protect\citeauthoryear{{Karovska} et~al.}{1999}]{1999SSRv...87..219K}
\begin{barticle}
\bauthor{\binits{M.} \bsnm{{Karovska}}},
\bauthor{\binits{B.E.} \bsnm{{Wood}}},
\bauthor{\binits{J.W.} \bsnm{{Cook}}}, \betal,
\batitle{{Study of Dynamical Properties of Coronal Structures in the Polar
  Regions}}.
\bjtitle{\ssr}
\bvolume{87},
\bfpage{219}
(\byear{1999}).
doi:\doiurl{10.1023/A:1005100618288}
\end{barticle}
\endbibitem


\bibitem[\protect\citeauthoryear{{Karpen} et~al.}{1995}]{1995ApJ...450..422K}
\begin{barticle}
\bauthor{\binits{J.T.} \bsnm{{Karpen}}},
\bauthor{\binits{S.K.} \bsnm{{Antiochos}}},
\bauthor{\binits{C.R.} \bsnm{{DeVore}}},
\batitle{{The Role of Magnetic Reconnection in Chromospheric Eruptions}}.
\bjtitle{\apj}
\bvolume{450},
\bfpage{422}
(\byear{1995}).
doi:\doiurl{10.1086/176152}
\end{barticle}
\endbibitem

\bibitem[\protect\citeauthoryear{{Karpen} et~al.}{1998}]{1998ApJ...495..491K}
\begin{barticle}
\bauthor{\binits{J.T.} \bsnm{{Karpen}}},
\bauthor{\binits{S.K.} \bsnm{{Antiochos}}},
\bauthor{\binits{C.R.} \bsnm{{DeVore}}},
\bauthor{\binits{L.} \bsnm{{Golub}}},
\batitle{{Dynamic Responses to Magnetic Reconnection in Solar Arcades}}.
\bjtitle{\apj}
\bvolume{495},
\bfpage{491}
(\byear{1998}).
doi:\doiurl{10.1086/305252}
\end{barticle}
\endbibitem

\bibitem[\protect\citeauthoryear{Karpen et~al.}{2016}]{Karpen15}
\begin{botherref}
\oauthor{\binits{J.T.} \bsnm{Karpen}},
\oauthor{\binits{C.R.} \bsnm{DeVore}},
\oauthor{\binits{S.K.} \bsnm{Antiochos}},
\oauthor{\binits{E.} \bsnm{Pariat}},
{Reconnection-Driven Coronal-Hole Jets with Gravity and Solar Wind}.
\bjtitle{\apj},
\textbf{submitted}
(\byear{2016}).
\end{botherref}
\endbibitem

\bibitem[\protect\citeauthoryear{{Kim} et~al.}{2007}]{2007PASJ...59S.763K}
\begin{barticle}
\bauthor{\binits{Y.-H.} \bsnm{{Kim}}},
\bauthor{\binits{Y.-J.} \bsnm{{Moon}}},
\bauthor{\binits{Y.-D.} \bsnm{{Park}}}, \betal,
\batitle{{Small-Scale X-Ray/EUV Jets Seen in Hinode XRT and TRACE}}.
\bjtitle{\pasj}
\bvolume{59},
\bfpage{763}
(\byear{2007}).
doi:\doiurl{10.1093/pasj/59.sp3.S763}
\end{barticle}
\endbibitem

%
%

\bibitem[\protect\citeauthoryear{{Ko} et~al.}{2005}]{2005ApJ...623..519K}
\begin{barticle}
\bauthor{\binits{Y.-K.} \bsnm{{Ko}}},
\bauthor{\binits{J.C.} \bsnm{{Raymond}}},
\bauthor{\binits{S.E.} \bsnm{{Gibson}}}, \betal,
\batitle{{Multialtitude Observations of a Coronal Jet during the Third Whole
  Sun Month Campaign}}.
\bjtitle{\apj}
\bvolume{623},
\bfpage{519}
(\byear{2005}).
doi:\doiurl{10.1086/428479}
\end{barticle}
\endbibitem

\bibitem[\protect\citeauthoryear{{Kohl} et~al.}{1995}]{1995SoPh..162..313K}
\begin{barticle}
\bauthor{\binits{J.L.} \bsnm{{Kohl}}},
\bauthor{\binits{R.} \bsnm{{Esser}}},
\bauthor{\binits{L.D.} \bsnm{{Gardner}}}, \betal,
\batitle{{The Ultraviolet Coronagraph Spectrometer for the Solar and
  Heliospheric Observatory}}.
\bjtitle{\solphys}
\bvolume{162},
\bfpage{313}
(\byear{1995}).
doi:\doiurl{10.1007/BF00733433}
\end{barticle}
\endbibitem

\bibitem[\protect\citeauthoryear{{Kosugi} et~al.}{2007}]{2007SoPh..243....3K}
\begin{barticle}
\bauthor{\binits{T.} \bsnm{{Kosugi}}},
\bauthor{\binits{K.} \bsnm{{Matsuzaki}}},
\bauthor{\binits{T.} \bsnm{{Sakao}}}, \betal,
\batitle{{The Hinode (Solar-B) Mission: An Overview}}.
\bjtitle{\solphys}
\bvolume{243},
\bfpage{3}
(\byear{2007}).
doi:\doiurl{10.1007/s11207-007-9014-6}
\end{barticle}
\endbibitem

\bibitem[\protect\citeauthoryear{{Krucker} et~al.}{2008}]{2008ApJ...681..644K}
\begin{barticle}
\bauthor{\binits{S.} \bsnm{{Krucker}}},
\bauthor{\binits{P.} \bsnm{{Saint-Hilaire}}},
\bauthor{\binits{S.} \bsnm{{Christe}}}, \betal,
\batitle{{Coronal Hard X-Ray Emission Associated with Radio Type III Bursts}}.
\bjtitle{\apj}
\bvolume{681},
\bfpage{644}
(\byear{2008}).
doi:\doiurl{10.1086/588549}
\end{barticle}
\endbibitem


\bibitem[\protect\citeauthoryear{{Kundu} et~al.}{1995}]{1995ApJ...447L.135K}
\begin{barticle}
\bauthor{\binits{M.R.} \bsnm{{Kundu}}},
\bauthor{\binits{J.P.} \bsnm{{Raulin}}},
\bauthor{\binits{N.} \bsnm{{Nitta}}}, \betal,
\batitle{{Detection of Nonthermal Radio Emission from Coronal X-Ray Jets}}.
\bjtitle{\apjl}
\bvolume{447},
\bfpage{135}
(\byear{1995}).
doi:\doiurl{10.1086/309567}
\end{barticle}
\endbibitem

\bibitem[\protect\citeauthoryear{{Kurokawa} et~al.}{1987}]{1987SoPh..108..251K}
\begin{barticle}
\bauthor{\binits{H.} \bsnm{{Kurokawa}}},
\bauthor{\binits{Y.} \bsnm{{Hanaoka}}},
\bauthor{\binits{K.} \bsnm{{Shibata}}},
\bauthor{\binits{Y.} \bsnm{{Uchida}}},
\batitle{{Rotating eruption of an untwisting filament triggered by the 3B flare
  of 25 April, 1984}}.
\bjtitle{\solphys}
\bvolume{108},
\bfpage{251}
(\byear{1987}).
doi:\doiurl{10.1007/BF00214165}
\end{barticle}
\endbibitem

\bibitem[\protect\citeauthoryear{{Lau} \& {Finn}}{1990}]{1990ApJ...350..672L}
\begin{barticle}
\bauthor{\binits{Y.-T.} \bsnm{{Lau}}},
\bauthor{\binits{J.M.} \bsnm{{Finn}}},
\batitle{{Three-dimensional kinematic reconnection in the presence of field
  nulls and closed field lines}}.
\bjtitle{\apj}
\bvolume{350},
\bfpage{672}
(\byear{1990}).
doi:\doiurl{10.1086/168419}
\end{barticle}
\endbibitem

\bibitem[\protect\citeauthoryear{{Leake} et~al.}{2013}]{leake13}
\begin{barticle}
\bauthor{\binits{J.E.} \bsnm{{Leake}}},
\bauthor{\binits{M.G.} \bsnm{{Linton}}},
\bauthor{\binits{T.} \bsnm{{T{\"o}r{\"o}k}}},
\batitle{{Simulations of Emerging Magnetic Flux. I. The Formation of Stable Coronal Flux Ropes}}.
\bjtitle{\apj}
\bvolume{778},
\bfpage{99}
(\byear{2013}).
doi:\doiurl{10.1088/0004-637X/778/2/99}
\end{barticle}
\endbibitem

\bibitem[\protect\citeauthoryear{{Lee} et~al.}{2013}]{2013ApJ...766....1L}
\begin{barticle}
\bauthor{\binits{K.-S.} \bsnm{{Lee}}},
\bauthor{\binits{D.E.} \bsnm{{Innes}}},
\bauthor{\binits{Y.-J.} \bsnm{{Moon}}}, \betal,
\batitle{{Fast Extreme-ultraviolet Dimming Associated with a Coronal Jet Seen
  in Multi-wavelength and Stereoscopic Observations}}.
\bjtitle{\apj}
\bvolume{766},
\bfpage{1}
(\byear{2013}).
doi:\doiurl{10.1088/0004-637X/766/1/1}
\end{barticle}
\endbibitem

\bibitem[\protect\citeauthoryear{{Lee} et~al.}{2015}]{2015ApJ...798L..10L}
\begin{barticle}
\bauthor{\binits{E.J.} \bsnm{{Lee}}},
\bauthor{\binits{V.} \bsnm{{Archontis}}},
\bauthor{\binits{A.W.} \bsnm{{Hood}}},
\batitle{{Helical Blowout Jets in the Sun: Untwisting and Propagation of Waves}}.
\bjtitle{\apjl}
\bvolume{798},
\bfpage{10}
(\byear{2015}).
doi:\doiurl{10.1088/2041-8205/798/1/L10}
\end{barticle}
\endbibitem

\bibitem[\protect\citeauthoryear{{Lemen} et~al.}{2012}]{2012SoPh..275...17L}
\begin{barticle}
\bauthor{\binits{J.R.} \bsnm{{Lemen}}},
\bauthor{\binits{A.M.} \bsnm{{Title}}},
\bauthor{\binits{D.J.} \bsnm{{Akin}}}, \betal,
\batitle{{The Atmospheric Imaging Assembly (AIA) on the Solar Dynamics
  Observatory (SDO)}}.
\bjtitle{\solphys}
\bvolume{275},
\bfpage{17}
(\byear{2012}).
doi:\doiurl{10.1007/s11207-011-9776-8}
\end{barticle}
\endbibitem

\bibitem[\protect\citeauthoryear{{Lin} et~al.}{2002}]{2002SoPh..210....3L}
\begin{barticle}
\bauthor{\binits{R.P.} \bsnm{{Lin}}},
\bauthor{\binits{B.R.} \bsnm{{Dennis}}},
\bauthor{\binits{G.J.} \bsnm{{Hurford}}}, \betal,
\batitle{{The Reuven Ramaty High-Energy Solar Spectroscopic Imager (RHESSI)}}.
\bjtitle{\solphys}
\bvolume{210},
\bfpage{3}
(\byear{2002}).
doi:\doiurl{10.1023/A:1022428818870}
\end{barticle}
\endbibitem

\bibitem[\protect\citeauthoryear{{Linton} et~al.}{2001}]{2001ApJ...553..905L}
\begin{barticle}
\bauthor{\binits{M.G.} \bsnm{{Linton}}},
\bauthor{\binits{R.B.} \bsnm{{Dahlburg}}},
\bauthor{\binits{S.K.} \bsnm{{Antiochos}}},
\batitle{{Reconnection of Twisted Flux Tubes as a Function of Contact Angle}}.
\bjtitle{\apj}
\bvolume{553},
\bfpage{905}
(\byear{2001}).
doi:\doiurl{10.1086/320974}
\end{barticle}
\endbibitem

\bibitem[\protect\citeauthoryear{Lionello et~al.}{2009}]{lionello09}
\begin{barticle}
\bauthor{\binits{R.} \bsnm{Lionello}},
\bauthor{\binits{J.A.} \bsnm{Linker}},
\bauthor{\binits{Z.} \bsnm{Mikic}},
\batitle{Multispectral emission of the sun during the first whole sun month:
  Magnetohydrodynamic simulations}.
\bjtitle{\apj}
\bvolume{690},
\bfpage{902}
(\byear{2009}).
doi:\doiurl{10.1088/0004-637X/690/1/902}
\end{barticle}
\endbibitem

\bibitem[\protect\citeauthoryear{Lionello et~al.}{2013}]{2013ApJ...777...76L}
\begin{barticle}
\bauthor{\binits{R.} \bsnm{Lionello}},
\bauthor{\binits{C.} \bsnm{Downs}},
\bauthor{\binits{J.~A.} \bsnm{Linker}}, \betal,
\batitle{Magnetohydrodynamic Simulations of Interplanetary Coronal Mass Ejections}.
\bjtitle{\apj}
\bvolume{777},
\bfpage{76}
(\byear{2013}).
doi:\doiurl{10.1088/0004-637X/777/1/76}
\end{barticle}
\endbibitem

\bibitem[\protect\citeauthoryear{{Liu} et~al.}{2004}]{2004ApJ...604..442L}
\begin{barticle}
\bauthor{\binits{C.} \bsnm{{Liu}}},
\bauthor{\binits{J.} \bsnm{{Qiu}}},
\bauthor{\binits{D.E.} \bsnm{{Gary}}}, \betal,
\batitle{{Studies of Microflares in RHESSI Hard X-Ray, Big Bear Solar
  Observatory H{$\alpha$}, and Michelson Doppler Imager Magnetograms}}.
\bjtitle{\apj}
\bvolume{604},
\bfpage{442}
(\byear{2004}).
doi:\doiurl{10.1086/381799}
\end{barticle}
\endbibitem

\bibitem[\protect\citeauthoryear{{Liu} \& {Kurokawa}}{2004}]{2004ApJ...610.1136L}
\begin{barticle}
\bauthor{\binits{Y.} \bsnm{{Liu}}},
\bauthor{\binits{H.} \bsnm{{Kurokawa}}},
\batitle{{On a Surge: Properties of an Emerging Flux Region}}.
\bjtitle{\apj}
\bvolume{610},
\bfpage{1136}
(\byear{2004}).
doi:\doiurl{10.1086/421715}
\end{barticle}
\endbibitem

\bibitem[\protect\citeauthoryear{{Liu} et~al.}{2009}]{2009ApJ...707L..37L}
\begin{barticle}
\bauthor{\binits{W.} \bsnm{{Liu}}},
\bauthor{\binits{T.E.} \bsnm{{Berger}}},
\bauthor{\binits{A.M.} \bsnm{{Title}}},
\bauthor{\binits{T.D.} \bsnm{{Tarbell}}},
\batitle{{An Intriguing Chromospheric Jet Observed by Hinode: Fine Structure
  Kinematics and Evidence of Unwinding Twists}}.
\bjtitle{\apjl}
\bvolume{707},
\bfpage{37}
(\byear{2009}).
doi:\doiurl{10.1088/0004-637X/707/1/L37}
\end{barticle}
\endbibitem

\bibitem[\protect\citeauthoryear{{Liu} et~al.}{2011a}]{2011ApJ...728..103L}
\begin{barticle}
\bauthor{\binits{W.} \bsnm{{Liu}}},
\bauthor{\binits{T.E.} \bsnm{{Berger}}},
\bauthor{\binits{A.M.} \bsnm{{Title}}}, \betal,
\batitle{{Chromospheric Jet and Growing ``Loop'' Observed by Hinode: New
  Evidence of Fan-spine Magnetic Topology Resulting from Flux Emergence}}.
\bjtitle{\apj}
\bvolume{728},
\bfpage{103}
(\byear{2011a}).
doi:\doiurl{10.1088/0004-637X/728/2/103}
\end{barticle}
\endbibitem

\bibitem[\protect\citeauthoryear{{Liu} et~al.}{2011b}]{2011ApJ...735L..18L}
\begin{barticle}
\bauthor{\binits{C.} \bsnm{{Liu}}},
\bauthor{\binits{N.} \bsnm{{Deng}}},
\bauthor{\binits{R.} \bsnm{{Liu}}}, \betal,
\batitle{{A Standard-to-blowout Jet}}.
\bjtitle{\apjl}
\bvolume{735},
\bfpage{18}
(\byear{2011b}).
doi:\doiurl{10.1088/2041-8205/735/1/L18}
\end{barticle}
\endbibitem

\bibitem[\protect\citeauthoryear{{Liu} et~al.}{2014}]{2014ApJ...782...94L}
\begin{barticle}
\bauthor{\binits{J.} \bsnm{{Liu}}},
\bauthor{\binits{Y.} \bsnm{{Wang}}},
\bauthor{\binits{R.} \bsnm{{Liu}}}, \betal,
\batitle{{When and how does a Prominence-like Jet Gain Kinetic Energy?}}
\bjtitle{\apj}
\bvolume{782},
\bfpage{94}
(\byear{2014}).
doi:\doiurl{10.1088/0004-637X/782/2/94}
\end{barticle}
\endbibitem

\bibitem[\protect\citeauthoryear{{Luhn} et~al.}{1987}]{1987ApJ...317..951L}
\begin{barticle}
\bauthor{\binits{A.} \bsnm{{Luhn}}},
\bauthor{\binits{B.} \bsnm{{Kleckler}}},
\bauthor{\binits{D.} \bsnm{{Hovestadt}}},
\bauthor{\binits{E.} \bsnm{{Moebius}}},
\batitle{{The mean ionic charge of silicon in He-3-rich solar flares}}.
\bjtitle{\apj}
\bvolume{317},
\bfpage{951}
(\byear{1987}).
doi:\doiurl{10.1088/0004-637X/782/2/94}
\end{barticle}
\endbibitem

%

\bibitem[\protect\citeauthoryear{{Madjarska}}{2011}]{madjarska11}
\begin{barticle}
\bauthor{\binits{M.S.} \bsnm{{Madjarska}}},
\batitle{{Dynamics and plasma properties of an X-ray jet from SUMER, EIS, XRT,
  and EUVI A {\&} B simultaneous observations}}.
\bjtitle{\aap}
\bvolume{526},
\bfpage{19}
(\byear{2011}).
doi:\doiurl{10.1051/0004-6361/201015269}
\end{barticle}
\endbibitem


\bibitem[\protect\citeauthoryear{{Manchester}
  et~al.}{2004}]{2004ApJ...610..588M}
\begin{barticle}
\bauthor{\binits{W.} \bsnm{{Manchester}} \bsuffix{IV}},
\bauthor{\binits{T.} \bsnm{{Gombosi}}},
\bauthor{\binits{D.} \bsnm{{DeZeeuw}}},
\bauthor{\binits{Y.} \bsnm{{Fan}}},
\batitle{{Eruption of a Buoyantly Emerging Magnetic Flux Rope}}.
\bjtitle{\apj}
\bvolume{610},
\bfpage{588}
(\byear{2004}).
doi:\doiurl{10.1086/421516}
\end{barticle}
\endbibitem

\bibitem[\protect\citeauthoryear{{Matsui} et~al.}{2012}]{matsui12}
\begin{barticle}
\bauthor{\binits{Y.} \bsnm{{Matsui}}},
\bauthor{\binits{T.} \bsnm{{Yokoyama}}},
\bauthor{\binits{N.} \bsnm{{Kitagawa}}},
\bauthor{\binits{S.} \bsnm{{Imada}}},
\batitle{{Multi-wavelength Spectroscopic Observation of Extreme-ultraviolet Jet
  in AR 10960}}.
\bjtitle{\apj}
\bvolume{759},
\bfpage{15}
(\byear{2012}).
doi:\doiurl{10.1088/0004-637X/759/1/15}
\end{barticle}
\endbibitem

\bibitem[\protect\citeauthoryear{{Meyer} et~al.}{2016}]{Meyer15}
\begin{botherref}
\oauthor{\binits{K.} \bsnm{{Meyer}}},
\oauthor{\binits{A.} \bsnm{{Savcheva}}},
\oauthor{\binits{E.E.} \bsnm{{DeLuca}}},
\oauthor{\binits{D.} \bsnm{{Mackay}}},
{Non-Linear Force-Free Field Modelling of Solar Coronal Jets in Theoretical Configurations}.
\bjtitle{\apj},
\textbf{{submitted}}
(\byear{2016}).
\end{botherref}
\endbibitem


\bibitem[\protect\citeauthoryear{{Michard}}{1974}]{1974IAUS...56....3M}
\begin{bchapter}
\bauthor{\binits{R.} \bsnm{{Michard}}},
\bctitle{{Spicules and Their Surroundings}},
in \bbtitle{Chromospheric Fine Structure},
ed. by \beditor{\binits{R.G.} \bsnm{{Athay}}}
\bsertitle{IAU Symp.},
vol. \bseriesno{56},
pp. \bfpage{3}--\blpage{22}
(\byear{1974}).
\end{bchapter}
\endbibitem

\bibitem[\protect\citeauthoryear{{Miyagoshi} \&
  {Yokoyama}}{2004}]{2004ApJ...614.1042M}
\begin{barticle}
\bauthor{\binits{T.} \bsnm{{Miyagoshi}}},
\bauthor{\binits{T.} \bsnm{{Yokoyama}}},
\batitle{{Magnetohydrodynamic Simulation of Solar Coronal Chromospheric
  Evaporation Jets Caused by Magnetic Reconnection Associated with Magnetic
  Flux Emergence}}.
\bjtitle{\apj}
\bvolume{614},
\bfpage{1042}
(\byear{2004}).
doi:\doiurl{10.1086/423731}
\end{barticle}
\endbibitem


\bibitem[\protect\citeauthoryear{{Moore} et~al.}{2010}]{2010ApJ...720..757M}
\begin{barticle}
\bauthor{\binits{R.L.} \bsnm{{Moore}}},
\bauthor{\binits{J.W.} \bsnm{{Cirtain}}},
\bauthor{\binits{A.C.} \bsnm{{Sterling}}},
\bauthor{\binits{D.A.} \bsnm{{Falconer}}},
\batitle{{Dichotomy of Solar Coronal Jets: Standard Jets and Blowout Jets}}.
\bjtitle{\apj}
\bvolume{720},
\bfpage{757}
(\byear{2010}).
doi:\doiurl{10.1088/0004-637X/720/1/757}
\end{barticle}
\endbibitem


\bibitem[\protect\citeauthoryear{{Moore} et~al.}{2013}]{2013ApJ...769..134M}
\begin{barticle}
\bauthor{\binits{R.L.} \bsnm{{Moore}}},
\bauthor{\binits{A.C.} \bsnm{{Sterling}}},
\bauthor{\binits{D.A.} \bsnm{{Falconer}}},
\bauthor{\binits{D.} \bsnm{{Robe}}},
\batitle{{The Cool Component and the Dichotomy, Lateral Expansion, and Axial
  Rotation of Solar X-Ray Jets}}.
\bjtitle{\apj}
\bvolume{769},
\bfpage{134}
(\byear{2013}).
doi:\doiurl{10.1088/0004-637X/769/2/134}
\end{barticle}
\endbibitem

\bibitem[\protect\citeauthoryear{{Moreno-Insertis}
  et~al.}{2008}]{2008ApJ...673L.211M}
\begin{barticle}
\bauthor{\binits{F.} \bsnm{{Moreno-Insertis}}},
\bauthor{\binits{K.} \bsnm{{Galsgaard}}},
\bauthor{\binits{I.} \bsnm{{Ugarte-Urra}}},
\batitle{{Jets in Coronal Holes: Hinode Observations and Three-dimensional
  Computer Modeling}}.
\bjtitle{\apjl}
\bvolume{673},
\bfpage{211}
(\byear{2008}).
doi:\doiurl{10.1086/527560}
\end{barticle}
\endbibitem

\bibitem[\protect\citeauthoryear{{Moreno-Insertis} \& {Galsgaard}}{2013}]{2013ApJ...771...20M}
\begin{barticle}
\bauthor{\binits{F.} \bsnm{{Moreno-Insertis}}},
\bauthor{\binits{K.} \bsnm{{Galsgaard}}},
\batitle{{Plasma Jets and Eruptions in Solar Coronal Holes: A Three-dimensional
  Flux Emergence Experiment}}.
\bjtitle{\apj}
\bvolume{771},
\bfpage{20}
(\byear{2013}).
doi:\doiurl{10.1088/0004-637X/771/1/20}
\end{barticle}
\endbibitem

\bibitem[\protect\citeauthoryear{{Morton} et~al.}{2012}]{2012A&A...542A..70M}
\begin{barticle}
\bauthor{\binits{R.J.} \bsnm{{Morton}}},
\bauthor{\binits{A.K.} \bsnm{{Srivastava}}},
\bauthor{\binits{R.} \bsnm{{Erd{\'e}lyi}}},
\batitle{{Observations of quasi-periodic phenomena associated with a large
  blowout solar jet}}.
\bjtitle{\aap}
\bvolume{542},
\bfpage{70}
(\byear{2012}).
doi:\doiurl{10.1051/0004-6361/201117218}
\end{barticle}
\endbibitem

\bibitem[\protect\citeauthoryear{{Moschou} et~al.}{2013}]{2013SoPh..284..427M}
\begin{barticle}
\bauthor{\binits{S.P.} \bsnm{{Moschou}}},
\bauthor{\binits{K.} \bsnm{{Tsinganos}}},
\bauthor{\binits{A.} \bsnm{{Vourlidas}}},
\bauthor{\binits{V.} \bsnm{{Archontis}}},
\batitle{{SDO Observations of Solar Jets}}.
\bjtitle{\solphys}
\bvolume{284},
\bfpage{427}
(\byear{2013}).
doi:\doiurl{10.1007/s11207-012-0190-7}
\end{barticle}
\endbibitem

\bibitem[\protect\citeauthoryear{{Neugebauer}
  et~al.}{1995}]{1995JGR...10023389N}
\begin{barticle}
\bauthor{\binits{M.} \bsnm{{Neugebauer}}},
\bauthor{\binits{B.E.} \bsnm{{Goldstein}}},
\bauthor{\binits{D.J.} \bsnm{{McComas}}}, \betal,
\batitle{{Ulysses observations of microstreams in the solar wind from coronal
  holes}}.
\bjtitle{\jgr}
\bvolume{100},
\bfpage{23389}
(\byear{1995}).
doi:\doiurl{10.1029/95JA02723}
\end{barticle}
\endbibitem

\bibitem[\protect\citeauthoryear{{Neugebauer}}{2012}]{2012ApJ...750...50N}
\begin{barticle}
\bauthor{\binits{M.} \bsnm{{Neugebauer}}},
\batitle{{Evidence for Polar X-Ray Jets as Sources of Microstream Peaks in the
  Solar Wind}}.
\bjtitle{\apj}
\bvolume{750},
\bfpage{50}
(\byear{2012}).
doi:\doiurl{10.1088/0004-637X/750/1/50}
\end{barticle}
\endbibitem

\bibitem[\protect\citeauthoryear{{Nishizuka}
  et~al.}{2008}]{2008ApJ...683L..83N}
\begin{barticle}
\bauthor{\binits{N.} \bsnm{{Nishizuka}}},
\bauthor{\binits{M.} \bsnm{{Shimizu}}},
\bauthor{\binits{T.} \bsnm{{Nakamura}}}, \betal,
\batitle{{Giant Chromospheric Anemone Jet Observed with Hinode and Comparison
  with Magnetohydrodynamic Simulations: Evidence of Propagating Alfv{\'e}n
  Waves and Magnetic Reconnection}}.
\bjtitle{\apjl}
\bvolume{683},
\bfpage{83}
(\byear{2008}).
doi:\doiurl{10.1086/591445}
\end{barticle}
\endbibitem

\bibitem[\protect\citeauthoryear{{Nistic{\`o}}
  et~al.}{2009}]{2009SoPh..259...87N}
\begin{barticle}
\bauthor{\binits{G.} \bsnm{{Nistic{\`o}}}},
\bauthor{\binits{V.} \bsnm{{Bothmer}}},
\bauthor{\binits{S.} \bsnm{{Patsourakos}}},
\bauthor{\binits{G.} \bsnm{{Zimbardo}}},
\batitle{{Characteristics of EUV Coronal Jets Observed with STEREO/SECCHI}}.
\bjtitle{\solphys}
\bvolume{259},
\bfpage{87}
(\byear{2009}).
doi:\doiurl{10.1007/s11207-009-9424-8}
\end{barticle}
\endbibitem

\bibitem[\protect\citeauthoryear{{Nistic{\`o}}
  et~al.}{2010}]{2010AnGeo..28..687N}
\begin{barticle}
\bauthor{\binits{G.} \bsnm{{Nistic{\`o}}}},
\bauthor{\binits{V.} \bsnm{{Bothmer}}},
\bauthor{\binits{S.} \bsnm{{Patsourakos}}},
\bauthor{\binits{G.} \bsnm{{Zimbardo}}},
\batitle{{Observational features of equatorial coronal hole jets}}.
\bjtitle{Ann. Geophys.}
\bvolume{28},
\bfpage{687}
(\byear{2010}).
doi:\doiurl{10.5194/angeo-28-687-2010}
\end{barticle}
\endbibitem

\bibitem[\protect\citeauthoryear{{Nistic{\`o}}
  et~al.}{2011}]{2011AdSpR..48.1490N}
\begin{barticle}
\bauthor{\binits{G.} \bsnm{{Nistic{\`o}}}},
\bauthor{\binits{S.} \bsnm{{Patsourakos}}},
\bauthor{\binits{V.} \bsnm{{Bothmer}}},
\bauthor{\binits{G.} \bsnm{{Zimbardo}}},
\batitle{{Determination of temperature maps of EUV coronal hole jets}}.
\bjtitle{Adv. Space Res.}
\bvolume{48},
\bfpage{1490}
(\byear{2011}).
doi:\doiurl{10.1016/j.asr.2011.07.003}
\end{barticle}
\endbibitem

\bibitem[\protect\citeauthoryear{{Nistic\`o}
  et~al.}{2015}]{2015arXiv150801072N}
\begin{barticle}
\oauthor{\binits{G.} \bsnm{{Nistic\`o}}},
\oauthor{\binits{G.} \bsnm{{Zimbardo}}},
\oauthor{\binits{S.} \bsnm{{Patsourakos}}}, \betal,
\batitle{{North-south asymmetry in the magnetic deflection of polar coronal hole jets}}.
\bjtitle{\apj}
\bvolume{583},
\bfpage{A127}
(\byear{2015}).
doi:\doiurl{10.1051/0004-6361/201525731}
\end{barticle}
\endbibitem

\bibitem[\protect\citeauthoryear{{Nitta} et~al.}{2006}]{2006ApJ...650..438N}
\begin{barticle}
\bauthor{\binits{N.V.} \bsnm{{Nitta}}},
\bauthor{\binits{D.V.} \bsnm{{Reames}}},
\bauthor{\binits{M.L.} \bsnm{{De Rosa}}}, \betal,
\batitle{{Solar Sources of Impulsive Solar Energetic Particle Events and Their
  Magnetic Field Connection to the Earth}}.
\bjtitle{\apj}
\bvolume{650},
\bfpage{438}
(\byear{2006}).
doi:\doiurl{10.1086/507442}
\end{barticle}
\endbibitem

\bibitem[\protect\citeauthoryear{{Nitta} et~al.}{2008}]{2008ApJ...675L.125N}
\begin{barticle}
\bauthor{\binits{N.V.} \bsnm{{Nitta}}},
\bauthor{\binits{G.M.} \bsnm{{Mason}}},
\bauthor{\binits{M.E.} \bsnm{{Wiedenbeck}}}, \betal,
\batitle{{Coronal Jet Observed by Hinode as the Source of a $^{3}$He-rich Solar
  Energetic Particle Event}}.
\bjtitle{\apjl}
\bvolume{675},
\bfpage{125}
(\byear{2008}).
doi:\doiurl{10.1086/533438}
\end{barticle}
\endbibitem

\bibitem[\protect\citeauthoryear{{Ogawara} et~al.}{1991}]{1991SoPh..136....1O}
\begin{barticle}
\bauthor{\binits{Y.} \bsnm{{Ogawara}}},
\bauthor{\binits{T.} \bsnm{{Takano}}},
\bauthor{\binits{T.} \bsnm{{Kato}}}, \betal,
\batitle{{The SOLAR-A Mission - An Overview}}.
\bjtitle{\solphys}
\bvolume{136},
\bfpage{1}
(\byear{1991}).
doi:\doiurl{10.1007/BF00151692}
\end{barticle}
\endbibitem

\bibitem[\protect\citeauthoryear{{\"Ohman} et~al.}{1968}]{Ohman1968}
\begin{barticle}
\bauthor{\binits{Y.} \bsnm{{\"Ohman}}},
\bauthor{\binits{G.} \bsnm{{Hosinsky}}},
\bauthor{\binits{U.} \bsnm{{Kusoffsky}}},
\batitle{{in Mass Motions in Solar Flares and Related Phenomena}}.
\bjtitle{Nobel Symp.}
\bvolume{9},
\bfpage{95}
(\byear{1968}).
\end{barticle}
\endbibitem

\bibitem[\protect\citeauthoryear{{Paraschiv}
  et~al.}{2010}]{2010SoPh..264..365P}
\begin{barticle}
\bauthor{\binits{A.R.} \bsnm{{Paraschiv}}},
\bauthor{\binits{D.A.} \bsnm{{Lacatus}}},
\bauthor{\binits{T.} \bsnm{{Badescu}}}, \betal,
\batitle{{Study of Coronal Jets During Solar Minimum Based on STEREO/SECCHI
  Observations}}.
\bjtitle{\solphys}
\bvolume{264},
\bfpage{365}
(\byear{2010}).
doi:\doiurl{10.1007/s11207-010-9584-6}
\end{barticle}
\endbibitem


\bibitem[\protect\citeauthoryear{{Pariat} et~al.}{2009}]{2009ApJ...691...61P}
\begin{barticle}
\bauthor{\binits{E.} \bsnm{{Pariat}}},
\bauthor{\binits{S.K.} \bsnm{{Antiochos}}},
\bauthor{\binits{C.R.} \bsnm{{DeVore}}},
\batitle{{A Model for Solar Polar Jets}}.
\bjtitle{\apj}
\bvolume{691},
\bfpage{61}
(\byear{2009}).
doi:\doiurl{10.1088/0004-637X/691/1/61}
\end{barticle}
\endbibitem

\bibitem[\protect\citeauthoryear{{Pariat} et~al.}{2010}]{2010ApJ...714.1762P}
\begin{barticle}
\bauthor{\binits{E.} \bsnm{{Pariat}}},
\bauthor{\binits{S.K.} \bsnm{{Antiochos}}},
\bauthor{\binits{C.R.} \bsnm{{DeVore}}},
\batitle{{Three-dimensional Modeling of Quasi-homologous Solar Jets}}.
\bjtitle{\apj}
\bvolume{714},
\bfpage{1762}
(\byear{2010}).
doi:\doiurl{10.1088/0004-637X/714/2/1762}
\end{barticle}
\endbibitem

\bibitem[\protect\citeauthoryear{{Pariat} et~al.}{2015}]{2015A&A...573A.130P}
\begin{barticle}
\bauthor{\binits{E.} \bsnm{{Pariat}}},
\bauthor{\binits{K.} \bsnm{{Dalmasse}}},
\bauthor{\binits{C.R.} \bsnm{{DeVore}}}, \betal,
\batitle{{Model for straight and helical solar jets. I. Parametric studies of
  the magnetic field geometry}}.
\bjtitle{\aap}
\bvolume{573},
\bfpage{130}
(\byear{2015}).
doi:\doiurl{10.1051/0004-6361/201424209}
\end{barticle}
\endbibitem

%
%

\bibitem[\protect\citeauthoryear{{Patsourakos}
  et~al.}{2008}]{2008ApJ...680L..73P}
\begin{barticle}
\bauthor{\binits{S.} \bsnm{{Patsourakos}}},
\bauthor{\binits{E.} \bsnm{{Pariat}}},
\bauthor{\binits{A.} \bsnm{{Vourlidas}}}, \betal,
\batitle{{STEREO SECCHI Stereoscopic Observations Constraining the Initiation
  of Polar Coronal Jets}}.
\bjtitle{\apjl}
\bvolume{680},
\bfpage{73}
(\byear{2008}).
doi:\doiurl{10.1086/589769}
\end{barticle}
\endbibitem

%

\bibitem[\protect\citeauthoryear{{Pesnell} et~al.}{2012}]{2012SoPh..275....3P}
\begin{barticle}
\bauthor{\binits{W.D.} \bsnm{{Pesnell}}},
\bauthor{\binits{B.J.} \bsnm{{Thompson}}},
\bauthor{\binits{P.C.} \bsnm{{Chamberlin}}},
\batitle{{The Solar Dynamics Observatory (SDO)}}.
\bjtitle{\solphys}
\bvolume{275},
\bfpage{3}
(\byear{2012}).
doi:\doiurl{10.1007/s11207-011-9841-3}
\end{barticle}
\endbibitem

\bibitem[\protect\citeauthoryear{{Pick} et~al.}{2006}]{2006ApJ...648.1247P}
\begin{barticle}
\bauthor{\binits{M.} \bsnm{{Pick}}},
\bauthor{\binits{G.M.} \bsnm{{Mason}}},
\bauthor{\binits{Y.-M.} \bsnm{{Wang}}}, \betal,
\batitle{{Solar Source Regions for $^{3}$He-rich Solar Energetic Particle
  Events Identified Using Imaging Radio, Optical, and Energetic Particle
  Observations}}.
\bjtitle{\apj}
\bvolume{648},
\bfpage{1247}
(\byear{2006}).
doi:\doiurl{10.1086/505926}
\end{barticle}
\endbibitem

\bibitem[\protect\citeauthoryear{{Pike} \& {Harrison}}{1997}]{pike97}
\begin{barticle}
\bauthor{\binits{C.D.} \bsnm{{Pike}}},
\bauthor{\binits{R.A.} \bsnm{{Harrison}}},
\batitle{{EUV Observations of a Macrospicule: Evidence for Solar Wind
  Acceleration?}}
\bjtitle{\solphys}
\bvolume{175},
\bfpage{457}
(\byear{1997}).
doi:\doiurl{10.1023/A:1004987505422}
\end{barticle}
\endbibitem

\bibitem[\protect\citeauthoryear{{Pike} \& {Mason}}{1998}]{pike98}
\begin{barticle}
\bauthor{\binits{C.D.} \bsnm{{Pike}}},
\bauthor{\binits{H.E.} \bsnm{{Mason}}},
\batitle{{Rotating Transition Region Features Observed with the SOHO Coronal
  Diagnostic Spectrometer}}.
\bjtitle{\solphys}
\bvolume{182},
\bfpage{333}
(\byear{1998}).
doi:\doiurl{10.1023/A:1005065704108}
\end{barticle}
\endbibitem

%

\bibitem[\protect\citeauthoryear{{Popescu} et~al.}{2007}]{popescu07}
\begin{barticle}
\bauthor{\binits{M.D.} \bsnm{{Popescu}}},
\bauthor{\binits{L.D.} \bsnm{{Xia}}},
\bauthor{\binits{D.} \bsnm{{Banerjee}}},
\bauthor{\binits{J.G.} \bsnm{{Doyle}}},
\batitle{{A study of a macro-spicule and a transition region explosive event in
  a solar coronal hole}}.
\bjtitle{Adv. Space Res.}
\bvolume{40},
\bfpage{1021}
(\byear{2007}).
doi:\doiurl{10.1016/j.asr.2007.06.068}
\end{barticle}
\endbibitem

\bibitem[\protect\citeauthoryear{{Powell} et~al.}{1999}]{1999JCoPh.154..284P}
\begin{barticle}
\bauthor{\binits{K.G.} \bsnm{{Powell}}},
\bauthor{\binits{P.L.} \bsnm{{Roe}}},
\bauthor{\binits{T.J.} \bsnm{{Linde}}}, \betal,
\batitle{{A Solution-Adaptive Upwind Scheme for Ideal Magnetohydrodynamics}}.
\bjtitle{J. Comput. Phys.}
\bvolume{154},
\bfpage{284}
(\byear{1999}).
doi:\doiurl{10.1006/jcph.1999.6299}
\end{barticle}
\endbibitem

\bibitem[\protect\citeauthoryear{{Pucci} et~al.}{2013}]{2013ApJ...776...16P}
\begin{barticle}
\bauthor{\binits{S.} \bsnm{{Pucci}}},
\bauthor{\binits{G.} \bsnm{{Poletto}}},
\bauthor{\binits{A.C.} \bsnm{{Sterling}}},
\bauthor{\binits{M.} \bsnm{{Romoli}}},
\batitle{{Physical Parameters of Standard and Blowout Jets}}.
\bjtitle{\apj}
\bvolume{776},
\bfpage{16}
(\byear{2013}).
doi:\doiurl{10.1088/0004-637X/776/1/16}
\end{barticle}
\endbibitem


\bibitem[\protect\citeauthoryear{{Rachmeler}
  et~al.}{2010}]{2010ApJ...715.1556R}
\begin{barticle}
\bauthor{\binits{L.A.} \bsnm{{Rachmeler}}},
\bauthor{\binits{E.} \bsnm{{Pariat}}},
\bauthor{\binits{C.E.} \bsnm{{DeForest}}}, \betal,
\batitle{{Symmetric Coronal Jets: A Reconnection-controlled Study}}.
\bjtitle{\apj}
\bvolume{715},
\bfpage{1556}
(\byear{2010}).
doi:\doiurl{10.1088/0004-637X/715/2/1556}
\end{barticle}
\endbibitem

\bibitem[\protect\citeauthoryear{{Raouafi} et~al.}{2008}]{2008ApJ...682L.137R}
\begin{barticle}
\bauthor{\binits{N.-E.} \bsnm{{Raouafi}}},
\bauthor{\binits{G.J.D.} \bsnm{{Petrie}}},
\bauthor{\binits{A.A.} \bsnm{{Norton}}}, \betal,
\batitle{{Evidence for Polar Jets as Precursors of Polar Plume Formation}}.
\bjtitle{\apjl}
\bvolume{682},
\bfpage{137}
(\byear{2008}).
doi:\doiurl{10.1086/591125}
\end{barticle}
\endbibitem

\bibitem[\protect\citeauthoryear{{Raouafi}}{2009}]{2009ASPC..415..144R}
\begin{bchapter}
\bauthor{\binits{N.-E.} \bsnm{{Raouafi}}},
\bctitle{{On the Relationship between Polar Coronal Jets and Plumes}},
\bsertitle{ASP Conf. Ser.},
vol. \bseriesno{415},
p. \bfpage{144}
(\byear{2009}).
\end{bchapter}
\endbibitem

\bibitem[\protect\citeauthoryear{{Raouafi} et~al.}{2010}]{2010ApJ...718..981R}
\begin{barticle}
\bauthor{\binits{N.-E.} \bsnm{{Raouafi}}},
\bauthor{\binits{M.K.} \bsnm{{Georgoulis}}},
\bauthor{\binits{D.M.} \bsnm{{Rust}}},
\bauthor{\binits{P.N.} \bsnm{{Bernasconi}}},
\batitle{{Micro-sigmoids as Progenitors of Coronal Jets: Is Eruptive Activity
  Self-similarly Multi-scaled?}}
\bjtitle{\apj}
\bvolume{718},
\bfpage{981}
(\byear{2010}).
doi:\doiurl{10.1088/0004-637X/718/2/981}
\end{barticle}
\endbibitem

\bibitem[\protect\citeauthoryear{{Raouafi} \&
  {Stenborg}}{2014}]{2014ApJ...787..118R}
\begin{barticle}
\bauthor{\binits{N.-E.} \bsnm{{Raouafi}}},
\bauthor{\binits{G.} \bsnm{{Stenborg}}},
\batitle{{Role of Transients in the Sustainability of Solar Coronal Plumes}}.
\bjtitle{\apj}
\bvolume{787},
\bfpage{118}
(\byear{2014}).
doi:\doiurl{10.1088/0004-637X/787/2/118}
\end{barticle}
\endbibitem

\bibitem[\protect\citeauthoryear{{Raulin} et~al.}{1996}]{1996A&A...306..299R}
\begin{barticle}
\bauthor{\binits{J.P.} \bsnm{{Raulin}}},
\bauthor{\binits{M.R.} \bsnm{{Kundu}}},
\bauthor{\binits{H.S.} \bsnm{{Hudson}}}, \betal,
\batitle{{Metric Type III bursts associated with soft X-ray jets.}}
\bjtitle{\aap}
\bvolume{306},
\bfpage{299}
(\byear{1996}).
\end{barticle}
\endbibitem

%

\bibitem[\protect\citeauthoryear{{Reames} et~al.}{1988}]{1988ApJ...327..998R}
\begin{barticle}
\bauthor{\binits{D.V.} \bsnm{{Reames}}},
\bauthor{\binits{B.R.} \bsnm{{Dennis}}},
\bauthor{\binits{R.G.} \bsnm{{Stone}}},
\bauthor{\binits{R.P.} \bsnm{{Lin}}},
\batitle{{X-ray and radio properties of solar (He-3) rich events}}.
\bjtitle{\apj}
\bvolume{327},
\bfpage{998}
(\byear{1988}).
doi:\doiurl{10.1086/166257}
\end{barticle}
\endbibitem

\bibitem[\protect\citeauthoryear{{Reames}}{1999}]{1999SSRv...90..413R}
\begin{barticle}
\bauthor{\binits{D.V.} \bsnm{{Reames}}},
\batitle{{Particle acceleration at the Sun and in the heliosphere}}.
\bjtitle{\ssr}
\bvolume{90},
\bfpage{413}
(\byear{1999}).
doi:\doiurl{10.1023/A:1005105831781}
\end{barticle}
\endbibitem


\bibitem[\protect\citeauthoryear{{Reames} \& {Ng}}{2004}]{2004ApJ...610..510R}
\begin{barticle}
\bauthor{\binits{D.V.} \bsnm{{Reames}}},
\bauthor{\binits{C.K.} \bsnm{{Ng}}},
\batitle{{Heavy-Element Abundances in Solar Energetic Particle Events}}.
\bjtitle{\apj}
\bvolume{610},
\bfpage{510}
(\byear{2004}).
doi:\doiurl{10.1086/421518}
\end{barticle}
\endbibitem


\bibitem[\protect\citeauthoryear{{Roy} \& {Tang}}{1975}]{1975SoPh...42..425R}
\begin{barticle}
\bauthor{\binits{J.-R.} \bsnm{{Roy}}},
\bauthor{\binits{F.} \bsnm{{Tang}}},
\batitle{{Slow X-ray bursts and flares with filament disruption}}.
\bjtitle{\solphys}
\bvolume{42},
\bfpage{425}
(\byear{1975}).
doi:\doiurl{10.1007/BF00149923}
\end{barticle}
\endbibitem

\bibitem[\protect\citeauthoryear{{Rust} \& {Kumar}}{1996}]{1996ApJ...464L.199R}
\begin{barticle}
\bauthor{\binits{D.M.} \bsnm{{Rust}}},
\bauthor{\binits{A.} \bsnm{{Kumar}}},
\batitle{{Evidence for Helically Kinked Magnetic Flux Ropes in Solar
  Eruptions}}.
\bjtitle{\apjl}
\bvolume{464},
\bfpage{199}
(\byear{1996}).
doi:\doiurl{10.1086/310118}
\end{barticle}
\endbibitem


\bibitem[\protect\citeauthoryear{{Saint-Hilaire}
  et~al.}{2009}]{2009ApJ...696..941S}
\begin{barticle}
\bauthor{\binits{P.} \bsnm{{Saint-Hilaire}}},
\bauthor{\binits{S.} \bsnm{{Krucker}}},
\bauthor{\binits{S.} \bsnm{{Christe}}},
\bauthor{\binits{R.P.} \bsnm{{Lin}}},
\batitle{{The X-ray Detectability of Electron Beams Escaping from the Sun}}.
\bjtitle{\apj}
\bvolume{696},
\bfpage{941}
(\byear{2009}).
doi:\doiurl{10.1088/0004-637X/696/1/941}
\end{barticle}
\endbibitem

\bibitem[\protect\citeauthoryear{{Sako} et~al.}{2013}]{2013ApJ...775...22S}
\begin{barticle}
\bauthor{\binits{N.} \bsnm{{Sako}}},
\bauthor{\binits{M.} \bsnm{{Shimojo}}},
\bauthor{\binits{T.} \bsnm{{Watanabe}}},
\bauthor{\binits{T.} \bsnm{{Sekii}}},
\batitle{{A Statistical Study of Coronal Active Events in the North Polar
  Region}}.
\bjtitle{\apj}
\bvolume{775},
\bfpage{22}
(\byear{2013}).
doi:\doiurl{10.1088/0004-637X/775/1/22}
\end{barticle}
\endbibitem

\bibitem[\protect\citeauthoryear{Sako}{2014}]{Sako2014}
\begin{botherref}
\oauthor{\binits{N.} \bsnm{Sako}},
Statistical Study of X-ray Jets using Hinode/XRT,
PhD thesis,
The Graduate University for Advanced Studies, Mitaka, Tokyo, Japan
(\byear{2014}).
\end{botherref}
\endbibitem

\bibitem[\protect\citeauthoryear{{Savcheva} et~al.}{2007}]{2007PASJ...59S.771S}
\begin{barticle}
\bauthor{\binits{A.} \bsnm{{Savcheva}}},
\bauthor{\binits{J.} \bsnm{{Cirtain}}},
\bauthor{\binits{E.E.} \bsnm{{Deluca}}}, \betal,
\batitle{{A Study of Polar Jet Parameters Based on Hinode XRT Observations}}.
\bjtitle{\pasj}
\bvolume{59},
\bfpage{771}
(\byear{2007}).
doi:\doiurl{10.1093/pasj/59.sp3.S771}
\end{barticle}
\endbibitem

\bibitem[\protect\citeauthoryear{{Savcheva} et~al.}{2009}]{Savcheva09}
\begin{barticle}
\bauthor{\binits{A.} \bsnm{{Savcheva}}},
\bauthor{\binits{J.W.} \bsnm{{Cirtain}}},
\bauthor{\binits{E.E.} \bsnm{{DeLuca}}},
\bauthor{\binits{L.} \bsnm{{Golub}}},
\batitle{{Does a Polar Coronal Hole's Flux Emergence Follow a Hale-Like Law?}}
\bjtitle{\apjl}
\bvolume{702},
\bfpage{32}
(\byear{2009}).
doi:\doiurl{10.1088/0004-637X/702/1/L32}
\end{barticle}
\endbibitem

\bibitem[\protect\citeauthoryear{{Savcheva} et~al.}{2012}]{2012ApJ...750...15S}
\begin{barticle}
\bauthor{\binits{A.} \bsnm{{Savcheva}}},
\bauthor{\binits{E.} \bsnm{{Pariat}}},
\bauthor{\binits{A.} \bsnm{{van Ballegooijen}}}, \betal,
\batitle{{Sigmoidal Active Region on the Sun: Comparison of a
  Magnetohydrodynamical Simulation and a Nonlinear Force-free Field Model}}.
\bjtitle{\apj}
\bvolume{750},
\bfpage{15}
(\byear{2012}).
doi:\doiurl{10.1088/0004-637X/750/1/15}
\end{barticle}
\endbibitem

\bibitem[\protect\citeauthoryear{{Savcheva} et~al.}{2016}]{Savcheva15}
\begin{botherref}
\oauthor{\binits{A.} \bsnm{{Savcheva}}},
\oauthor{\binits{K.} \bsnm{{Meyer}}},
\oauthor{\binits{H.} \bsnm{{Tian}}},  \betal,
{Interpreting IRIS Jet Observations with a Magnetofrictional Simulation}.
\bjtitle{\apj},
\textbf{in preparation}
(\byear{2016}).
\end{botherref}
\endbibitem

\bibitem[\protect\citeauthoryear{{Scherrer} et~al.}{1995}]{1995SoPh..162..129S}
\begin{barticle}
\bauthor{\binits{P.H.} \bsnm{{Scherrer}}},
\bauthor{\binits{R.S.} \bsnm{{Bogart}}},
\bauthor{\binits{R.I.} \bsnm{{Bush}}}, \betal,
\batitle{{The Solar Oscillations Investigation - Michelson Doppler Imager}}.
\bjtitle{\solphys}
\bvolume{162},
\bfpage{129}
(\byear{1995}).
doi:\doiurl{10.1007/BF00733429}
\end{barticle}
\endbibitem

\bibitem[\protect\citeauthoryear{{Scherrer} et~al.}{2012}]{2012SoPh..275..207S}
\begin{barticle}
\bauthor{\binits{P.H.} \bsnm{{Scherrer}}},
\bauthor{\binits{J.} \bsnm{{Schou}}},
\bauthor{\binits{R.I.} \bsnm{{Bush}}}, \betal,
\batitle{{The Helioseismic and Magnetic Imager (HMI) Investigation for the
  Solar Dynamics Observatory (SDO)}}.
\bjtitle{\solphys}
\bvolume{275},
\bfpage{207}
(\byear{2012}).
doi:\doiurl{10.1007/s11207-011-9834-2}
\end{barticle}
\endbibitem

\bibitem[\protect\citeauthoryear{{Schmieder}
  et~al.}{2013}]{2013A&A...559A...1S}
\begin{barticle}
\bauthor{\binits{B.} \bsnm{{Schmieder}}},
\bauthor{\binits{Y.} \bsnm{{Guo}}},
\bauthor{\binits{F.} \bsnm{{Moreno-Insertis}}}, \betal,
\batitle{{Twisting solar coronal jet launched at the boundary of an active
  region}}.
\bjtitle{\aap}
\bvolume{559},
\bfpage{1}
(\byear{2013}).
doi:\doiurl{10.1051/0004-6361/201322181}
\end{barticle}
\endbibitem


\bibitem[\protect\citeauthoryear{{Shen} et~al.}{2011}]{2011ApJ...735L..43S}
\begin{barticle}
\bauthor{\binits{Y.} \bsnm{{Shen}}},
\bauthor{\binits{Y.} \bsnm{{Liu}}},
\bauthor{\binits{J.} \bsnm{{Su}}},
\bauthor{\binits{A.} \bsnm{{Ibrahim}}},
\batitle{{Kinematics and Fine Structure of an Unwinding Polar Jet Observed by
  the Solar Dynamic Observatory/Atmospheric Imaging Assembly}}.
\bjtitle{\apjl}
\bvolume{735},
\bfpage{43}
(\byear{2011}).
doi:\doiurl{10.1088/2041-8205/735/2/L43}
\end{barticle}
\endbibitem

\bibitem[\protect\citeauthoryear{{Shen} et~al.}{2012}]{2012ApJ...745..164S}
\begin{barticle}
\bauthor{\binits{Y.} \bsnm{{Shen}}},
\bauthor{\binits{Y.} \bsnm{{Liu}}},
\bauthor{\binits{J.} \bsnm{{Su}}},
\bauthor{\binits{Y.} \bsnm{{Deng}}},
\batitle{{On a Coronal Blowout Jet: The First Observation of a Simultaneously
  Produced Bubble-like CME and a Jet-like CME in a Solar Event}}.
\bjtitle{\apj}
\bvolume{745},
\bfpage{164}
(\byear{2012}).
doi:\doiurl{10.1088/0004-637X/745/2/164}
\end{barticle}
\endbibitem

\bibitem[\protect\citeauthoryear{{Shibata} \& {Uchida}}{1986}]{1986SoPh..103..299S}
\begin{barticle}
\bauthor{\binits{K.} \bsnm{{Shibata}}},
\bauthor{\binits{Y.} \bsnm{{Uchida}}},
\batitle{{Sweeping-magnetic-twist mechanism for the acceleration of jets in the
  solar atmosphere}}.
\bjtitle{\solphys}
\bvolume{103},
\bfpage{299}
(\byear{1986}).
doi:\doiurl{10.1007/BF00147831}
\end{barticle}
\endbibitem

\bibitem[\protect\citeauthoryear{{Shibata} et~al.}{1992}]{1992PASJ...44L.173S}
\begin{barticle}
\bauthor{\binits{K.} \bsnm{{Shibata}}},
\bauthor{\binits{Y.} \bsnm{{Ishido}}},
\bauthor{\binits{L.W.} \bsnm{{Acton}}}, \betal,
\batitle{{Observations of X-ray jets with the YOHKOH Soft X-ray Telescope}}.
\bjtitle{\pasj}
\bvolume{44},
\bfpage{173}
(\byear{1992}).
\end{barticle}
\endbibitem

\bibitem[\protect\citeauthoryear{{Shibata} et~al.}{1994}]{1994ApJ...431L..51S}
\begin{barticle}
\bauthor{\binits{K.} \bsnm{{Shibata}}},
\bauthor{\binits{N.} \bsnm{{Nitta}}},
\bauthor{\binits{K.T.} \bsnm{{Strong}}}, \betal,
\batitle{{A gigantic coronal jet ejected from a compact active region in a coronal hole}}.
\bjtitle{\apjl}
\bvolume{431},
\bfpage{51}
(\byear{1994}).
doi:\doiurl{10.1086/187470}
\end{barticle}
\endbibitem

\bibitem[\protect\citeauthoryear{{Shibata}}{2001}]{shibata01}
\begin{botherref}
\oauthor{\binits{K.} \bsnm{{Shibata}}},
\batitle{{Solar X-ray Jets}},
\bsertitle{Encyclopedia of Astronomy and Astrophysics}
(\byear{2000}).
\end{botherref}
\endbibitem

\bibitem[\protect\citeauthoryear{{Shimojo} et~al.}{1996}]{1996PASJ...48..123S}
\begin{barticle}
\bauthor{\binits{M.} \bsnm{{Shimojo}}},
\bauthor{\binits{S.} \bsnm{{Hashimoto}}},
\bauthor{\binits{K.} \bsnm{{Shibata}}}, \betal,
\batitle{{Statistical Study of Solar X-Ray Jets Observed with the YOHKOH Soft X-Ray Telescope}}.
\bjtitle{\pasj}
\bvolume{48},
\bfpage{123}
(\byear{1996}).
doi:\doiurl{10.1093/pasj/48.1.123}
\end{barticle}
\endbibitem

\bibitem[\protect\citeauthoryear{{Shimojo} et~al.}{1998}]{1998SoPh..178..379S}
\begin{barticle}
\bauthor{\binits{M.} \bsnm{{Shimojo}}},
\bauthor{\binits{K.} \bsnm{{Shibata}}},
\bauthor{\binits{K.L.} \bsnm{{Harvey}}},
\batitle{{Magnetic Field Properties of Solar X-Ray Jets}}.
\bjtitle{\solphys}
\bvolume{178},
\bfpage{379}
(\byear{1998}).
doi:\doiurl{10.1023/A:1005091905214}
\end{barticle}
\endbibitem

\bibitem[\protect\citeauthoryear{{Shimojo} \& {Shibata}}{2000}]{2000ApJ...542.1100S}
\begin{barticle}
\bauthor{\binits{M.} \bsnm{{Shimojo}}},
\bauthor{\binits{K.} \bsnm{{Shibata}}},
\batitle{{Physical Parameters of Solar X-Ray Jets}}.
\bjtitle{\apj}
\bvolume{542},
\bfpage{1100}
(\byear{2000}).
doi:\doiurl{10.1086/317024}
\end{barticle}
\endbibitem

\bibitem[\protect\citeauthoryear{{Shimojo} et~al.}{2001}]{2001ApJ...550.1051S}
\begin{barticle}
\bauthor{\binits{M.} \bsnm{{Shimojo}}},
\bauthor{\binits{K.} \bsnm{{Shibata}}},
\bauthor{\binits{T.} \bsnm{{Yokoyama}}},
\bauthor{\binits{K.} \bsnm{{Hori}}},
\batitle{{One-dimensional and Pseudo-Two-dimensional Hydrodynamic Simulations
  of Solar X-Ray Jets}}.
\bjtitle{\apj}
\bvolume{550},
\bfpage{1051}
(\byear{2001}).
doi:\doiurl{10.1086/319788}
\end{barticle}
\endbibitem

\bibitem[\protect\citeauthoryear{{Shimojo} et~al.}{2007}]{2007PASJ...59S.745S}
\begin{barticle}
\bauthor{\binits{M.} \bsnm{{Shimojo}}},
\bauthor{\binits{N.} \bsnm{{Narukage}}},
\bauthor{\binits{R.} \bsnm{{Kano}}}, \betal,
\batitle{{Fine Structures of Solar X-Ray Jets Observed with the X-Ray Telescope
  aboard Hinode}}.
\bjtitle{\pasj}
\bvolume{59},
\bfpage{745}
(\byear{2007}).
doi:\doiurl{10.1093/pasj/59.sp3.S745}
\end{barticle}
\endbibitem

\bibitem[\protect\citeauthoryear{{Srivastava} \& {Murawski}}{2011}]{2011A&A...534A..62S}
\begin{barticle}
\bauthor{\binits{A.K.} \bsnm{{Srivastava}}},
\bauthor{\binits{K.} \bsnm{{Murawski}}},
\batitle{{Observations of a pulse-driven cool polar jet by SDO/AIA}}.
\bjtitle{\aap}
\bvolume{534},
\bfpage{62}
(\byear{2011}).
doi:\doiurl{10.1051/0004-6361/201117359}
\end{barticle}
\endbibitem

\bibitem[\protect\citeauthoryear{{Sterling}}{2000}]{2000SoPh..196...79S}
\begin{barticle}
\bauthor{\binits{A.C.} \bsnm{{Sterling}}},
\batitle{{Solar Spicules: A Review of Recent Models and Targets for Future Observations - (Invited Review)}}.
\bjtitle{\solphys}
\bvolume{196},
\bfpage{79}
(\byear{2000}).
doi:\doiurl{10.1023/A:1005213923962}
\end{barticle}
\endbibitem

\bibitem[\protect\citeauthoryear{{Sterling} \& {Moore}}{2005}]{2005ApJ...630.1148S}
\begin{barticle}
\bauthor{\binits{A.C.} \bsnm{{Sterling}}},
\bauthor{\binits{R.L.} \bsnm{{Moore}}},
\batitle{{Slow-Rise and Fast-Rise Phases of an Erupting Solar Filament, and Flare Emission Onset}}.
\bjtitle{\apj}
\bvolume{630},
\bfpage{1148}
(\byear{2005}).
doi:\doiurl{10.1086/432044}
\end{barticle}
\endbibitem


\bibitem[\protect\citeauthoryear{{Sterling} et~al.}{2015}]{sterling.et15}
\begin{barticle}
\bauthor{\binits{A.C.} \bsnm{{Sterling}}},
\bauthor{\binits{R.L.} \bsnm{{Moore}}},
\bauthor{\binits{D.A.} \bsnm{{Falconer}}},
\bauthor{\binits{M.} \bsnm{{Adams}}},
\batitle{{Small-scale filament eruptions as the driver of X-ray jets in solar coronal holes}}.
\bjtitle{Nature}
\bvolume{523},
\bfpage{437}
(\byear{2015}).
doi:\doiurl{10.1038/nature14556}
\end{barticle}
\endbibitem

\bibitem[\protect\citeauthoryear{{Strong} et~al.}{1992}]{1992PASJ...44L.161S}
\begin{barticle}
\bauthor{\binits{K.T.} \bsnm{{Strong}}},
\bauthor{\binits{K.} \bsnm{{Harvey}}},
\bauthor{\binits{T.} \bsnm{{Hirayama}}}, \betal,
\batitle{{Observations of the variability of coronal bright points by the Soft X-ray Telescope on YOHKOH}}.
\bjtitle{\pasj}
\bvolume{44},
\bfpage{161}
(\byear{1992}).
\end{barticle}
\endbibitem

\bibitem[\protect\citeauthoryear{{Subramanian}
  et~al.}{2010}]{2010A&A...516A..50S}
\begin{barticle}
\bauthor{\binits{S.} \bsnm{{Subramanian}}},
\bauthor{\binits{M.S.} \bsnm{{Madjarska}}},
\bauthor{\binits{J.G.} \bsnm{{Doyle}}},
\batitle{{Coronal hole boundaries evolution at small scales. II. XRT view. Can
  small-scale outflows at CHBs be a source of the slow solar wind}}.
\bjtitle{\aap}
\bvolume{516},
\bfpage{50}
(\byear{2010}).
doi:\doiurl{10.1051/0004-6361/200913624}
\end{barticle}
\endbibitem


\bibitem[\protect\citeauthoryear{{Tian} et~al.}{2014}]{2014Sci...346A.315T}
\begin{barticle}
\bauthor{\binits{H.} \bsnm{{Tian}}},
\bauthor{\binits{E.~E.} \bsnm{{DeLuca}}},
\bauthor{\binits{S.~R.} \bsnm{{Cranmer}}}, \betal,
\batitle{{Prevalence of small-scale jets from the networks of the solar transition region and chromosphere}}.
\bjtitle{Science}
\bvolume{346},
\bfpage{1255711 }
(\byear{2014}).
doi:\doiurl{10.1126/science.1255711}
\end{barticle}
\endbibitem


\bibitem[\protect\citeauthoryear{{Titov} \& {D{\'e}moulin}}{1999}]{titov99}
\begin{barticle}
\bauthor{\binits{V.S.} \bsnm{{Titov}}},
\bauthor{\binits{P.} \bsnm{{D{\'e}moulin}}},
\batitle{{Basic topology of twisted magnetic configurations in solar flares}}.
\bjtitle{\aap}
\bvolume{351},
\bfpage{707}
(\byear{1999}).
\end{barticle}
\endbibitem

\bibitem[\protect\citeauthoryear{{T{\"o}r{\"o}k} \& {Kliem}}{2005}]{2005ApJ...630L..97T}
\begin{barticle}
\bauthor{\binits{T.} \bsnm{{T{\"o}r{\"o}k}}},
\bauthor{\binits{B.} \bsnm{{Kliem}}},
\batitle{{Confined and Ejective Eruptions of Kink-unstable Flux Ropes}}.
\bjtitle{\apjl}
\bvolume{630},
\bfpage{97}
(\byear{2005}).
doi:\doiurl{10.1086/462412}
\end{barticle}
\endbibitem

\bibitem[\protect\citeauthoryear{{T{\"o}r{\"o}k}
  et~al.}{2009}]{2009ApJ...704..485T}
\begin{barticle}
\bauthor{\binits{T.} \bsnm{{T{\"o}r{\"o}k}}},
\bauthor{\binits{G.} \bsnm{{Aulanier}}},
\bauthor{\binits{B.} \bsnm{{Schmieder}}}, \betal,
\batitle{{Fan-Spine Topology Formation Through Two-Step Reconnection Driven by
  Twisted Flux Emergence}}.
\bjtitle{\apj}
\bvolume{704},
\bfpage{485}
(\byear{2009}).
doi:\doiurl{10.1088/0004-637X/704/1/485}
\end{barticle}
\endbibitem

\bibitem[\protect\citeauthoryear{{T{\"o}r{\"o}k} et~al.}{2016}]{torok15}
\begin{botherref}
\bauthor{\binits{T.} \bsnm{{T{\"o}r{\"o}k}}},
\bauthor{\binits{R.} \bsnm{{Lionello}}},
\bauthor{\binits{V.S.} \bsnm{{Titov}}}, et al.,
\batitle{Modeling Jets in the Corona and Solar Wind},
\bsertitle{in {\it{Ground-based Solar Observations in the Space Instrumentation}}. ASP Conf. Ser.},
vol. \bseriesno{504 (2016)},
p. \bfpage{187}
\end{botherref}
\endbibitem

\bibitem[\protect\citeauthoryear{{T{\'o}th} et~al.}{2012}]{2012JCoPh.231..870T}
\begin{barticle}
\bauthor{\binits{G.} \bsnm{{T{\'o}th}}},
\bauthor{\binits{B.} \bsnm{{van der Holst}}},
\bauthor{\binits{I.V.} \bsnm{{Sokolov}}}, \betal,
\batitle{{Adaptive numerical algorithms in space weather modeling}}.
\bjtitle{J. Comput. Phys.}
\bvolume{231},
\bfpage{870}--\blpage{903}
(\byear{2012}).
doi:\doiurl{10.1016/j.jcp.2011.02.006}
\end{barticle}
\endbibitem

\bibitem[\protect\citeauthoryear{{Tsiropoula}
  et~al.}{2012}]{2012SSRv..169..181T}
\begin{barticle}
\bauthor{\binits{G.} \bsnm{{Tsiropoula}}},
\bauthor{\binits{K.} \bsnm{{Tziotziou}}},
\bauthor{\binits{I.} \bsnm{{Kontogiannis}}}, \betal,
\batitle{{Solar Fine-Scale Structures. I. Spicules and Other Small-Scale,
  Jet-Like Events at the Chromospheric Level: Observations and Physical
  Parameters}}.
\bjtitle{\ssr}
\bvolume{169},
\bfpage{181}
(\byear{2012}).
doi:\doiurl{10.1007/s11214-012-9920-2}
\end{barticle}
\endbibitem

\bibitem[\protect\citeauthoryear{{Tsuneta} et~al.}{1991}]{1991SoPh..136...37T}
\begin{barticle}
\bauthor{\binits{S.} \bsnm{{Tsuneta}}},
\bauthor{\binits{L.} \bsnm{{Acton}}},
\bauthor{\binits{M.} \bsnm{{Bruner}}}, \betal,
\batitle{{The Soft X-ray Telescope for the SOLAR-A Mission}}.
\bjtitle{\solphys}
\bvolume{136},
\bfpage{37}
(\byear{1991}).
doi:\doiurl{10.1007/BF00151694}
\end{barticle}
\endbibitem


\bibitem[\protect\citeauthoryear{{van Ballegooijen} et~al.}{2000}]{2000ApJ...539..983V}
\begin{barticle}
\bauthor{\binits{A.A.} \bsnm{{van Ballegooijen}}},
\bauthor{\binits{E.R.} \bsnm{{Priest}}},
\bauthor{\binits{D.H.} \bsnm{{Mackay}}},
\batitle{{Mean Field Model for the Formation of Filament Channels on the Sun}}.
\bjtitle{\apj}
\bvolume{539},
\bfpage{983}
(\byear{2000}).
doi:\doiurl{10.1086/309265}
\end{barticle}
\endbibitem

\bibitem[\protect\citeauthoryear{{van Ballegooijen}}{2004}]{2004ApJ...612..519V}
\begin{barticle}
\bauthor{\binits{A.A.} \bsnm{{van Ballegooijen}}},
\batitle{{Observations and Modeling of a Filament on the Sun}}.
\bjtitle{\apj}
\bvolume{612},
\bfpage{519}
(\byear{2004}).
doi:\doiurl{10.1086/422512}
\end{barticle}
\endbibitem

\bibitem[\protect\citeauthoryear{{von Rosenvinge}
  et~al.}{1995}]{1995SSRv...71..155V}
\begin{barticle}
\bauthor{\binits{T.T.} \bsnm{{von Rosenvinge}}},
\bauthor{\binits{L.M.} \bsnm{{Barbier}}},
\bauthor{\binits{J.} \bsnm{{Karsch}}}, \betal,
\batitle{{The Energetic Particles: Acceleration, Composition, and Transport
  (EPACT) investigation on the WIND spacecraft}}.
\bjtitle{\ssr}
\bvolume{71},
\bfpage{155}
(\byear{1995}).
doi:\doiurl{10.1007/BF00751329}
\end{barticle}
\endbibitem

\bibitem[\protect\citeauthoryear{{Wang}}{1994}]{1994ApJ...435L.153W}
\begin{barticle}
\bauthor{\binits{Y.-M.} \bsnm{{Wang}}},
\batitle{{Polar plumes and the solar wind}}.
\bjtitle{\apjl}
\bvolume{435},
\bfpage{153}
(\byear{1994}).
doi:\doiurl{10.1086/187617}
\end{barticle}
\endbibitem

\bibitem[\protect\citeauthoryear{{Wang}}{1998}]{1998ApJ...501L.145W}
\begin{barticle}
\bauthor{\binits{Y.-M.} \bsnm{{Wang}}},
\batitle{{Network Activity and the Evaporative Formation of Polar Plumes}}.
\bjtitle{\apjl}
\bvolume{501},
\bfpage{145}
(\byear{1998}).
doi:\doiurl{10.1086/311445}
\end{barticle}
\endbibitem

\bibitem[\protect\citeauthoryear{{Wang} et~al.}{1998}]{1998ApJ...508..899W}
\begin{barticle}
\bauthor{\binits{Y.-M.} \bsnm{{Wang}}},
\bauthor{\binits{N.R.} \bsnm{{Sheeley}} \bsuffix{Jr.}},
\bauthor{\binits{D.G.} \bsnm{{Socker}}}, \betal,
\batitle{{Observations of Correlated White-Light and Extreme-Ultraviolet Jets
  from Polar Coronal Holes}}.
\bjtitle{\apj}
\bvolume{508},
\bfpage{899}
(\byear{1998}).
doi:\doiurl{10.1086/306450}
\end{barticle}
\endbibitem

\bibitem[\protect\citeauthoryear{{Wang} \& {Sheeley}}{2002}]{2002ApJ...575..542W}
\begin{barticle}
\bauthor{\binits{Y.-M.} \bsnm{{Wang}}},
\bauthor{\binits{N.R.} \bsnm{{Sheeley}} \bsuffix{Jr.}},
\batitle{{Coronal White-Light Jets near Sunspot Maximum}}.
\bjtitle{\apj}
\bvolume{575},
\bfpage{542}
(\byear{2002}).
doi:\doiurl{10.1086/341145}
\end{barticle}
\endbibitem

\bibitem[\protect\citeauthoryear{{Wang} et~al.}{2006}]{2006ApJ...639..495W}
\begin{barticle}
\bauthor{\binits{Y.-M.} \bsnm{{Wang}}},
\bauthor{\binits{M.} \bsnm{{Pick}}},
\bauthor{\binits{G.M.} \bsnm{{Mason}}},
\batitle{{Coronal Holes, Jets, and the Origin of $^{3}$He-rich Particle Events}}.
\bjtitle{\apj}
\bvolume{639},
\bfpage{495}
(\byear{2006}).
doi:\doiurl{10.1086/499355}
\end{barticle}
\endbibitem

\bibitem[\protect\citeauthoryear{{Wang} \& {Liu}}{2012}]{2012ApJ...760..101W}
\begin{barticle}
\bauthor{\binits{H.} \bsnm{{Wang}}},
\bauthor{\binits{C.} \bsnm{{Liu}}},
\batitle{{Circular Ribbon Flares and Homologous Jets}}.
\bjtitle{\apj}
\bvolume{760},
\bfpage{101}
(\byear{2012}).
doi:\doiurl{10.1088/0004-637X/760/2/101}
\end{barticle}
\endbibitem

\bibitem[\protect\citeauthoryear{{Wilhelm} et~al.}{1995}]{1995SoPh..162..189W}
\begin{barticle}
\bauthor{\binits{K.} \bsnm{{Wilhelm}}},
\bauthor{\binits{W.} \bsnm{{Curdt}}},
\bauthor{\binits{E.} \bsnm{{Marsch}}}, \betal,
\batitle{{SUMER - Solar Ultraviolet Measurements of Emitted Radiation}}.
\bjtitle{\solphys}
\bvolume{162},
\bfpage{189}
(\byear{1995}).
doi:\doiurl{10.1007/BF00733430}
\end{barticle}
\endbibitem

\bibitem[\protect\citeauthoryear{{Wilhelm} et~al.}{2002a}]{wilhelm02}
\begin{barticle}
\bauthor{\binits{K.} \bsnm{{Wilhelm}}},
\bauthor{\binits{I.E.} \bsnm{{Dammasch}}},
\bauthor{\binits{D.M.} \bsnm{{Hassler}}},
\batitle{{Transition region and coronal plasmas: instrumentation and spectral analysis}}.
\bjtitle{\apss}
\bvolume{282},
\bfpage{189}
(\byear{2002}).
doi:\doiurl{10.1023/A:1021158705329}
\end{barticle}
\endbibitem

%
\bibitem[\protect\citeauthoryear{{Wilhelm} et~al.}{2002b}]{2002A&A...382..328W}
\begin{barticle}
\bauthor{\binits{K.} \bsnm{{Wilhelm}}},
\bauthor{\binits{B.} \bsnm{{Inhester}}},
\bauthor{\binits{J.S.} \bsnm{{Newmark}}},
\batitle{{The inner solar corona seen by SUMER, LASCO/C1, and EIT: Electron densities and temperatures during the rise of the new solar cycle}}.
\bjtitle{\aap}
\bvolume{382},
\bfpage{328}
(\byear{2002}).
doi:\doiurl{10.1051/0004-6361:20011608}
\end{barticle}
\endbibitem
%

\bibitem[\protect\citeauthoryear{{Wilhelm}}{2006}]{2006A&A...455..697W}
\begin{barticle}
\bauthor{\binits{K.} \bsnm{{Wilhelm}}},
\batitle{{Solar coronal-hole plasma densities and temperatures}}.
\bjtitle{\aap}
\bvolume{455},
\bfpage{697}
(\byear{2006}).
doi:\doiurl{10.1051/0004-6361:20054693}
\end{barticle}
\endbibitem

\bibitem[\protect\citeauthoryear{{Wilhelm} et~al.}{2011}]{wilhelm11}
\begin{barticle}
\bauthor{\binits{K.} \bsnm{{Wilhelm}}},
\bauthor{\binits{L.} \bsnm{{Abbo}}},
\bauthor{\binits{F.} \bsnm{{Auchere}}}, \betal,
\batitle{{Morphology, dynamics and plasma parameters of plumes and inter-plume regions in solar coronal holes}}.
\bjtitle{\aapr}
\bvolume{19},
\bfpage{35}
(\byear{2011}).
doi:\doiurl{10.1007/s00159-011-0035-7}
\end{barticle}
\endbibitem

\bibitem[\protect\citeauthoryear{{Withbroe} et~al.}{1976}]{1976ApJ...203..528W}
\begin{barticle}
\bauthor{\binits{G.L.} \bsnm{{Withbroe}}},
\bauthor{\binits{D.T.} \bsnm{{Jaffe}}},
\bauthor{\binits{P.V.} \bsnm{{Foukal}}}, \betal,
\batitle{{Extreme-ultraviolet transients observed at the solar pole}}.
\bjtitle{\apj}
\bvolume{203},
\bfpage{528}
(\byear{1976}).
doi:\doiurl{10.1086/154108}
\end{barticle}
\endbibitem

\bibitem[\protect\citeauthoryear{{Wood} et~al.}{1999}]{1999ApJ...523..444W}
\begin{barticle}
\bauthor{\binits{B.E.} \bsnm{{Wood}}},
\bauthor{\binits{M.} \bsnm{{Karovska}}},
\bauthor{\binits{J.W.} \bsnm{{Cook}}}, \betal,
\batitle{{Kinematic Measurements of Polar Jets Observed by the Large-Angle Spectrometric Coronagraph}}.
\bjtitle{\apj}
\bvolume{523},
\bfpage{444}
(\byear{1999}).
doi:\doiurl{10.1086/307721}
\end{barticle}
\endbibitem

\bibitem[\protect\citeauthoryear{{Wuelser} et~al.}{2004}]{2004SPIE.5171..111W}
\begin{bchapter}
\bauthor{\binits{J.-P.} \bsnm{{Wuelser}}},
\bauthor{\binits{J.R.} \bsnm{{Lemen}}},
\bauthor{\binits{T.D.} \bsnm{{Tarbell}}}, \betal,
\bctitle{{EUVI: the STEREO-SECCHI extreme ultraviolet imager}},
\bsertitle{SPIE Conf. Ser.},
vol. \bseriesno{5171},
pp. \bfpage{111}--\blpage{122}
(\byear{2004}).
doi:\doiurl{10.1117/12.506877}
\end{bchapter}
\endbibitem

\bibitem[\protect\citeauthoryear{{Xu} et~al.}{1984}]{1984ChA&A...8..294X}
\begin{barticle}
\bauthor{\binits{A.-A.} \bsnm{{Xu}}},
\bauthor{\binits{J.-P.} \bsnm{{Ding}}},
\bauthor{\binits{S.-Y.} \bsnm{{Yin}}},
\batitle{{Rotating motion in solar surges}}.
\bjtitle{ChA\&A}
\bvolume{8},
\bfpage{294}
(\byear{1984}).
doi:\doiurl{10.1016/0275-1062(84)90056-0}
\end{barticle}
\endbibitem

\bibitem[\protect\citeauthoryear{{Yang} et~al.}{2011}]{yang11}
\begin{barticle}
\bauthor{\binits{S.} \bsnm{{Yang}}},
\bauthor{\binits{J.} \bsnm{{Zhang}}},
\bauthor{\binits{T.} \bsnm{{Li}}},
\bauthor{\binits{Y.} \bsnm{{Liu}}},
\batitle{{SDO Observations of Magnetic Reconnection At Coronal Hole
  Boundaries}}.
\bjtitle{\apjl}
\bvolume{732},
\bfpage{7}
(\byear{2011}).
doi:\doiurl{10.1088/2041-8205/732/1/L7}
\end{barticle}
\endbibitem

\bibitem[\protect\citeauthoryear{{Yang} et~al.}{2013}]{2013ApJ...777...16Y}
\begin{barticle}
\bauthor{\binits{L.} \bsnm{{Yang}}},
\bauthor{\binits{J.} \bsnm{{He}}},
\bauthor{\binits{H.} \bsnm{{Peter}}}, \betal,
\batitle{{Numerical Simulations of Chromospheric Anemone Jets Associated with
  Moving Magnetic Features}}.
\bjtitle{\apj}
\bvolume{777},
\bfpage{16}
(\byear{2013}).
doi:\doiurl{10.1088/0004-637X/777/1/16}
\end{barticle}
\endbibitem


\bibitem[\protect\citeauthoryear{{Yashiro} et~al.}{2004}]{2004ASPC..325..401Y}
\begin{bchapter}
\bauthor{\binits{S.} \bsnm{{Yashiro}}},
\bauthor{\binits{N.} \bsnm{{Gopalswamy}}},
\bauthor{\binits{E.W.} \bsnm{{Cliver}}}, \betal,
\bctitle{{Association of Coronal Mass Ejections and Type II Radio Bursts with
  Impulsive Solar Energetic Particle Events}},
\bsertitle{ASP Conf. Ser.},
vol. \bseriesno{325},
p. \bfpage{401}
(\byear{2004}).
\end{bchapter}
\endbibitem

\bibitem[\protect\citeauthoryear{{Yokoyama} \& {Shibata}}{1994}]{1994ApJ...436L.197Y}
\begin{barticle}
\bauthor{\binits{T.} \bsnm{{Yokoyama}}},
\bauthor{\binits{K.} \bsnm{{Shibata}}},
\batitle{{What is the condition for fast magnetic reconnection?}}
\bjtitle{\apjl}
\bvolume{436},
\bfpage{197}
(\byear{1994}).
doi:\doiurl{10.1086/187666}
\end{barticle}
\endbibitem

\bibitem[\protect\citeauthoryear{{Yokoyama} \& {Shibata}}{1995}]{1995Natur.375...42Y}
\begin{barticle}
\bauthor{\binits{T.} \bsnm{{Yokoyama}}},
\bauthor{\binits{K.} \bsnm{{Shibata}}},
\batitle{{Magnetic reconnection as the origin of X-ray jets and H{$\alpha$}
  surges on the Sun}}.
\bjtitle{\nat}
\bvolume{375},
\bfpage{42}
(\byear{1995}).
doi:\doiurl{10.1038/375042a0}
\end{barticle}
\endbibitem

\bibitem[\protect\citeauthoryear{{Yokoyama} \& {Shibata}}{1996}]{1996PASJ...48..353Y}
\begin{barticle}
\bauthor{\binits{T.} \bsnm{{Yokoyama}}},
\bauthor{\binits{K.} \bsnm{{Shibata}}},
\batitle{{Numerical Simulation of Solar Coronal X-Ray Jets Based on the
  Magnetic Reconnection Model}}.
\bjtitle{\pasj}
\bvolume{48},
\bfpage{353}
(\byear{1996}).
doi:\doiurl{10.1093/pasj/48.2.353}
\end{barticle}
\endbibitem

\bibitem[\protect\citeauthoryear{{Yokoyama} \& {Shibata}}{2001}]{2001ApJ...549.1160Y}
\begin{barticle}
\bauthor{\binits{T.} \bsnm{{Yokoyama}}},
\bauthor{\binits{K.} \bsnm{{Shibata}}},
\batitle{{Magnetohydrodynamic Simulation of a Solar Flare with Chromospheric
  Evaporation Effect Based on the Magnetic Reconnection Model}}.
\bjtitle{\apj}
\bvolume{549},
\bfpage{1160}
(\byear{2001}).
doi:\doiurl{10.1086/319440}
\end{barticle}
\endbibitem

\bibitem[\protect\citeauthoryear{{Young} et~al.}{2007}]{young07-eis}
\begin{barticle}
\bauthor{\binits{P.R.} \bsnm{{Young}}},
\bauthor{\binits{G.} \bsnm{{Del Zanna}}},
\bauthor{\binits{H.E.} \bsnm{{Mason}}}, \betal,
\batitle{{EUV Emission Lines and Diagnostics Observed with Hinode/EIS}}.
\bjtitle{\pasj}
\bvolume{59},
\bfpage{857}
(\byear{2007}).
\end{barticle}
\endbibitem

\bibitem[\protect\citeauthoryear{{Young} \& {Muglach}}{2014a}]{2014SoPh..289.3313Y}
\begin{barticle}
\bauthor{\binits{P.R.} \bsnm{{Young}}},
\bauthor{\binits{K.} \bsnm{{Muglach}}},
\batitle{{Solar Dynamics Observatory and Hinode Observations of a Blowout Jet
  in a Coronal Hole}}.
\bjtitle{\solphys}
\bvolume{289},
\bfpage{3313}
(\byear{2014}a).
doi:\doiurl{10.1007/s11207-014-0484-z}
\end{barticle}
\endbibitem

\bibitem[\protect\citeauthoryear{{Young} \& {Muglach}}{2014b}]{2014PASJ...66S..12Y}
\begin{barticle}
\bauthor{\binits{P.R.} \bsnm{{Young}}},
\bauthor{\binits{K.} \bsnm{{Muglach}}},
\batitle{{A coronal hole jet observed with Hinode and the Solar Dynamics
  Observatory}}.
\bjtitle{\pasj}
\bvolume{66},
\bfpage{12}
(\byear{2014}b).
doi:\doiurl{10.1093/pasj/psu088}
\end{barticle}
\endbibitem

\bibitem[\protect\citeauthoryear{{Young}}{2015}]{2015ApJ...801..124Y}
\begin{barticle}
\bauthor{\binits{P.R.} \bsnm{{Young}}},
\batitle{{Dark Jets in Solar Coronal Holes}}.
\bjtitle{\apj}
\bvolume{801},
\bfpage{124}
(\byear{2015}).
doi:\doiurl{10.1088/0004-637X/801/2/124}
\end{barticle}
\endbibitem

\bibitem[\protect\citeauthoryear{{Yu} et~al.}{2014}]{2014ApJ...784..166Y}
\begin{barticle}
\bauthor{\binits{H.-S.} \bsnm{{Yu}}},
\bauthor{\binits{B.V.} \bsnm{{Jackson}}},
\bauthor{\binits{A.} \bsnm{{Buffington}}}, \betal,
\batitle{{The Three-dimensional Analysis of Hinode Polar Jets using Images from
  LASCO C2, the Stereo COR2 Coronagraphs, and SMEI}}.
\bjtitle{\apj}
\bvolume{784},
\bfpage{166}
(\byear{2014}).
doi:\doiurl{10.1088/0004-637X/784/2/166}
\end{barticle}
\endbibitem

\bibitem[\protect\citeauthoryear{{Zhang} et~al.}{2012}]{2012ApJ...746...19Z}
\begin{barticle}
\bauthor{\binits{Q.M.} \bsnm{{Zhang}}},
\bauthor{\binits{P.F.} \bsnm{{Chen}}},
\bauthor{\binits{Y.} \bsnm{{Guo}}}, \betal,
\batitle{{Two Types of Magnetic Reconnection in Coronal Bright Points and the
  Corresponding Magnetic Configuration}}.
\bjtitle{\apj}
\bvolume{746},
\bfpage{19}
(\byear{2012}).
doi:\doiurl{10.1088/0004-637X/746/1/19}
\end{barticle}
\endbibitem


\bibitem[\protect\citeauthoryear{{Zhang} \& {Ji}}{2014a}]{2014A&A...567A..11Z}
\begin{barticle}
\bauthor{\binits{Q.M.} \bsnm{{Zhang}}},
\bauthor{\binits{H.S.} \bsnm{{Ji}}},
\batitle{{Blobs in recurring extreme-ultraviolet jets}}.
\bjtitle{\aap}
\bvolume{567},
\bfpage{11}
(\byear{2014}a).
doi:\doiurl{10.1051/0004-6361/201423698}
\end{barticle}
\endbibitem

\bibitem[\protect\citeauthoryear{{Zhang} \& {Ji}}{2014b}]{2014A&A...561A.134Z}
\begin{barticle}
\bauthor{\binits{Q.M.} \bsnm{{Zhang}}},
\bauthor{\binits{H.S.} \bsnm{{Ji}}},
\batitle{{A swirling flare-related EUV jet}}.
\bjtitle{\aap}
\bvolume{561},
\bfpage{134}
(\byear{2014}b).
doi:\doiurl{10.1051/0004-6361/201322616}
\end{barticle}
\endbibitem


\end{thebibliography}

\end{document}